\documentclass[12pt]{article}

\usepackage{amsmath,cite} 

\def\elabel#1{\label{#1}}

\DeclareSymbolFont{AMSb}{U}{msb}{m}{n}
\DeclareSymbolFontAlphabet{\Bbb}{AMSb}

\renewcommand{\thesection}{\Roman{section}}
\renewcommand{\theequation}{\arabic{section}.\arabic{equation}}
\def\rsen{\setcounter{equation}{0}}

\newcommand{\startappendix}{
\setcounter{section}{0}
\renewcommand{\thesection}{\Alph{section}}
\renewcommand{\theequation}{\Alph{section}.\arabic{equation}}}

\newcommand{\Appendix}[1]{
\refstepcounter{section}
\begin{flushleft}
{\large\bf Appendix \thesection: #1}
\end{flushleft}}

\newcommand{\aD}{{\dot\alpha}}

\newcommand{\bD}{{\dot\beta}}
\newcommand{\U}{{U}}
\newcommand{\SU}{{SU}}
\newcommand{\SO}{{SO}}
\newcommand{\BX}{\boldsymbol{X}}

\newcommand{\BL}{{\bf L}}
\newcommand{\Bk}{{\boldsymbol{k}}}
\newcommand{\BN}{{\boldsymbol{N}}}

\def\com{X}
\def\rmf{{\rm f}}
\def\rmb{{\rm b}}
\def\sfc{\hat\Omega}
\def\grp{\Upsilon}

\def\N{{\cal N}}
\def\det{{\rm det}}
\def\susic{supersymmetric}
\def\susy{supersymmetry}
\def\adss{AdS_5\times S^5}

\def\G{{G}}
\def\M{{\cal M}}
\def\tr{{\rm tr}}
\def\trtwo{\tr^{}_2\,}
\def\Mbar{\bar{\cal M}}
\def\dalpha{{\dot\alpha}}
\def\dbeta{{\dot\beta}}
\def\dgamma{{\dot\gamma}}

\def\Skinst{S^k_{\rm inst}}
\def\dmuphys{d\mu^{k}_{\rm phys}}
\def\Skquad{S^k_{\rm quad}}
\def\sqrtwo{\sqrt{2}\,}
\def\hf{{\textstyle{1\over2}}}
\def\wbar{\bar w}
\def\mubar{\bar\mu}
\def\abar{\bar a}
\def\sigmabar{\bar\sigma}
\def\etabar{\bar\eta}
\def\zetabar{\bar\zeta}
\def\mubar{\bar\mu}
\def\nubar{\bar\nu}

\def\calG{\cal G}
\newcommand{\Rowspace}{\phantom{$\Big($}}

\begin{document}

\addtolength{\baselineskip}{2pt}
\thispagestyle{empty}

\begin{flushright}
{\tt hep-th/9901128}\\
January 1999
\end{flushright}


\begin{center}
{\scshape\Large  Multi-Instanton~Calculus\\
\vspace{0.15cm}
and~the~A{\rm d}S/CFT~Correspondence\\
\vspace{0.15cm}
in~${\cal N}{=}4$~Superconformal~Field~Theory\\}

\vspace{1.2cm}

{\scshape Nicholas Dorey$^{1,4}$, Timothy J.~Hollowood$^{2,4}$,
Valentin V.~Khoze$^3$,\\ Michael
P.~Mattis$^2$ and Stefan Vandoren$^4$}

\vspace{0.2cm}
$^1${\sl Physics Department, University of Washington,\\
Seattle, WA 98195, USA}\hspace{0.6cm} {\tt
dorey@phys.washington.edu}\\

\vspace{0.15cm}
$^2${\sl Theoretical Division T-8, Los Alamos National Laboratory,\\
Los Alamos, NM 87545, USA} \\
{\tt mattis@lanl.gov}, {\tt pyth@schwinger.lanl.gov}\\

\vspace{0.15cm}
$^3${\sl Department of Physics, University of Durham,\\
Durham, DH1 3LE, UK}\hspace{0.6cm} {\tt valya.khoze@durham.ac.uk}\\

\vspace{0.15cm}
$^4${\sl Department of Physics, University of Wales Swansea,\\
Swansea, SA2 8PP, UK}\hspace{0.6cm} {\tt pysv@swan.ac.uk}\\

\end{center}
\vspace{0.3cm}
\noindent We present a self-contained study  of ADHM multi-instantons
in $SU(N)$ gauge theory, especially  the novel interplay with
supersymmetry and the large-$N$ limit. We give both field- and
string-theoretic derivations of the $\N=4$ supersymmetric
 multi-instanton action and
collective coordinate integration measure. As a central application,
 we focus on
certain $n$-point functions $G_n$, $n=16,$ 8 or 4,  in ${\cal N}=4$ 
$SU(N)$ gauge theory at the conformal point (as well as on related
higher-partial-wave correlators); these are correlators in which the
16 exact supersymmetric and superconformal fermion zero modes are
saturated.   In the large-$N$ limit,
for the first time in any 4-dimensional  theory,
we are able to evaluate all leading-order multi-instanton contributions
exactly. We find compelling evidence for Maldacena's conjecture:
\hbox{(1) The} large-$N$ $k$-instanton collective coordinate space has
the geometry of a single copy of $AdS_5\times S^5\,$.  
(2) The integration measure on this space includes the partition
function of 10-dimensional ${\cal N}=1$  $SU(k)$ gauge theory 
 dimensionally reduced to 0 dimensions, matching the description of
D-instantons in Type IIB string theory.
\hbox{(3) In}
exact agreement with Type IIB string calculations, at the
$k$-instanton level, $G_n =
\sqrt{N}\,g^8\,k^{n-7/2}e^{2\pi ik\tau}\sum_{d|k}\,d^{-2}\cdot
F_n(x_1,\ldots,x_n)$, where $F_n$ is identical to a convolution of $n$
bulk-to-boundary supergravity propagators.

\newpage

\section*{Contents}
\def\SP{\phantom{a}}

\contentsline {section}{\numberline {\uppercase {i}}\SP Introduction}{3}
\contentsline {subsection}{\numberline {\uppercase {i}.1}\SP Overview of the paper}{3}
\contentsline {subsection}{\numberline {\uppercase {i}.2}\SP Review of the superstring prediction}{11}
\contentsline {subsection}{\numberline {\uppercase {i}.3}\SP Review of the Yang-Mills calculation at the one-instanton level}{15}
\contentsline {section}{\numberline {\uppercase {ii}}\SP The ${\cal N}=4$ Multi-Instanton Supermultiplet}{18}
\contentsline {subsection}{\numberline {\uppercase {ii}.1}\SP Construction of the classical gauge field}{19}
\contentsline {subsection}{\numberline {\uppercase {ii}.2}\SP Constraints, collective coordinates and canonical forms}{21}
\contentsline {subsection}{\numberline {\uppercase {ii}.3}\SP Asymptotics of the multi-instanton}{24}
\contentsline {subsection}{\numberline {\uppercase {ii}.4}\SP Connection to the usual one-instanton collective coordinates, and the dilute instanton gas limit}{25}
\contentsline {subsection}{\numberline {\uppercase {ii}.5}\SP Construction of the adjoint fermion zero modes}{27}
\contentsline {subsection}{\numberline {\uppercase {ii}.6}\SP Classification and overlap formula for the fermion zero modes}{28}
\contentsline {subsection}{\numberline {\uppercase {ii}.7}Construction of the adjoint Higgs bosons}{31}
\contentsline {section}{\numberline {\uppercase {iii}}\SP Construction of the Multi-Instanton Action}{33}
\contentsline {section}{\numberline {\uppercase {iv}}\SP The Multi-Instanton Collective Coordinate Integration Measure}{37}
\contentsline {subsection}{\numberline {\uppercase {iv}.1}\SP The ADHM multi-instanton measure}{38}
\contentsline {subsection}{\numberline {\uppercase {iv}.2}\SP D-instantons and the ADHM Measure}{41}
\contentsline {subsection}{\numberline {\uppercase {iv}.3}\SP The gauge-invariant measure}{51}
\contentsline {section}{\numberline {\uppercase {v}}\SP The Large-$N$ Limit in a Saddle-Point Approximation}{57}
\contentsline {subsection}{\numberline {\uppercase {v}.1}\SP The one-instanton measure revisited, and the emergence of the $S^5$}{57}
\contentsline {subsection}{\numberline {\uppercase {v}.2}\SP Na\"\i vely the $k$-instanton moduli space contains $k$ copies of $AdS_5\times S^5$}{58}
\contentsline {subsection}{\numberline {\uppercase {v}.3}\SP The exact $k$-instanton moduli space collapses to one copy of $AdS_5\times S^5$}{60}
\contentsline {subsection}{\numberline {\uppercase {v}.4}\SP Comments on the ten-dimensional $SU(k)$ partition function}{67}
\contentsline {section}{\numberline {\uppercase {vi}}\SP Large-$N$ Instanton Correlation Functions}{69}
\contentsline {subsection}{\numberline {\uppercase {vi}.1}\SP Multi-instanton contributions to the correlators $G_n$}{70}
\contentsline {subsection}{\numberline {\uppercase {vi}.2}\SP Kaluza-Klein modes on $S^5$ and Yang-Mills correlators}{72}
\contentsline {section}{\numberline {\uppercase {vii}}\SP Comments and Conclusions}{75}
\contentsline {subsection}{\numberline {\uppercase {vii}.1}\SP Living large in a large-$N$ world}{75}
\contentsline {subsection}{\numberline {\uppercase {vii}.2}\SP Comments on the nonrenormalization theorem, and on higher-order corrections in $1/N$ and in $g^2$}{77}
\contentsline {section}{\numberline \SP Acknowledgments}{79}
\contentsline {section}{\numberline \SP Appendix A: Clifford Algebras and fermions in 4, 6 and 10 dimensions}{79}
\contentsline {section}{\numberline \SP References}{81}

\def\gst{g_{st}}
\def\Skinst{S^k_{\rm inst}}
\def\dmuphys{d\mu^k_{\rm phys}}
\def\N{{\cal N}}

\section{Introduction}

\subsection{Overview of the paper}

The $1/N$ expansion proposed by 't Hooft twenty-five years ago
\cite{TH1} continues to offer the tantalizing prospect of a tractable
approximation scheme for QCD in which confinement is visible at
leading order. Although this prospect is still a long way out of reach
for QCD itself, the last year has seen significant progress in
understanding the large-$N$ limit in the more controlled context of
non-abelian gauge theories with extended supersymmetry. In particular,
Maldacena has conjectured \cite{MAL} that the large-$N$ limit of
${\cal N}=4$ supersymmetric Yang-Mills with gauge group $SU(N)$, at
the conformal point of vanishing Higgs VEVs, is dual to weakly
coupled Type IIB superstring theory on an $AdS_{5}\times S^{5}$
background. This
provides a realization of the long-anticipated connection between the
$1/N$ expansion, which is organized as a sum over Feynman diagrams of
different topology, on the one hand, and string perturbation theory,
which is a sum over string world-sheets of different topology, on the
other.  In particular, the gauge coupling $g$ and vacuum angle $\theta$
of the four-dimensional theory are given in terms of the string
parameters by 
\begin{equation}
g\ =\
\sqrt{4\pi\gst}\ =\ \sqrt{4\pi e^\phi}\ ,\qquad \theta\ =\ 2\pi
c^{(0)}\ .
\elabel{corresp}
\end{equation}
Here $g_{st}$ is the string coupling while $c^{(0)}$ is the
expectation value of the Ramond-Ramond scalar of Type IIB string
theory. Also $N$ appears explicitly, through the relation
\begin{equation}
{L^2\over\alpha'} \ =\ \sqrt{g^2\,N}
\elabel{alphadef}
\end{equation}
 where
$(\alpha')^{-1}$ is the string tension and $L$ is the radius of both
the $AdS_5$ and $S^5$ factors of the background. 

More generally, there is a precise correspondence between the 
finite mass string states in an $AdS_{5}\times S^{5}$ background
and the gauge-invariant composite operators on the Yang-Mills side,
and hence, a proposed equivalence between all correlators in the two
theories that are built from these operators
\cite{GKP,WIT150,FFone,AFone,HO,FZ1}. In this paper we will focus on a class
of such correlators which are known to receive contributions from all
orders in D-instantons, on the superstring side, and from
all orders in
Yang-Mills instantons, on the gauge theory side. Moreover, on the
superstring side, these contributions are known to sum to a specific modular
form of the  complexified coupling constants $\tau$ and $\bar\tau$,
with
\begin{equation}
\tau\ =\ ie^{-\phi}+c^{(0)}\ \equiv\ {4\pi i\over
g^2}+{\theta\over2\pi}\ .
\elabel{taudef}
\end{equation}
Below we will calculate these contributions from first principles in
the $\N=4$ theory at large $N$.  Specifically, for each topological
number $k$, we will extract the leading semiclassical contribution to
these correlators. This is the first time in a four-dimensional theory
that the instanton series has been explicitly evaluated to all
orders. Our calculation involves a novel interplay between
supersymmetry, the large-$N$ limit, and the multi-instanton formalism
of Atiyah, Drinfeld, Hitchin and Manin (ADHM) \cite{ADHM}. For each
$k$, we obtain perfect quantitative and qualitative agreement with the
supergravity result, although, as described below, there is an
outstanding puzzle about the differing regimes of validity of the
supergravity and gauge theory calculations.  Quantitatively, we
precisely recover the $k^{\rm th}$ Taylor coefficient of the predicted
modular form.  And qualitatively, we discover how the ten-dimensional
$AdS_5\times S^5$ space emerges, in a surprising way, from the
four-dimensional picture: this space describes the geometry of the
subspace of multi-instanton collective coordinates which dominates the
path integral in the large-$N$ limit.  Moreover, the integration
measure on this space includes the partition function of
ten-dimensional ${\cal N}=1$ supersymmetric $SU(k)$ gauge theory
dimensionally reduced to zero dimensions, matching the description of
D-instantons in Type IIB string theory.  In our view, this
multi-faceted agreement constitutes compelling circumstantial evidence
in favor of Maldacena's conjecture.

An abbreviated account of our calculation was presented in
Ref.~\cite{LETT}.  The present paper not only presents the
calculational details of Ref.~\cite{LETT} in a self-contained manner,
but several additional results as well, described below.

Let us comment further on both sides of the proposed correspondence.
Unfortunately, little is known about the string theory on
$AdS_{5}\times S^{5}$, even at leading order in perturbation theory,
due to the presence of background Ramond-Ramond fields. To make
progress, it is necessary to focus on the regime where the radius of
curvature of the background is large compared to the string length
scale (i.e., small $\alpha'$) in addition to weak coupling (i.e.,
small $\gst$). In this case the IIB string theory on $AdS_{5}\times
S^{5}$ is well approximated by classical IIB supergravity in the same
background.  On the gauge theory side of the conjecture, per
Eqs.~\eqref{corresp}-\eqref{alphadef}, this regime corresponds to a large-$N$
limit with $g^2\ll1$ but $g^{2}N\gg1$. Recall that the 't~Hooft
coupling $g^2N$ is the effective expansion parameter of the large-$N$
gauge theory \cite{TH1}.  Thus, via Maldacena's conjecture, classical
supergravity yields predictions for the strong-coupling behavior of
the four-dimensional theory. Stringy and quantum corrections to classical
supergravity correspond to corrections in powers of $(g^{2}N)^{-1}$
and $g^{2},$ respectively, on the gauge theory side.

On the four-dimensional side, the $\N=4$ models constitute a particularly
interesting class of quantum field theories in their own right. For
any value of $N$ and of the Higgs VEVs, they are finite theories, with
vanishing chiral anomaly and $\beta$-function \cite{N4finite}. They are also the
original setting for Olive-Montonen duality \cite{MOOS}; consequently,
the spectrum of monopoles and dyons can in some cases be computed exactly
\cite{SEN}.  Moreover, in the absence of such
VEVs, the symmetry group of the theory is enlarged both at the
classical and quantum levels: the model becomes a highly nontrivial
superconformal field theory, in an interesting and ill-understood
non-abelian Coulomb phase. 

Yet apart from the well-studied constraints imposed by superconformal
invariance for $n\leq3$, little is known about $n$-point functions in
this model for $n>3$, beyond the regime $g^2N\ll1$ where they are
amenable to standard perturbative and/or semiclassical
analysis.\footnote{The fact that planar diagram corrections correspond
to an expansion in $g^2N$ is true in instanton backgrounds as well as
around the perturbative vacuum.}  Maldacena's conjecture is of special
interest to field theory, precisely because (at least for large $N$)
it provides a quantitative means of extrapolating these correlators
into a domain where perturbation theory fails.  However this also
raises an important problem: it is very hard to find quantitative
tests of the conjecture against our existing knowledge of the ${\cal
N}=4$ theory. One way around this difficulty is to isolate a special
class of correlators which are protected against quantum corrections
by supersymmetry. The results of a weak coupling calculation can then
legitimately be extrapolated to strong coupling and compared with the
corresponding supergravity predictions.

The simplest examples of such supersymmetric nonrenormalization
theorems involve two- and three-point functions of certain chiral primary
operators \cite{WIT150,GKP,EFS,HW,OSII,INT}; these are highly constrained by the
superconformal invariance of the ${\cal N}=4$ theory and apparently are given
exactly by their classical free field values. However, experience has
shown that many supersymmetric theories contain a family of
correlators which, though similarly protected, are far less trivial.
Although the details vary from theory to theory, the characteristic
(but not sufficient)
property of these correlators is that they contain precisely the
correct number of fermions to saturate the exact zero modes of an
instanton. It is well-known that instanton 
contributions to the infra-red safe correlators of chiral superfields
are protected against further quantum corrections \cite{SV,Seib,Shifman,IS}.
Classic examples of exact results in $\N=1$ supersymmetry
include instanton derivation of the Affleck--Dine--Seiberg
superpotential \cite{Ads,Cordes} and of the NSVZ beta-function 
\cite{NSVZbeta}.
In theories with eight supercharges (both in $D=4$
\cite{Seib,SeibWitt,MO-I} and $D=3$ \cite{SW3D,DKMTV}), the quantities which
are one-loop-exact in each multi-instanton sector are
typically four-fermion correlators which come from the
two-derivative/four-fermion terms in a low-energy action. 
For the three-dimensional theory
with sixteen supercharges studied in \cite{PP,DKM3D} the relevant
correlator is related to an eight-fermion term in the effective action
for which a non-renormalization theorem has recently been proved
\cite{PSS}.  Other relevant examples involve the correlators of chiral
operators which are the observables of a topological quantum field
theory obtained by twisting the original supersymmetric gauge theory
(see \cite{VW}, in particular, for an application to the ${\cal N}=4$
theory, which may be related to the calculation presented below). 

In this paper we will study the analogous correlators in the $\N=4$
superconformal theory, although, unlike the previous examples, it is
not yet known to what extent these correlators are constrained by
supersymmetry.  In particular we will calculate the leading
semiclassical contribution of ADHM multi-instantons to a
sixteen-fermion correlator $G_{16}(x_1,\ldots,x_{16})$ in ${\cal N}=4$
supersymmetric $SU(N)$ gauge theory at large $N$ \cite{BG,BGKR} (see
also the review \cite{BK}).  The number of fermion insertions is
dictated by the sixteen exact zero modes of the instanton which are
protected by the supersymmetric and superconformal invariance of the
${\cal N}=4$ theory with vanishing VEVs (turning
on the VEVs reduces this number to eight). In addition, supersymmetry
relates the fermionic correlator $G_{16}(x_1,\ldots,x_{16})$ to
certain bosonic 8-point and 4-point functions $G_8(x_1,\ldots,x_8)$
and $G_4(x_1,\ldots,x_4)$ \cite{BGKR} in which the instanton zero
modes are saturated, respectively, by fermion-bilinear and
fermion-quadrilinear parts of bosonic fields.

By synthesizing Maldacena's conjecture with earlier results
\cite{GG1,GG2,GG3,GG4,GGK,KP1,KP2} about
D-instanton contributions to the IIB effective action, Banks and Green \cite{BG}
have obtained a closed-form supergravity prediction for the 
$4$-point correlator of the stress-tensor operators in the $\N=4$ 
superconformal Yang-Mills theory. Predictions for correlators
$G_n$ ($n=16,8$ or 4) of related operators\footnote{These operators
as well as the stress-tensor operator
correspond to different components of the Noether current
superfield associated with the superconformal and chiral $SU(4)$ 
transformations of the $\N=4$  theory written down in \cite{BRW,BGKR}. }
 were obtained in 
\cite{BGKR}.  The $G_n$ are
expressed in terms of a non-holomorphic modular form
$f_n(\tau,\bar\tau)$, where the modular group $SL(2,\Bbb Z)$ is the
$S$-duality symmetry group of the Type IIB string theory.  
Suggestively, the $f_n$ are amenable to a
simultaneous Taylor expansion in $g^2,$ $e^{2\pi i\tau}$ and $e^{-2\pi
i\bar\tau}$.  On the supergravity side, this amounts, respectively, to
perturbative, D-instanton, and anti-D-instanton effects. The
dictionary \eqref{corresp} then suggests that (anti-)D-instantons are
mapped onto Yang-Mills (anti-)instantons by the correspondence.  We
stress once again that the supergravity prediction was derived by
assuming large $g^{2}N$, while the semiclassical instanton calculation
presented below is valid only for small $g^{2}N$.  Despite this
caveat, as stated earlier, for each topological number $k$ we find
precise accord between our weak-coupling results and the corresponding
D-instanton terms in the Banks-Green prediction.  Above, we asserted
that this agreement for the $G_n$ (which extends to the whole tower
of Kaluza-Klein states of supergravity compactified on $S^5$, see Sec.~VI.2) constitutes
convincing evidence in favor of Maldacena's conjecture. In
addition---though this is admittedly circular reasoning---it can be
seen as equally compelling circumstantial evidence that a
supersymmetric nonrenormalization theorem does, in fact, apply to
these correlators too.

This paper is organized as follows. In the remainder of this
introductory section, we outline the supergravity
prediction for the $G_n$ (Sec.~I.2) \cite{BG,BGKR}. 
In particular, a Taylor expansion of the exact solution reveals a surprising and
under-appreciated aspect of the D-instanton moduli space; namely that
at any topological level $k$, it effectively contains only one copy---and not $k$
copies---of $AdS_5\times S^5$. This is contrary to what one usually
expects from the physics of D-branes: the collective dynamics of a
multi-charged configuration of D-branes is described by a non-abelian
$U(k)$ Yang-Mills theory on the brane world-volume. The Coulomb branch
of this gauge theory describes the freedom for the branes to separate
from one another. However, it is more appropriate to interpret the 
D-instanton contribution to the correlation functions $G_n$, as being
due to a charge $k$ D-instanton ``bound state''. This is due to the
surprising fact that the integrals over the coordinates, which one
would ordinarily identify as the relative positions of the
singly-charged D-objects, are actually convergent \cite{GG2,GG3,GG4}. This
fact will be seen to play a similarly crucial r\^ole on the Yang-Mills
side of the story, as we elucidate in Sec.~V. In Sec.~I.3, we review
the corresponding Yang-Mills calculation at the one-instanton level,
and show how it matches the supergravity result, both in its
space-time dependence \cite{BGKR}, and in the overall coupling
strength $(\alpha')^{-1}\sim\sqrt{N}$ \cite{DKMV}. Section I ends with
a puzzle: na\"\i vely, this agreement appears to be an accident of the
one-instanton level, for only at this level does an instanton 
field-strength squared
resemble a Euclidean supergravity bulk-to-boundary 
propagator. Indeed, even if a dilute
instanton gas approximation were justified---and generally it is
not---the $k$-instanton Yang-Mills calculation would naturally lead to
$k$ copies of $AdS_5$ (i.e., $k$ independent instanton 4-positions and
scale sizes) rather than just one copy as the supergravity side of the
correspondence implies.

Motivated in part by this puzzle, in Sec.~II we first define
what we mean precisely by the multi-instanton supermultiplet
in the superconformal case and then
give a thorough,
self-contained review of multi-instanton calculus in supersymmetric
$SU(N)$ gauge theory.  Most of this review material is borrowed from
Ref.~\cite{KMS} (written in collaboration with M. Slater) and is included
herein for the reader's convenience.  Sections II.1-II.4 are devoted
to the ADHM construction of the general self-dual gauge field
configuration $v_n(x)$ of arbitrary topological number $k$
\cite{ADHM}. Subsequently we flesh out the remaining components of the
$\N=4$ instanton supermultiplet: Secs.~II.5-II.6 review the
construction of the four  gauginos $\lambda^A_\alpha(x)$
($A=1,2,3,4$) \cite{CGTone}, and Sec.~II.7 reviews the construction of
the six adjoint scalars $A^{AB}(x)$ \cite{MO-I,MO-II,KMS}.\footnote{In
our convention, upper (lower) $SU(4)$ indices indicate a ${\bf 4}$
($\bar{\bf 4}$).}

Semiclassical physics requires knowledge, not only of the relevant
saddle-point configurations that dominate a given process (in this
case, the multi-instanton supermultiplet of Sec.~II), but also, of how
properly to weight these configurations in the path
integral. Specifically, one needs to construct the weighting factor 
\begin{equation}
\dmuphys e^{-\Skinst}\ ,
\end{equation}
where $\Skinst$ is the $k$-instanton action, and $\dmuphys$ stands
for the $k$-instanton bosonic and fermionic collective coordinate
measure. These two quantities are studied in detail in Secs.~III and
IV, respectively. $\Skinst$ has a crucial role to play: it is
responsible for lifting the non-exact fermion zero modes, thereby
ensuring a nonzero result for the Grassmann integrations.  For the
gauge group $SU(N)$ with $\N=4$ supersymmetry, the $k$-instanton ADHM
configuration has $8kN$ adjoint fermion zero modes.  As mentioned
earlier, precisely sixteen of these modes are exact; these are
saturated by the explicit insertions of the external operators in the
correlators $G_n$. The remaining $8kN-16$ lifted modes must be
saturated instead by bringing down the appropriate power of $\Skinst.$
In previous work \cite{DKMn4}, we constructed $\Skinst$ in the
$\N=4$ model in a somewhat indirect way, first by implementing the
supersymmetry algebra directly on the multi-instanton collective
coordinates, and then by requiring that $\Skinst$ be a supersymmetric
invariant quantity (among other properties). In Sec.~III, we present
an alternative construction of $\Skinst$ which is much more direct,
albeit calculationally intensive, and leads to the same result. In
brief, the $k$-instanton supermultiplet is inserted into the component
Lagrangian, and the space-time integrations are carried out
explicitly using Gauss's law. In the
absence of Higgs VEVs, we find that $\Skinst$ is a pure fermion
quadrilinear term (see Eqs.~\eqref{Skinstdef}-\eqref{Skquadef} below),
 with one fermion collective coordinate drawn from
each of the four gaugino sectors $A=1,2,3,4.$

Section IV is a detailed multi-purpose study of the $k$-instanton
collective coordinate measure, $\dmuphys.$ Section IV.1 reviews the
construction of this measure given previously in Ref.~\cite{DHKM}; the
requirements of $\N=4$ supersymmetry, cluster decomposition, and
renormalization group flow to the measures of lower supersymmetry
(among other constraints) suffice to fix its form uniquely. The
remainder of Sec.~IV is completely new. In Sec.~IV.2, we offer an
alternative construction of $\dmuphys\exp-\Skinst$ which is also
directly relevant to the physics of the anti-deSitter/superconformal field 
theory (AdS/CFT) correspondence.
Specifically, we consider ${\cal N}=4$ supersymmetric $U(N)$ gauge
theory as realized on a set of $N$ parallel D3-branes.  Of course,
this is the starting point for the analysis leading to Maldacena's
conjecture. Here, however, we will not take the large-$N$ supergravity
limit which leads to the strong-coupled ${\cal N}=4$ theory, but
rather stay at weak coupling and take the ordinary decoupling limit
$\alpha'\rightarrow 0$ while keeping $g^{2}$ and $N$ fixed. In fact it
is well known that D-instantons located on the D3-branes are
equivalent to Yang-Mills instantons in the world-volume gauge theory
\cite{D1,D2}.  In a similar context, Witten has shown that the ADHM
construction emerges in a natural way from standard facts about
D-branes \cite{W1}.  This viewpoint turns out to be quite powerful
and, in particular, it provides a simple rederivation of the ADHM
formalism developed in our previous papers. While this work was being
completed, a paper which has some overlap with Sec.~IV.2 has appeared
\cite{AKH}.

The form of the measure derived in Secs.~IV.1-IV.2 can be termed the
``flat measure,'' as the ADHM collective coordinates are integrated
over as Cartesian variables. However the number of such integration
variables grows with $N$ and one must choose a more appropriate set of
coordinates before a traditional saddle-point
approximation can be undertaken. To this end, we 
transform to a smaller,
gauge-invariant set of collective coordinates, which is done in
Sec.~IV.3.  In the resulting gauge-invariant measure
(Eq.~\eqref{gaugeinvmeas} below), the number of integration variables
is independent of $N$, and only grows with the topological number $k$.
This form of the measure is then the appropriate
starting point for the large-$N$ saddle-point treatment that we
develop beginning in Sec.~V.

Section V has the largest overlap with the preliminary version of our
paper, Ref.~\cite{LETT}. It is devoted to marrying the large-$N$
limit, on the one hand, with supersymmetric ADHM calculus on the
other. As a warm-up, in Sec.~V.1 we revisit the one-instanton sector
already discussed in Sec.~I.3, and develop a formal saddle-point
approximation to the gauge-invariant measure as $N\rightarrow\infty.$
To leading order in $1/N$, the bosonic collective coordinate
integration turns out to be dominated by a ten-dimensional submanifold
with the geometry of $AdS_5\times S^5.$ Thus we find the intriguing
result that, even at weak coupling, the large-$N$ instanton ``sees''
the metric of the supergravity background.  The correspondence between
the instanton scale-size $\rho$ and the radial direction on $AdS_{5}$
has been previously noted in \cite{WIT112,BGKR}. Indeed the power of
$\rho$ in the factor $d^4X\, d\rho\,\rho^{-5}$ which enters the
measure (with $X_n$ the instanton 4-position) is forced by the scale
invariance of the $\N=4$ model, and necessarily agrees with the volume
form of the conformally invariant space $AdS_5$.  Rather, the main
novelty in our analysis is the explanation of the $S^5\,$: it arises
from a scalar field  $\chi_a$ transforming in the vector
representation of the $SO(6)$ $R$-symmetry which is introduced in order to
bilinearize the fermion quadrilinear action $\Skinst$. In particular,
as $N\rightarrow\infty,$ the $\chi_a$ variables are increasingly
peaked around an $S^5$ subspace of ${\Bbb R}^6$. The use of such auxiliary
Gaussian variables is familiar from the standard large-$N$ treatment of
theories with fermions in the fundamental representation of $SU(N)$,
such as the Gross-Neveu model \cite{GN}. It is remarkable that they
are not just mathematical artifacts, but  play such a
central r\^ole in the AdS/CFT correspondence. 
In fact, as explained in Section IV.2, 
these auxiliary fields have their origin as the gauge fields 
of a six-dimensional $U(k)$ gauge theory whose dimensional reduction describes 
$k$ D-instantons in the presence of D3 branes. 

Sections V.2-V.3 extend the saddle-point method to the $k$-instanton
case. In Sec.~V.2 we solve the coupled saddle-point equations at the
strictly classical level, i.e., neglecting (unjustifiably) the
prefactor contribution due to small fluctuations. In this simple
approximation scheme, we find that the $k$ instantons live in mutually
commuting $SU(2)$ subgroups of the $SU(N)$ gauge group, and are
described by $k$ independent copies of $AdS_5\times S^5.$ This is the
puzzle discussed earlier, and it has to be reconciled 
with the supergravity side
of the correspondence.  The puzzle is resolved in Sec.~V.3, in which
the small fluctuations are carefully taken into account. These
generate a singular attractive effective potential which draws the $k$
instantons to a common point, thereby reducing the moduli space to a
\it single \rm copy of $AdS_5\times S^5,$ precisely as required by the
AdS/CFT correspondence.  (However the $k$ instantons remain in
mutually commuting $SU(2)$'s.) And there is a further surprise: the
Lagrangian which describes small fluctuations in the vicinity of this
reduced moduli space turns out to be identical to ten-dimensional
${\cal N}=1$ supersymmetric $SU(k)$ Yang-Mills theory, dimensionally
reduced to $0+0$ dimensions. The collective coordinate $k$-instanton-measure 
thus factorizes into the measure on $AdS_{5}\times S^{5}$ times the
partition function $\hat{\cal Z}_k$ for this $SU(k)$ theory. 
Thus, the $k$-instanton semiclassical measure,
$\int \dmuphys \exp-\Skinst,$ in the $\N=4$ Yang-Mills theory
is reduced to the partition function of the multi-D-instanton
matrix model. We then
use the recent explicit evaluation of $\hat{\cal Z}_{k}$
\cite{KNS,MNS,part} to complete the specification of the effective
large-$N$ $k$-instanton measure; the final expression is given in
Eq.~\eqref{endexp}. Intriguingly, this matrix model is precisely the one posited by
Ref.~\cite{IKKT} as a definition of Type IIB string theory.

In Sec.~V.4 we comment further on $\hat{\cal Z}_{k}$. In particular,
it is remarkable that $\hat{\cal Z}_k$ is precisely the partition
function that describes $k$-D-instantons in {\it flat\/} space
\cite{GG2,GG3,GG4}. It is also noteworthy how, having started with an $SU(N)$
gauge theory, one ends up with an $SU(k)$ gauge theory; this type of
``duality'' is well known from a string theory context
\cite{W1,D1,D2}, and is apparent from our discussion in Sec.~IV.2.  We
also discuss the expected relation between $\hat{\cal Z}_{k}$ and the
Euler character of the charge-$k$ ADHM moduli space.

In Sec.~VI we use this large-$N$ effective measure to evaluate the
correlators of interest. Section VI.1 is devoted to the correlators
$G_n$, with $n=16,$ 8 or 4; as advertised above, we find exact
agreement with the supergravity prediction of \cite{BGKR}.
However the $n$ insertions in these correlators all correspond, on the
supergravity side, to fields which have no dependence on $S^5$,
i.e.~which are ``$s$-wave''. As such, they do
not probe this part of the string theory background. Accordingly, in Sec.~VI.2 we
extend the analysis to correlators corresponding to supergravity
fields which are higher partial waves on $S^5$, i.e.~massive
Kaluza-Klein modes on $S^5$,
and verify the proposed AdS/CFT correspondence for a tower of such
operators. This analysis highlights the novel r\^ole played by the
auxiliary scalar variables $\chi_a$ in our formalism.

 Section~VII contains some concluding thoughts. Section VII.1
gives a slanted historical overview of the ADHM multi-instanton
program: largely stymied in nonsupersymmetric applications due to the
intractability of the nonlinear constraints and the lack of knowledge
of the measure; recently rejuvenated in the context of supersymmetric
theories; and, as the present paper demonstrates, simplified even more
dramatically in the large-$N$ limit. And Sec.~VII.2 briefly discusses 
the source of both $1/N$ corrections and (despite the aforementioned
nonrenormalization theorem) $g^2$ corrections to the Yang-Mills calculations.

\def\bigZ{Z\!\!\!Z}
\def\lambdabar{\bar\lambda}
\def\Pinfty{{\cal P}_{\sst\infty}}
\def\P{{\cal P}}
\def\J{{\cal J}}
\def\G{{\cal G}}
\def\calG{{\cal G}}
\def\Skquad{S^k_{\rm quad}}
\def\rmv{{\rm v}}
\def\Skinst{S^k_{\rm inst}}
\def\sst{\scriptscriptstyle}
\def\susy{supersymmetry}
\def\susic{supersymmetric}
\def\homoa{\langle {\cal A}\rangle }
\def\calL{{\cal L}}
\def\Lgauge{S_{\rm gauge}}
\def\Lyuk{S_{\rm yuk}}
\def\Lhiggs{S_{\rm higgs}}
\def\barhomoa{\bar{\langle {\cal A}\rangle} }
\def\Mbar{\bar{\M}}
\def\Nbar{\bar{\N}}
\def\xbar{\bar{x}}
\def\D{{\cal D}}
\def\Dbarslash{\,\,{\raise.15ex\hbox{/}\mkern-12mu {\bar\D}}}
\def\delslash{\,\,{\raise.15ex\hbox{/}\mkern-9mu \partial}}
\def\Dslash{\,\,{\raise.15ex\hbox{/}\mkern-12mu \D}}
\def\adss{AdS_5\times S^5}
\def\bigR{{\rm I}\!{\rm R}}
\def\sigmabar{\bar\sigma}
\def\psibar{\bar\psi}
\def\mubar{\bar\mu}
\def\dalpha{{\dot\alpha}}
\def\dgamma{{\dot\gamma}}
\def\etabar{\bar\eta}
\def\Sinst{S_{\rm inst}}
\def\Squad{S_{\rm quad}}
\def\Deltabar{\bar\Delta}
\def\Ubar{\bar U}
\def\N{{\cal N}}
\def\M{{\cal M}}
\def\A{{\cal A}}
\def\hf{{\textstyle{1\over2}}}
\def\quarter{{\textstyle{1\over4}}}
\def\Vbar{\bar V}
\def\eighth{{\textstyle{1\over8}}}
\def\fourth{\quarter}
\def\dbeta{{\dot\beta}}
\def\abar{\bar a}
\def\wbar{\bar w}
\def\trtwo{\tr^{}_2\,}
\def\trN{\tr^{}_N\,}
\def\Xsd{X^{\sst\rm SD}}
\def\vasd{v^{\sst\rm ASD}}
\def\nubar{\bar{\nu}}
\def\vhiggs{{\rm v}}
\def\vbarhiggs{\bar{\rm v}}
\def\vhiggsbar{\bar{\rm v}}
\def\bbar{\bar b}
\def\bigL{{\bf L}}
\def\Xasd{X^{\sst\rm ASD}}
\def\Atot{{\cal A}_{\rm tot}}
\def\sqrtwo{\sqrt{2}\,}
\def\gst{g_{st}}
\def\CO{{\cal O}}
\def\CM{{\cal M}}
\def\CN{{\cal N}}

\subsection{Review of the superstring prediction}

Before we describe in detail the superstring prediction, some general
comments are in order. As we have alluded to in the last section,
Maldacena's conjecture \cite{MAL} relates Type IIB supergravity
on $AdS_5\times S^{5}$ to the large-$N$ limit of
${\cal N}=4$ supersymmetric Yang-Mills theory in four dimensions.  The
refinement of this correspondence presented in \cite{WIT150,GKP}
identifies the ${\cal N}=4$ theory as living on the four-dimensional
boundary of $AdS_{5}$. In particular, each chiral primary operator
${\cal O}$ in the boundary conformal field theory is identified with a
particular Kaluza-Klein mode of the supergravity fields which we
denote as $\Phi_{\cal O}$. The generating function for the correlation
functions of ${\cal O}$ is then given in terms of the supergravity
action $S_{\rm IIB}[\phi_{\cal O}]$ according to
\begin{equation}
\left\langle \exp\,\int\, d^{4}x \, J_{\cal O}(x){\cal O}(x)\,
\right \rangle = \exp-S_{\rm IIB}\left[\Phi_{\cal O};
J\right]\, .
\elabel{corr}
\end{equation}  
The IIB action on the right-hand side of the equation is evaluated on a configuration
which solves the classical field equations subject to the condition
$\Phi_{\cal O}(x)=J(x)$ on the four-dimensional boundary. 

In most
applications considered so far, the relation \eqref{corr} has
primarily been applied at the level of classical supergravity, which
corresponds to $N\rightarrow \infty$, with $g^{2}N$ fixed and large,
in the boundary theory
\cite{WIT150,GKP,HW,AFGJ,FMMR,EFS,LMRS,CNSS,LT,MV,BGut,AF1,SOLOD,GHEZ}.  
However, the full equivalence of IIB
superstrings on $AdS_{5}\times S^{5}$ and ${\cal N}=4$ supersymmetric
Yang-Mills theory  conjectured
in \cite{MAL} suggests that \eqref{corr} should hold more generally,
with quantum and stringy corrections to the classical supergravity
action corresponding to $g^2$ and $1/g^{2}N$ corrections in the ${\cal
N}=4$ theory, respectively. The particular comparison that concerns us
here is between Yang-Mills instanton contributions to the correlators
of ${\cal O}$ generated by the left-hand side of \eqref{corr} and
D-instanton corrections to the IIB effective action on the right-hand
side.  For future reference, the standard metric on $AdS_{5}\times
S^{5}$, in coordinates $x_\mu=(\com_n,\rho,\sfc)$, is given as,
\begin{equation}
ds^{2}= g_{\mu\nu}dx^{\mu}dx^{\nu} = 
\frac{1}{\rho^{2}}\left(d\com_nd\com_n + d\rho^{2}\right) + 
d\hat{\Omega}^2\ .
\elabel{metric}
\end{equation}
Here, $\sfc$ is a unit $SO(6)$ vector denoting a point on the unit
5-sphere. The four-dimensional theory in question lives on the
boundary of $AdS_{5}$ which is located at $\rho=0$.

The relevant D-instanton contributions arise
as $(\alpha^{\prime})^3$ corrections \cite{BG} 
to the classical IIB theory on
$AdS_{5}\times S^{5}$. In particular, as in more conventional field
theory settings, the corrections take the form of instanton-induced
vertices in a local effective action. The D-instanton contribution to
an $n$-point correlator $G_n$ comes from a tree level Feynman diagram
with one vertex, located at some point $(X_n,\rho,\sfc)$ in the bulk
of $AdS_{5}\times S^{5}$.  The diagram also has $n$ external legs
corresponding to operator insertions on the boundary.
 The position-space Feynman rules associate an overall constant with the
vertex and a bulk-to-boundary propagator, to be described below, with
each external leg \cite{GKP,WIT150,FMMR}. 
The selection rule which specifies the necessary
fermionic insertions is conveniently encoded by integrating over
Grassmann variables $\varepsilon$ which correspond to the sixteen
supersymmetries of the IIB theory which are broken by the
D-instanton. These variables are associated to the sixteen supersymmetric
and superconformal zero modes of an instanton configuration on the
Yang-Mills side. The final amplitude is then obtained by integrating
over the position of the vertex on $AdS_{5}\times S^{5}$. Hence the
combined bosonic and fermionic integration measure for the vertex is ,
\begin{equation}
\int\, \sqrt{g}\, d^{10}x \,d^{16}\varepsilon = \int\, {d^{4}\com\, 
{d\rho}\over{\rho^{5}}}\, d^5\sfc\, d^{16}\varepsilon\, . 
\elabel{int}
\end{equation}
For instance, the bulk-to-boundary propagator for an $SO(6)$ singlet scalar 
free field of mass $m$ on $AdS_{5}$ is,  
\begin{equation}
K_{\Delta}(X,\rho;x,0)=\frac{\rho^{\Delta}}{(\rho^{2}+(x-X)^{2})^{\Delta}}\, ,
\elabel{prop}
\end{equation}  
where $(mL)^2=\Delta(\Delta-4)$.

We now turn in detail to the IIB superstring prediction \cite{BG}
(closely following the more detailed treatment in \cite{BGKR}). In
\cite{GG1}, Green and Gutperle conjectured an exact form for certain
non-perturbative (in $g_{st}$) corrections to certain terms in the
Type IIB supergravity effective action. In the present application,
where the string theory is compactified on $AdS_5\times S^5$, it is
important for the overall consistency of the Banks-Green prediction
that the non-perturbative terms in the effective action do not alter
the $AdS_5\times S^5$ background, since the latter is conformally flat
\cite{BG}.  In particular, at leading order beyond the
Einstein-Hilbert term in the derivative expansion, the IIB effective
action is expected to contain a totally antisymmetric 16-dilatino
effective vertex of the form \cite{GGK,KP2}
\begin{equation}(\alpha')^{-1}\int d^{10}x\,\sqrt{\det g}\,e^{-\phi/2}\,
f_{16}(\tau,\bar\tau)\,\Lambda^{16}\ + \ \hbox{H.c.}
\elabel{effvert}\end{equation}
Here $\Lambda$ is a complex chiral $SO(9,1)$ spinor, and $f_{16}$ is a certain
weight $(12,-12)$ modular form under $SL(2,{\Bbb Z})$. At the same
order in the derivative expansion there are other terms related to
\eqref{effvert} by supersymmetry and involving other modular forms
\cite{GGK,KP1,KP2} and, in particular, $f_n(\tau,\bar\tau)$, with $n=8$ and 4.
This modular symmetry is precisely $S$-duality of the Type IIB superstring, and
although this does not completely determine the modular
forms $f_n$, for the $n=4$ term Green and Gutperle \cite{GG1} were strongly attracted to
the following conjecture, later proved in
Ref.~\cite{GSI} and generalized to $n\neq4$ in \cite{KP1,KP2}:
\begin{equation}
f_n(\tau,\bar\tau)=({\rm Im}\tau)^{3/2}
\sum_{(p,q)\neq(0,0)}(p+q\bar\tau)^{n-11/2}(p+q\tau)^{-n+5/2}.
\end{equation}
These rather arcane expressions turn out to have the right modular
properties, i.e. weight $(n-4,-n+4)$, and also have very suggestive
weak coupling expansions \cite{GG1,GG2,GGK,KP2}:
\begin{equation}e^{-\phi/2}f_n\ =\
32\pi^2\zeta(3)g^{-4}-{2\pi^2\over3(9-2n)(7-2n)}+\sum_{k=1}^\infty\calG_{k,n}\
, \elabel{fexpand}\end{equation} where
\begin{equation}\begin{split}
\calG_{k,n}\ =\ \left({8\pi^2 k\over
g^2}\right)^{n-7/2}\Big(\sum_{d|k}{1\over d^2}\Big)&
\Big[e^{-(8\pi^2/g^2-i\theta)k} \sum_{j=0}^\infty
c_{4-n,j-n+4}\left({g^2\over8\pi^2k}\right)^{j} \\
&+e^{-(8\pi^2/g^2+i\theta)k} \sum_{j=0}^\infty
c_{n-4,j+n-4}\left({g^2\over8\pi^2k}\right)^{j+2n-8}\Big]\, ,
\elabel{Gdef}\end{split}\end{equation}
 and the numerical coefficients are
\begin{equation}
c_{n,r}\ =\ {(-1)^n\sqrt{8\pi}\,\Gamma(3/2)\Gamma(r-1/2)\over2^r\Gamma(r-n+1)
\Gamma(n+3/2)\Gamma(-r-1/2)}\ .
\elabel{defcs}
\end{equation}
The summation over $d$ in \eqref{Gdef} runs over the positive integral
divisors of $k$.  Notice that, having taken into account the
conjectured correspondence \eqref{corresp} to the couplings of
four-dimensional Yang-Mills theory, the expansion \eqref{fexpand} has
the structure of a semiclassical expansion: the first two terms
correspond to the tree and one-loop pieces, while the sum on $k$ is
interpretable as a sum on Yang-Mills instanton number, the first and
second terms in the square bracket being instantons and
anti-instantons, respectively (this is dictated by the $\theta$
dependence). Each of these terms includes a perturbative expansion
around the instantons, although notice that the leading order
anti-instanton contributions are suppressed by a factor of $g^{4n-16}$
(so not suppressed for $n=4$) relative to the leading order instanton
contributions.

In the present paper, our focus will be on the leading
semiclassical contributions to the $f_n$; by this we mean, for each
value of the topological number $k$, the leading-order contribution in
$g^2$. For $f_{16}$ and $f_8$ the leading semiclassical contributions
come from instantons only and have
the form
\begin{equation}
e^{-\phi/2}f_n\Big\vert_{k\hbox{-}\rm instanton}\ =
{\rm const}\cdot\left({k\over g^2}\right)^{n-7/2}
e^{2\pi ik\tau}\sum_{d|k}{1\over d^2}\ ,\elabel{foksh}
\end{equation}
neglecting $g^2$ corrections.  For the special case of $f_4$ there is
an identical anti-instanton contribution with
$i\tau\rightarrow-i\bar\tau.$ Later, we shall find it very significant
that $e^{-\phi/2}f_n$ has a dependence on the instanton number of the
form $k^{n-7/2}$.  In  Type IIB superstring theory, the terms in
the square bracket in \eqref{Gdef}, which are non-perturbative in the
string coupling, are interpreted as being due to D-instantons.

{}From the effective vertex \eqref{effvert} one can construct Green's functions
$G_{16}(x_1,\ldots,x_{16})$ for sixteen 
dilatinos $\Lambda(x_i),$ $1\le i\le 16,$ which live on the boundary
of $AdS_5$:
\begin{equation}
G_{16}\ =\ \langle\,\Lambda(x_1)\cdots\Lambda(x_{16})\rangle\ \sim\
(\alpha')^{-1}\,e^{-\phi/2}\,f_{16}\,t_{16}\int{d^4\com\,d{\rho}\over{\rho}^5}\,
\prod_{i=1}^{16}\,K_{7/2}^F(\com,{\rho};x_i,0)
\elabel{grnfcn}\end{equation}
suppressing spinor indices.
Here $K_{7/2}^F$ is the bulk-to-boundary propagator for a spin-$\hf$ Dirac
fermion of mass $m=-\tfrac32L^{-1}$ and scaling dimension $\Delta=\tfrac72\,$
\cite{HS1,GKP,WIT150,FMMR}:
\begin{equation}K_{7/2}^F(\com,{\rho};x,0)\ =\ 
K_{4}(\com,{\rho};x,0)\,\big(\rho^{1/2}\gamma_5-\rho^{-1/2}(x-\com)_n
\gamma^n\big)
\elabel{KFdef}\end{equation}
with
\begin{equation}K_{4}(\com,{\rho};x,0)\ =\ {{\rho}^4\over\big({\rho}^2+
(x-\com)^2\big)^4}\ .
\elabel{Kfourdef}\end{equation}
In these expressions the $x_i$ are four-dimensional space-time
coordinates for the boundary of $AdS_5$ while $\rho$ is the fifth,
radial, coordinate.  The quantity $t_{16}$ in Eq.~\eqref{grnfcn} is
(in the notation of Ref.~\cite{BGKR}) a 16-index antisymmetric invariant
tensor which enforces Fermi statistics and ensures, {\it inter alia\/},
that precisely 8 factors of $\rho^{1/2}\gamma_5$ and 8 factors of
$\rho^{-1/2}\gamma^n$ are picked out in the product over $K^F_{7/2}$.
Related by supersymmetry to $G_{16}$, are correlation functions $G_8$
and $G_4$ which have an analogous structure to \eqref{grnfcn}, but
involving different bulk-to-boundary propagators and with overall factors
that we can summarize as
\begin{equation}
G_n\propto (\alpha')^{-1}e^{-\phi/2}f_n(\tau,\bar\tau)\ .
\elabel{yetae}\end{equation}

Notice that there is no explicit dependence in these expressions on
 the coordinates on $S^5$; in particular, the propagator does not
 depend on them (save through an overall multiplicative factor which
 we drop). This is because $G_{16}$, $G_8$ and $G_4$ are correlators of operators whose
 supergravity associates are constant on $S^5$.
In Sec.~VI.2 below, after we have elucidated the meaning
 of the $S^5$ in the four-dimensional picture, we will broaden the
 discussion to include operators that correspond to fields on the
 supergravity side that vary over $S^5$,
 and verify that the correct dependence on these coordinates
 ensues.

Notice further that in the expression \eqref{grnfcn}, the
five-dimensional space-time
dependence does not mix in any way with the $\tau$ dependence, where
the latter is carried solely by the modular form $f_{16}.$ In
particular this means that for arbitrary topological charge $k$, the
D-instanton contribution to $G_{16}$ only contains a single copy---and
not $k$ copies---of $AdS_5$, as evinced by the single copy of the
$d^4X\,d\rho\,\rho^{-5}$ volume form in Eq.~\eqref{grnfcn}. The
rational for this, as we alluded to earlier, is that when only 16
fermion zero-modes are saturated, as is apparent in the term in the
effective action \eqref{effvert}, it is more appropriate to think of
a configuration of $k$ D-instantons as a single charged $k$
bound state since the integrals over the relative positions are
actually convergent \cite{GG2,GG3,GG4}. However, it will be
an interesting challenge (successfully resolved in Sec.~V.3 below) to
reproduce this feature from the Yang-Mills side, rather than $k$
distinct copies which one would normally expect from a dilute
instanton gas approximation (if it could be justified).  Beyond this
qualitative agreement, it is the principal goal of this paper to
demonstrate that the identical expression for $G_{16}$ emerges from
the gauge theory side of the correspondence, to leading semiclassical
order (meaning, for each topological number $k$, to leading order in
$g^2$). But before developing the ADHM machinery necessary to treat
the case of general $k$, let us review the rather simpler situation
for $k=1.$

\subsection{Review of the Yang-Mills calculation at the one-instanton level}

According to Maldacena's conjecture, the correlator \eqref{grnfcn} in
the IIB theory should correspond to a certain 16-fermion correlator in
four-dimensional large-$N$ supersymmetric Yang-Mills theory.  The
fermion operator in the four-dimensional Yang-Mills picture with the right
transformation properties to correspond to the dilatino is the
gauge-invariant composite operator \cite{GKP,WIT150,BGKR}
\begin{equation}\Lambda_\alpha^A\ =\ g^{-2}\sigma^{mn}{}_\alpha^{\
\beta}\,{\rm tr}_N\,
v_{mn}\,\lambda_\beta^A\ ,
\elabel{compop}\end{equation}
which is a spin-$1/2$ fermionic component of the superfield
Noether current detailed in \cite{BRW,BGKR}. Here $v_{mn}$ is the  $SU(N)$
gauge field strength while the $\lambda_\beta^A$ are the 
Weyl gauginos, with the index $A=1,2,3,4$ labeling the four 
supersymmetries. The numerical tensor $\sigma^{mn}$ projects out the
self-dual component of the field strength,\footnote{We use Wess and
Bagger conventions throughout \cite{WB}, continued to Euclidean space
via $(x^0,\vec x)\rightarrow(x^0,i\vec x)$, with
$\eta^{nm}a_na_m\rightarrow-a_na_n$, except that our convention for
integrating Weyl spinors is $\int d^2\lambda\, \lambda^2=2$, rather than
Wess and Bagger's 1.} so that, at leading order, only  instantons
rather than  anti-instantons can contribute.

To begin to see how, in the Yang-Mills picture, a 16-point correlator of the
operator \eqref{compop} might possibly reproduce the structure of
Eq.~\eqref{grnfcn}, let us recall the suggestive observation made by
\cite{BGKR}. These authors focused on the one-instanton sector of
$\N=4$ supersymmetric Yang-Mills theory with gauge group $SU(2).$ In
this particularly transparent case, the single instanton contains
precisely sixteen adjoint fermion zero modes: a supersymmetric plus a
superconformal zero mode, each of which is a Weyl 2-spinor, times four
supersymmetries.  A non-vanishing 16-fermion correlator is therefore
obtained by saturating each of the fermion insertions with a distinct
such zero mode.  The semiclassical Yang-Mills calculation then proceeds by
replacing $v_{mn}$ in the composite operator \eqref{compop} by the
standard one-instanton field strength $v_{mn}(x-\com),$ with $\com$
the center of the instanton; likewise for $\lambda_\beta^A$ one
substitutes into Eq.~\eqref{compop} the well-known expression for the
gaugino:\footnote{See for instance Eqs.~(4.3a) and (A.5) of
\cite{MO-I}.}
\begin{equation}\lambda_\beta^A(x)\ =\ 
-\big(\xi_\alpha^A-\sigma^n_{\alpha\aD}\etabar^{\aD A}
\cdot(x-\com)_n\big)
\sigma^{kl\ \alpha}_{\phantom{kl}\beta}\,v_{kl}(x-\com)\ .
\elabel{gnomode}\end{equation}
Here $\xi^{\alpha A}$ and $\etabar_{\aD}^A$ are Grassmann parameters
that multiply the supersymmetric and superconformal zero modes,
respectively. Using the fact that
\begin{equation}\tr_N\,v_{mn}v_{kl}\,{\Big|}_{\rm 1\hbox{-}inst}\ 
 =\ {1\over3}\,{\cal P}_{mn,kl}^{\rm SD}\,
\tr_N\,v_{pq}^2\,{\Big|}_{\rm 1\hbox{-}inst}\ 
 =\ 
 32{\cal P}_{mn,kl}^{\rm SD}\,K_4(\com,\rho;x,0)\ ,
\elabel{projid}\end{equation}
where 
\begin{equation}{\cal P}_{mn,kl}^{\rm SD}\ =\
 \quarter(\delta_{mk}\delta_{nl}
-\delta_{ml}\delta_{nk}+\epsilon_{mnkl})
\elabel{projdef}\end{equation}
is the projector onto self-dual antisymmetric tensors, the composite
fermion \eqref{compop} becomes
\begin{equation}\Lambda^{A}_\alpha(x)\,{\Big|}_{\rm 1\hbox{-}inst}\  =\ 
-{96\over
g^2}\,\big(\xi^{A}_\alpha-\sigma^n_{\alpha\aD}\etabar^{\aD A}
\cdot(x-\com)_n\big)
K_4(\com,\rho;x,0)\ .
\elabel{insertid}\end{equation}
We therefore find for
the one-instanton contribution to the 16-fermion correlator in
$\N=4$ $SU(2)$ gauge theory \cite{BGKR}:
\begin{equation}\begin{split}
\langle\Lambda_{\alpha_1}^1(x_1)\cdots
\Lambda_{\alpha_{16}}^4(x_{16})\rangle
\,{\Big|}_{\rm 1\hbox{-}inst}
\ =\ 
&C_2\,g^{-24}\,e^{2\pi i\tau}\,
\int {d^4\com d\rho\over\rho^5}\,\prod_{A=1,2,3,4}d^2\xi^A\,d^2\etabar^A\,
\\&\times\ 
\big(\xi^1_{\alpha_1}-\sigma^n_{\alpha_1\aD}\etabar^{\aD 1}
\cdot(x_1-\com)_n\big)K_4(\com,\rho;x_1,0)
\\\times\cdots&\times\ 
\big(\xi^4_{\alpha_{16}}-\sigma^n_{\alpha_{16}\aD}\etabar^{\aD 4}
\cdot(x_{16}-\com)_n\big)K_4(\com,\rho;x_{16},0)\ .
\elabel{YManswer}\end{split}\end{equation}
Here, the power of $\rho$ in the integration measure is fixed by the
fact that the $\N=4$ model has vanishing $\beta$-function; the power
of $g$ comes from combining the explicit dependence in Eq.~\eqref{insertid}
with  factors of $g^{-8}$ and 
$g^{16}$ from the bosonic and fermionic integrations, respectively
\cite{tHooft}; and $C_2$ is an overall numerical constant.

As the authors of Ref.~\cite{BGKR} point out, the space-time
dependence of the one-instanton Yang-Mills expression \eqref{YManswer}
precisely matches the supergravity expression \eqref{grnfcn}. Indeed,
performing the sixteen Grassmann integrations over $\xi_\alpha^A$ and
$\etabar^{\dalpha A}$ in Eq.~\eqref{YManswer} produces (by definition)
the totally antisymmetric tensor $t_{16}$ tied into sixteen propagators
$K^F_{7/2}$. Furthermore, the remaining integration measure for the
five bosonic collective coordinates $\{\rho,\com_n\}$ in
Eq.~\eqref{YManswer} is identical to the volume form of $AdS_5$ that
appears in Eq.~\eqref{grnfcn}; there, the scale-size $\rho$ is
re-interpreted as the inverse of the radial position of the
D-instanton in $AdS_5$ \cite{BGKR,WIT112}.

Beyond the spatial dependence, one also finds agreement at the
one-instanton level in the overall strength of the couplings in front
of the two expressions.  On the supergravity side, from
Eqs.~\eqref{fexpand}-\eqref{grnfcn} one extracts the one-instanton prefactor
\begin{equation}(\alpha')^{-1}\,e^{-\phi/2}\,f_{16}{\Big|}_{\rm 1\hbox{-}inst}
\ \sim\ g^{-24}\,e^{-8\pi^2/g^2+i\theta}\,\sqrt{N} \ ,
\elabel{IIBconst}\end{equation} up to an overall numerical
constant. This $g$ dependence is already present in the Yang-Mills expression
\eqref{YManswer} \cite{BGKR}.  Reproducing this $\sqrt{N}$ dependence,
too, means generalizing the above Yang-Mills calculation from $SU(2)$ to
$SU(N)$, which was done in Ref.~\cite{DKMV}.  In $\N=4$ $SU(N)$ gauge
theory a single instanton has a total of $8N$ adjoint fermion zero
modes so that for $N>2$ there are additional zero modes beyond the
sixteen $\xi_\alpha^A$ and $\etabar^{\aD A}$ modes which must be
lifted in order to obtain a non-zero result.  As in several other
cases in three \cite{DKM3D} and four \cite{DKMn4} dimensions, this
lifting is due to the presence of a specific Grassmann quadrilinear
term in the instanton action (see Sec.~III below).  Pulling down
$2N-4$ powers of this quadrilinear and performing the Grassmann
integrations over the lifted modes turns out to be straightforward,
and is reviewed below in Sec.~V.1.  The chief result of
Ref.~\cite{DKMV} is that the Yang-Mills expression \eqref{YManswer} continues
to hold for the gauge group $SU(N)$, up to the replacement
$C_2\rightarrow C_N$ where\footnote{The constant $C_N$ differs by a
factor of $2^{-8}$ from that quoted in \cite{DKMV}. The difference is
due to the present conventions of footnote 4 
for integrating Weyl spinors.}
\begin{equation}C_N\ =\ {(2N-2)!\over(N-1)!(N-2)!}\,
2^{-2N+49}\,3^{16}\,\pi^{-10}
\ =\  2^{47}3^{16}\pi^{-21/2}\sqrt{N}\big(\,1-\tfrac58N^{-1}+\CO(N^{-2})\,\big),
\elabel{CNdef}\end{equation}
the final equality following from Stirling's formula. In this
algebraically nontrivial way, the overall $\sqrt{N}$ dependence
predicted by supergravity is recovered in the large-$N$ limit of the
gauge theory. Moreover, since
\begin{equation}{1\over N}\ =\ {g^2\over g^2N}\ =\
{4\pi\gst\,(\alpha')^2\over L^4} 
\elabel{oneoverN}\end{equation}
from Eqs.~\eqref{corresp}-\eqref{alphadef}, the
explicit form of the $1/N$ series implied by the exact Yang-Mills one-instanton
result \eqref{CNdef} gives an infinite number of predictions for
 corrections in powers of $g_{st}(\alpha')^2$
on the IIB side of the correspondence.

Upon reflection, however, one might suspect that the agreement
described above between the supergravity and Yang-Mills
 expressions for $G_{16}$ is
an accident of the one-instanton sector. For, only in that sector is the
instanton action density $\tr_N\,v_{mn}^2$ proportional to the propagator
$K_4$, as per Eq.~\eqref{projid}. As reviewed in Sec.~II.1 below, for
topological number $k$ the instanton action density is a rational function of
$x$ whose denominator generically involves a polynomial in $x$ of
degree $2k$, raised to a power.  In particular, this means that for
$k>1,$ the $k$-instanton density generically looks nothing like a
classical propagator. One plausible way around this problem is to
posit that the dilute instanton gas approximation somehow becomes
valid in the Yang-Mills $k$-instanton calculation. If this were the case,
however, the $k$-instanton density would resemble a \it sum \rm of $k$
distinct one-instanton densities (hence a sum of supergravity
propagators). Moreover, the moduli space would generically contain $k$
distinct copies (rather than a single copy) of $AdS_5$, meaning
integrations over $k$ distinct instanton 4-positions and sizes. In
other words, even if the dilute instanton gas approximation were
valid, the $k$-instanton Yang-Mills expression for $k>1$ can
reasonably be expected to look nothing like the supergravity result
\eqref{grnfcn}!

Actually we will find below that precisely the right propagator-like
structure of the instanton density, accompanied by the collapse of $k$
copies of $AdS_5$ to just one copy, is recovered in the Yang-Mills
calculation as $N\rightarrow\infty.$ This is one of several pleasing
simplifications that occur in the multi-instanton formalism when one
passes to the large-$N$ limit.

\rsen
\section{The $\N=4$ Multi-Instanton Supermultiplet}

We now review the formalism of
ADHM instantons, suitably supersymmetrized. 
There are two philosophically distinct ways to introduce the
multi-instanton supermultiplet.  One way is to view it as the exact
solution of the coupled classical Euler-Lagrange equations in Euclidean
space; in the absence of all Higgs VEVs such exact solutions do of
course exist. 
Our ultimate goal, however, is not 
just to find classical solutions, but rather to calculate their quantum
contributions to correlators $G_n$, which includes the effects of
a perturbative expansion in the instanton background. An efficient and
elegant way to take the leading  perturbations into account
automatically is to modify the background configuration itself as 
explained below; however, the instanton supermultiplet is then no longer an
exact solution to the coupled equations of motion. This is the essence
of the second approach which we follow in this paper.

The key distinction between the two methods is whether or not all of the
fermion zero modes are included in the  multi-instanton
multiplet. 
In $\N=4$ $SU(N)$ gauge theory there are $8kN$ fermion zero modes 
of the covariant Weyl operator $\Dbarslash=\sigmabar^n\D_n$
in the background of the gauge-field instanton
of topological charge $k$. However, most of these are  flat directions of the
action only at the {\it Gaussian\/} level and are lifted when the 
action is expanded to  higher order in fluctuations.
The  zero modes which remain exact to all orders are those
 protected by symmetry, namely the eight
 supersymmetric and eight superconformal fermion zero modes
\eqref{susymo}-\eqref{suconmo}.
If one insists on working with only exact solutions, then only these
sixteen fermion modes can be included
in the multi-instanton super-multiplet;
the $k$-instanton action  is then simply a constant,
$ 8\pi^2k /g^2 -ik\theta$.
The remaining $8kN - 16$ fermion zero modes are lifted and are treated
in this method as  fluctuations around the background. If one
nevertheless attaches Grassmann collective coordinates to the lifted
modes, one finds that
the resulting $k$-instanton {\it effective\/} action
will depend on these collective coordinates. In particular there will
be a Grassmann quadrilinear term (Eq.~\eqref{Skquadef} below)
generated by scalar exchange through the Yukawa couplings; since this
is a tree-level diagram it is not forbidden by a perturbative
nonrenormalization theorem.

Instead, in the alternative approach adopted here, we will 
include all $8kN$ Dirac operator 
fermion zero modes in the multi-instanton super-multiplet from the outset.
The aforementioned scalar exchange tree graph is then already built in
at the classical level, so that the quadrilinear action
\eqref{Skquadef} is generated without our having to calculate
separate diagrams. However, in this more efficient approach, the
coupled Euler-Lagrange equations must be solved iteratively, order by
order in $g$; this is not surprising since the
multi-instanton configuration is no longer an exact solution. 
In this way, our definition of the multi-instanton supermultiplet 
is similar to the more general case of the instanton calculus
in the non-conformal cases developed in \cite{MO-I,MO-II,DKMn4,KMS}.
Recall that for a non-zero VEV non-trivial exact solutions of 
the coupled Euler-Lagrange equations cannot exist
and instead one is instructed to solve
{\it constrained} equations \cite{Affleck} iteratively
in $g$; the two formalisms are the same in the VEV$\to 0$ limit.

As
stated earlier, in this paper we restrict our attention to leading
semiclassical order, meaning the first non-vanishing order in $g$ at
each topological level. For convenience
we will rescale all the fields so that
the only $g$ dependence in the action is through the overall
coefficient $g^{-2}$; the explicit $g$ dependence in the Euler-Lagrange
equations can be trivially restored by undoing this rescaling.

The hallmark of
ADHM calculus is that, with the correct ansatz for the classical
component fields, their defining Euler-Lagrange equations are mapped
into finite-dimensional matrix equations. Thus, for the gauge field
$v_m(x),$ the Yang-Mills equation (a nonlinear differential equation)
is mapped into the nonlinear ADHM constraint equations
\eqref{conea}-\eqref{conec} below. Similarly, for the classical
gauginos $\lambda^A_\alpha(x),$ the covariant Weyl equation (a linear
homogeneous differential equation) is mapped into the homogeneous
linear constraint equations \eqref{zmcona}-\eqref{zmconb} below. And
finally, for the classical adjoint Higgs fields $A^{AB}(x),$ the
covariant Klein-Gordon equation with a Yukawa source term (an
inhomogeneous linear differential equation) is mapped into the
inhomogeneous linear constraint equation \eqref{thirtysomething}
below. Note that at leading semiclassical order, the anti-gauginos
$\bar\lambda_{\dalpha A}(x)$ and the auxiliary superfield components $F$
and $D$ are all turned off; they do, however, turn on at a higher
order in $g$, in the process of solving by iteration the coupled
approximate Euler-Lagrange equations. 

\subsection{Construction of the classical gauge field}

In Secs.~II.1-II.4
we concern ourselves with pure  $SU(N)$ gauge theory,
without fermions or scalars.
Gauge fields $v_m$ are traceless anti-Hermitian $N\times N$ matrices
and 
\begin{equation}
v_{mn}\ =\ \partial_m v_n -\partial_n v_m +[v_m , v_n]
\elabel{vmndef}
\end{equation}
 is the field-strength.
The ADHM multi-instanton is the general solution
of the self-duality equation 
\begin{equation}v_{mn}\ =\ ^*v_{mn}\ \equiv\ \hf\epsilon_{mnkl}v_{kl}
\elabel{selfdual}\end{equation}
in the sector of topological number (equivalently, winding or
instanton number) $k$, where
\begin{equation}
k\ =\ -{1 \over 16 \pi^2} \int d^4 x \ \tr_{N}\,{v_{mn} \
}^*v^{mn} \ . 
\elabel{tch}\end{equation}
The ADHM construction of such multi-instantons is discussed in
Refs.~\cite{ADHM,CGTone,CWS,OSB}. Here we follow, with minor
modifications, the $SU(N)$ formalism of Refs.~\cite{CGTone,KMS}.

 The basic object in the ADHM construction
is the $(N+2k)\times 2k$ complex-valued matrix 
$\Delta_{\sst [N+2k] \times [2k]}$ which is taken to be linear in the 
space-time variable $x_n\,$:\footnote{For 
clarity, in Sec.~II we will occasionally show matrix sizes explicitly,
e.g. the $SU(N)$ gauge field will be denoted
$v^m_{\sst [N] \times [N]}$. To
represent matrix multiplication in this notation
we will underline contracted indices: 
$(AB)_{\sst [a] \times [c]} = \ A_{\sst [a] \times \underline{\sst [b]}} \ 
B_{ \underline{\sst [b]}\times [c]}$. Also we adopt the shorthand
$X_{\,[m}Y_{n]}=X_mY_n-X_nY_m$. 
}
\begin{equation}
\Delta_{\sst [N+2k] \times [2k]} \ (x) \ \equiv
\ \Delta_{\sst [N+2k] \times [k] \times [2]} \ (x) 
\ =\  a_{\sst [N+2k] \times [k] \times [2]} \ + 
\ b_{\sst [N+2k] \times [k] \times \underline{\sst [2]}} \ 
x_{\sst \underline{[2]} \times [2]} \ .
\elabel{dlt}\end{equation}
Here we have represented the $[2k]$ index set as a product of two index sets
$[k] \times [2]$ and have used a quaternionic representation of $x_n$,
\begin{equation}
x_{\sst [2] \times [2]}\  =\ x_{\alpha\dalpha}\ =\
 \ x_n \ \sigma^n_{\alpha\dalpha}\ .
\elabel{quat}\end{equation}
It follows that $\partial_n\Delta = b\sigma_n$.  By counting the number
of bosonic and fermionic zero modes, we will soon verify that $k$ in
Eq.~\eqref{dlt} is indeed the instanton number while $N$ is the
parameter in the gauge group $SU(N)$.  As discussed below, the
complex-valued constant matrices $a$ and $b$ in Eq. \eqref{dlt}
constitute a (highly over complete) set of $k$-instanton collective
coordinates.

For generic $x$, the null-space of the Hermitian conjugate matrix
 $\bar\Delta(x)$
is $N$-dimensional, as it has $N$ fewer rows than columns. 
The basis vectors for this null-space can be assembled
into an $(N+2k)\times N$ dimensional  complex-valued matrix $U(x)$, 
\begin{equation}
\Deltabar_{\sst [2k] \times \underline{\sst [N+2k]}} \ 
U_{\sst \underline{[N+2k]} \times [N]} 
= \ 0 = \
\Ubar_{\sst [N] \times \underline{\sst [N+2k]}} \ 
\Delta_{\sst \underline{[N+2k]} \times [2k]} \ ,
\elabel{uan}\end{equation}
where $U$ is orthonormalized according to
\begin{equation}
\Ubar_{\sst [N] \times \underline{\sst [N+2k]}} 
\ U_{\underline{\sst [N+2k]} \times {\sst [N]}}\ = \ 1_{{\sst
[N]}\times{\sst [N]}}\,.
\elabel{udef}\end{equation}
In turn, the classical ADHM gauge field $v_m$ is constructed from $U$ as
follows. Note first that
for the special case $k=0,$  the antisymmetric
gauge configuration $v_n$  defined by
\begin{equation}v_n{}_{\sst [N] \times [N]}\  = \ 
\Ubar_{\sst [N] \times \underline{\sst [N+2k]}} \ \partial_{n}
\ U_{\underline{\sst [N+2k]} \times [N]} 
\elabel{vdef}\end{equation}
is ``pure gauge'' (i.e., it is a gauge transformation of the vacuum),
 so that it automatically
solves the self-duality equation \eqref{selfdual}
in the vacuum sector. The ADHM ansatz is that Eq.~\eqref{vdef} continues
to give a solution to Eq.~\eqref{selfdual}, even for nonzero $k$. As we
shall see, this requires
the additional condition
\begin{equation}\Deltabar_{\sst [2] \times[k] \times \underline{\sst [N+2k]}} \
\Delta_{\sst \underline{[N+2k]} \times [k] \times[2]} \ = \ 
1_{\sst [2] \times [2]} \ f^{-1}_{\sst [k]\times [k]}  \ ,
\elabel{dbd}\end{equation}
where $f$ is an arbitrary $x$-dependent $k\times k$ dimensional
Hermitian matrix. 

To check the validity of the ADHM ansatz, we first observe
that  Eq.~\eqref{dbd} combined with the null-space condition \eqref{uan}
imply the completeness relation
\begin{equation}\begin{split}
{\cal P}_{[N+2k]\times[N+2k]}\ 
&\equiv\ 
U_{\sst [N+2k] \times \underline{\sst [N]}} \ 
\Ubar_{\sst \underline{[N]} \times [N+2k]}\ 
\\&=\ 
1_{\sst [N+2k] \times[N+2k]} \ -\ 
\Delta_{\sst [N+2k] \times \underline{\sst [k]}
 \times \underline{\sst [2]} } \ 
f_{\underline{\sst [k]} \times \underline{\sst [k]} }\,
\Deltabar_{\sst \underline{[2]} \times \underline{[k]} \times [N+2k]} 
\ .   
\elabel{cmpl}\end{split}\end{equation}
Note that $\cal P$, as defined, is actually a projection operator; the
fact that one can write $\cal P$ in these two equivalent ways turns
out to be a useful trick in ADHM algebra, used pervasively below.
With the above relations together with integrations by parts, the expression
for the field strength $v_{mn}$ may  be massaged as follows:
\begin{equation}\begin{split}
v_{mn}\ &\equiv \ \partial_{\,[m}v_{n]} + \ v_{\,[m} v_{n]} \ = \ 
\partial_{\,[m}(\Ubar\partial_{n]}U)+
(\Ubar\partial_{\,[m}U)(\Ubar\partial_{n]}U)\ =\ 
\partial_{\,[m}\Ubar(1-U\Ubar)\partial_{n]}U    
\\&=\ \partial_{\,[m}\Ubar \Delta f \Deltabar\partial_{n]}U \ = \ 
\Ubar\partial_{\,[m}\Delta f \partial_{n]}\Deltabar U\ =\ 
\Ubar b \sigma_{[m}\sigmabar_{n]}f \bar{b} U \ =\
4\Ubar b \sigma_{mn}f\bbar U\ .
\elabel{sdu}\end{split}\end{equation}
Self-duality of the field strength then follows automatically from the
 well-known self-duality property of the numerical tensor
 $\sigma_{mn}$, defined by \cite{WB}
\begin{equation}
\sigma_{mn\alpha}^{\phantom{mn\alpha}\beta}\ =\ \quarter(\sigma_{m\alpha\dalpha}^{}
\sigmabar_n^{\dalpha\beta}-\sigma^{}_{n\alpha\dalpha}\sigmabar_m^{\dalpha\beta})
\ .
\elabel{sigmamndef}\end{equation}
 A technical point: the above construction
 does not actually distinguish between the gauge group $U(N)$ and
 $SU(N)$, i.e., the classical gauge field constructed in this way is
 not automatically traceless (unlike the field strength). However, it
 can be made so by a gauge transformation $U \to U g^{\dagger}$, where
 $g^\dagger\in U(1)$.

Already at this stage we can infer the behavior of $\trN\,v_{mn}^2$ as
a rational function of $x$. From Eq.~\eqref{dbd} we see that the
$k\times k$ matrix $f^{-1}$ is a quadratic polynomial in
$x$. Consequently the denominator of $f$ itself is a polynomial in $x$
of degree $2k$ (i.e., like $\det f^{-1}$); the denominator of
$\trN\,v_{mn}^2$ then generically goes like the fourth power
 of this polynomial, as follows from Eqs.~\eqref{cmpl}-\eqref{sdu}. This
explains our comments at the end of Sec.~I.3, that only at the
one-instanton level can this quantity look like a classical supergravity
propagator, which only involves  quadratic forms in the
denominator. We will
see in Sec.~V below how this pessimistic observation is averted when
the ADHM formalism is combined with the large-$N$ limit.

In the next subsection we will count the independent degrees of freedom
of the ADHM configuration and confirm that it has precisely the
number of collective coordinates needed to describe a $k$-instanton
solution.

\subsection{Constraints, collective coordinates and canonical forms}

We have seen that the ADHM construction for $SU(N)$ makes essential use
of matrices of various sizes:
$(N+2k)\times N$ matrices 
$U$,  $(N+2k) \times 2k$ matrices 
$\Delta$, $a$ and $b$, $k\times k$ matrices $f$, and
$2\times2$ matrices
 $\sigma^n_{ \alpha \dalpha}$, 
 $\sigmabar^{n \, \dalpha \alpha},$ $x_{\alpha\dalpha},$ etc. 
(Notice that when $N=2,$ the dimensionalities of $U$ and $\Delta$ differ
from the ``$SU(2)$ as $Sp(1)$'' formalism reviewed in Ref.~\cite{MO-I}.)
Accordingly, we introduce a variety
of index assignments:
\begin{align}\hbox{Instanton number indices\ }[k]:\qquad&1\le i,j,l\cdots\le
k&\notag \\
\hbox{Color indices\ }[N]:\qquad&1\le u,v\cdots\le N&\notag \\
\hbox{ADHM indices\ }[N+2k]:\qquad&1\le \lambda,\mu\cdots\le N+2k&\notag \\
\hbox{Quaternionic (Weyl) indices\ }[2]:\qquad&\alpha,\beta,\dalpha,
\dbeta\cdots=1,2&\notag \\
\hbox{Lorentz indices\ }[4]:\qquad&m,n\cdots=0,1,2,3&\notag 
\end{align}
No extra notation is required for the $2k$ dimensional column index attached
to $\Delta,$ $a$ and $b$, since it can be factored as $[2k]=[k]\times[2]
=j\dbeta,$ etc., as in Eq.~\eqref{dlt}. With these index conventions, Eq.~\eqref{dlt} reads
\begin{equation}
\Delta_{\lambda \, i \dalpha}(x)
\ =\ a_{\lambda \, i \dalpha}\ +\
b_{\lambda \, i}^{\alpha}\,x_{\alpha\dalpha}\ ,\qquad
\Deltabar^{\dalpha\lambda}_{ i  }(x)
\ =\ \abar^{\dalpha\lambda}_{ i }\ +\
\bar{x}^{\dalpha \alpha} \, \bbar^\lambda_{\alpha i}\ ,
\elabel{del}\end{equation}
while the factorization condition \eqref{dbd} becomes
\begin{equation}
\Deltabar^{\dbeta\lambda}_{ i  }\,
\Delta^{}_{\lambda \, j \dalpha}\ =
 \  \delta^{\dbeta}_{\ \dalpha} \, {(f^{-1})}_{ij}\ .
\elabel{fac}\end{equation}
Combining Eqs.~\eqref{del}-\eqref{fac}, and noting that $f_{ij}(x)$ is arbitrary,
one extracts the three $x$-independent conditions on $a$ and $b$:
\begin{subequations}
\begin{align}
\abar^{\dalpha\lambda}_{ i }\, a_{\lambda  j \dbeta}\
&= \ (\hf \abar a)^{}_{ij} \ \delta^{\dalpha}_{\ \dbeta} 
\ ,
\elabel{conea}\\
\abar^{\dalpha\lambda}_{ i  }\, b_{\lambda  j}^{\beta}\
&= \ 
\bbar^{\beta\lambda}_{ i} \, a_{\lambda  j}^\dalpha\ ,
\elabel{coneb}\\
\bbar^\lambda_{\alpha i} \, b_{\lambda  j}^{\beta}\
&= \ (\hf \bbar b)^{}_{ij} \ \delta_{\alpha}^{\ \beta}\ .
\elabel{conec}\end{align}
\end{subequations}
These three conditions are known as the ADHM constraints
\cite{CGTone,CWS}. They define a set of coupled quadratic conditions
on the matrix elements of $a,$ $\abar,$ $b$ and $\bbar$. As such, for
$k>3,$ they cannot be solved in closed form in terms of algebraic
functions. This unfortunate fact is the single biggest historical
impediment that has hindered progress in multi-instanton calculus.

The elements of the matrices $a$ and $b$ comprise the collective coordinates
for the \hbox{$k$-instanton} gauge configuration. Clearly the number
of independent such elements grows as $k^2,$ even after accounting
for the constraints \eqref{conea}-\eqref{conec}. In contrast, the number of physical
collective coordinates should equal $4kN$ which scales
 linearly with $k$.{$\,$}\footnote{To see this, consider the limit of $k$
far-separated (distinguishable) instantons; each individual
instanton is then described
by four positions, one scale size, and $4N-5$ iso-orientations, totaling
$4N$ collective coordinates. Usually, as in \cite{BCGW,CWS}, 
the number of collective coordinates of the \hbox{$k$-instanton}
is quoted  as $4kN-N^2+1$ for $k\geq N/2$ and
$4k^2+1$ for $k< N/2$. These formulae represent only true collective
coordinates, that is,  excluding the global gauge rotations
of the $k$-instanton configuration. In contrast,
in our counting we include such global rotations since they appear
 in our $k$-instanton measure. Then the total number of 
collective coordinates is $4kN$.}
 It follows that
$a$ and $b$ constitute a highly redundant set. Much of this redundancy
can be eliminated by noting that the ADHM construction is
unaffected by $x$-independent transformations of the form
\begin{equation}\begin{split}\Delta_{\sst [N+2k] \times [k] \times [2]}
 \ &\to \ \Lambda_{\sst [N+2k] \times \underline{[N+2k]} } 
 \ 
\Delta_{\sst \underline{[N+2k]} \times \underline{\sst [k]} 
\times [2] }
 \ B^{-1}_{\sst \underline{[k]} \times {[k]} }\ ,
\\
U_{\sst [N+2k] \times [N]} \ &\to \
\Lambda_{\sst [N+2k] \times \underline{[N+2k]} } \ 
U_{\sst \underline{[N+2k]} \times [N]}\ ,
\\
f_{\sst [k] \times [k]} \ &\to \
B_{\sst [k] \times \underline{[k]} } \ 
f_{\sst \underline{[k]} \times \underline{[k]}} \
B^\dagger_{\sst \underline{[k]} \times [k] }\ ,
\elabel{tra}\end{split}\end{equation}
provided $\Lambda \in U(N+2k)$ and $B \in Gl(k, {\bf C})$.
(These are in addition to the usual space-time gauge symmetries
reviewed in Sec.~6 of \cite{MO-I}.)
Exploiting these symmetries, one can choose
a representation in which $b$ assumes a simple  canonical
form \cite{CGTone}: 
\begin{equation}
b_{\sst [N+2k]\times [2k]} \ = \ 
\begin{pmatrix} 0_{\sst [N]\times [2k]} \\  1_{\sst [2k]\times [2k]}\end{pmatrix}   
\ , \qquad
a_{\sst [N+2k]\times [2k]} \ = \ 
\begin{pmatrix} w_{\sst [N]\times [2k]} \\  a'_{\sst [2k]\times
[2k]}\end{pmatrix}\ .
\elabel{canform}\end{equation}

We can make this canonical form a little more explicit with a 
convenient representation of the index set  $[N+2k]$.
We decompose each ADHM index $\lambda\in[N+2k]$  into\footnote{The Weyl
index $\beta$ in this decomposition is raised and lowered with the
$\epsilon$ tensor as always \cite{WB}, whereas for the $[N]$ and $[k]$ indices
$u$ and $l$ there is no distinction between upper and lower indices.}
\begin{equation}
\lambda = u + l\beta\ ,\quad1\le u\le N\ ,\quad1\le l\le k\ ,
\quad\beta=1,2\ .  
\elabel{rplam}\end{equation}
In other words, the top $N\times2k$ submatrices in Eq.~\eqref{canform} have rows
indexed by $u\in[N],$ whereas the bottom $2k\times2k$ submatrices have
rows indexed by the pair $l\beta\in[k]\times[2].$ Equation \eqref{canform}
then becomes
%
%
\begin{subequations}
\begin{align}
a_{\lambda \, i \dalpha}\,&=
 \, a_{(u+l\beta) \, i \dalpha}\, = \, 
\begin{pmatrix}  w_{u \, i \dalpha}\\ 
(a'_{\beta \dalpha})^{ }_{li}\end{pmatrix}_{\phantom{q}}
\ ,\elabel{aaa}\\
\abar^{\dalpha\lambda}_{i} \, &= \, \abar_{i}^{\dalpha\,
(u+l\beta)}
 \, = \,
\big(\wbar^\dalpha_{i u}\ ,\ (\abar^{\prime\dalpha\beta})^{}_{il} 
\big)^{\phantom{T} }_{\phantom{q}}\ ,\elabel{aab}
\\
b_{\lambda \, i}^\alpha\,&=\,b_{(u+l\beta)\, i}^\alpha\, =\,
\begin{pmatrix}  0 \\ \delta_{\beta}^{\ \alpha}\, 
\delta_{li}^{}\end{pmatrix}^{\phantom{T}}_{\phantom{q}}
\ , \elabel{aac}\\
\bbar_{\alpha i}^\lambda \, &= \,\bbar^{u+l\beta}_{\alpha i}\,=\,
\big(0 \ ,\ \delta^{\ \beta}_{\alpha}\, \delta_{il}^{} 
\big)^{\phantom{T}}\ .
\elabel{aad}\end{align}
\end{subequations}
With $a$ and $b$ in the canonical form \eqref{aaa}, 
the third ADHM constraint of \eqref{conec} is
satisfied automatically, while the remaining constraints \eqref{conea}
and \eqref{coneb}
boil down to:
\begin{subequations}
\begin{align}
\trtwo\,\tau^c \abar a \ &= \ 0\elabel{fconea}\\
(a^{\prime}_n)^\dagger  \ &= \ a^{\prime}_n\ .
\elabel{fconeb}\end{align}
\end{subequations}
In Eq.~\eqref{fconea} we have contracted $\abar^\bD a_\aD$ with the three Pauli matrices
$({\tau^c})_{\ \dbeta}^{\dalpha}$, while in Eq.~\eqref{fconeb} we have decomposed
 $(a'_{\alpha\dalpha})_{li}$ and $(\abar^{\prime\dalpha\alpha})_{il}$
in our usual quaternionic basis of spin matrices: 
\begin{equation}(a'_{\alpha \dalpha})^{}_{li} \ = \ 
(a'_n)^{}_{li} \ \sigma^n_{ \alpha\dalpha} \ , \quad
(\abar^{\prime\dalpha \alpha})^{}_{il} \ = \
(a'_n)^{}_{il} \ \sigmabar^{n \, \dalpha \alpha}\ .
\elabel{dec}\end{equation}

Note that the canonical form for $b$ given in Eq.~\eqref{aaa} 
is  preserved by a residual $U(k)$ subgroup
of the \hbox{$U(N+2k)\times Gl(k,{\bf C})$} symmetry group
\eqref{tra}, namely:
\begin{equation}
\Delta_{\sst [N+2k]\times [2k]} \ \to \ 
\begin{pmatrix} 1_{\sst [N]\times [N]} & 0_{\sst [2k]\times [N]} \\ 
0_{\sst [N]\times [2k]}  & {\cal R}^\dagger_{\sst [2k]\times [2k]}\end{pmatrix} 
\ \Delta_{\sst [N+2k]\times [2k]} \ {\cal R}_{\sst [2k]\times [2k]}
\elabel{res}\end{equation}
where ${\cal R}_{\sst [2k]\times [2k]} 
 =  R_{ij} \ \delta_{\ \dalpha}^\dbeta$
and $R_{ij} \in U(k)$. In terms of $w$ and $a'$, this residual
transformation acts as
\begin{equation}w_{u  i\dalpha}\, \ \to \ w_{u  j\dalpha}\, R_{ji}
\ , \quad 
(a'_{\alpha\dalpha})_{ij} \ \to \ R^\dagger_{il} \ 
(a'_{\alpha \dalpha})_{lp}\ R_{pj}\ .
\elabel{restw}\end{equation}
It follows that
 the physical moduli space, ${\rm M}^k_{\rm phys}$, 
of inequivalent self-dual gauge configurations in the topological
sector $k$ is the
quotient of the space ${\rm M}^k$ of all 
solutions of the ADHM canonical constraints \eqref{fconea} and \eqref{fconeb},
by this residual symmetry group $U(k)$: 
\begin{equation}{\rm M}^k_{\rm phys} \  =\ {{\rm M}^k \over U(k)}\ .
\elabel{modspace}\end{equation}

Finally we can count the 
independent collective coordinate degrees of freedom of the ADHM 
multi-instanton. A general complex matrix 
$a_{\sst [N+2k]\times [2k]}$ has $4k(N+2k)$ real degrees of freedom.
The two ADHM conditions \eqref{fconea} and \eqref{fconeb} impose
$3k^2$ and $4k^2$ real constraints, respectively, while modding out
by the residual $U(k)$ symmetry removes another
 $k^2$ degrees of freedom. In total we therefore have
\begin{equation}
4k(N+2k)-3k^2-4k^2- k^2 \, = \, 4kN 
\elabel{dofb}\end{equation}
real degrees of freedom, precisely as required. Of these, the four real
degrees of freedom $\com_{\alpha\dalpha}=\com_n\sigma^n_{\alpha\dalpha}$
corresponding to
\begin{equation}a_{\lambda i\dalpha}\ =\ -b^\alpha_{\lambda i}\,\com_{\alpha\dalpha}
\elabel{transmo}\end{equation}
are the translational collective coordinates, as is obvious from Eq.~\eqref{del}.

\subsection{Asymptotics of the multi-instanton}

Let us determine the instanton $v_n$ more explicitly, in a
 particularly useful gauge.  This entails solving
 for $U$, and hence $v_n$ itself via \eqref{vdef}, in terms of $\Delta$.  It is
 convenient to make the decomposition:
\begin{equation}U_{\sst [N+2k]\times [N]} \ = \
\begin{pmatrix} V_{\sst [N]\times [N]} \\  U'_{\sst [2k]\times [N]}\end{pmatrix}
 \ , \quad
\Delta_{\sst [N+2k]\times [2k]} \ = \ 
\begin{pmatrix} w_{\sst [N]\times [2k]} \\ 
\Delta'_{\sst [2k]\times [2k]}\end{pmatrix}\ .
\elabel{dcmp}\end{equation}
Then from the completeness condition \eqref{cmpl} one finds
\begin{equation}
 V_{\sst [N]\times [N]} \ \Vbar_{\sst [N]\times [N]} \ = \ 
1_{\sst [N]\times [N]} \ - \ 
w_{\sst [N] \times \underline{[k]} \times \underline{[2]}} \ 
f_{\sst \underline{[k]} \times \underline{[k]}} \ 
\wbar_{\sst \underline{[2]} \times \underline{[k]} \times [N]}\ .
\elabel{cmpag}\end{equation}
For any $V$ that solves this equation, one can find another by
right-multiplying it by an $x$-dependent $U(N)$ matrix. A specific choice of 
$V$ corresponds to fixing the space-time gauge. The ``instanton singular
gauges'' correspond to taking any one of the $2^N$ choices of
matrix square roots:
\begin{equation}V \ = \ (1 \ - \ w f \wbar)^{1/2} \ .
\elabel{sga}\end{equation}
Next, $U'$ in Eq. \eqref{dcmp} is determined in terms of
$V$ via
\begin{equation}
U' \ = \ -\Delta' f \wbar \Vbar^{-1} 
\elabel{prim}\end{equation}
which likewise follows from Eq.~\eqref{cmpl}.

Equations \eqref{sga} and \eqref{prim} determine $U$ in \eqref{dcmp},
and hence the gauge field $v_n$ via Eq. \eqref{vdef}.
We list for later use the leading large-$|x|$ asymptotic behavior of
several key ADHM quantities, assuming instanton singular gauge
\eqref{sga}:
\begin{equation}
\Delta\ \rightarrow\ bx\ ,\qquad f_{kl}\ \rightarrow\
{1\over|x|^2}\,\delta_{kl}\ , \qquad U'\ \rightarrow\
-{1\over|x|^2}\,x\,\wbar\ ,\qquad V\ \rightarrow\ 1_{\sst [N] \times
[N]} \ .  \elabel{asymadhm}\end{equation}

\subsection{Connection to the usual one-instanton collective coordinates, and
 the dilute instanton gas limit}

It is illuminating to compare the ADHM collective coordinates
contained in the matrix $a$, to the more familiar variables describing
the instanton position, size, and iso-orientation. Let us first focus on
the one-instanton sector. In particular,
let us verify that the usual one instanton solution
\cite{BPST,tHooft} follows from the general ADHM formalism reviewed above.
 We adopt the canonical
 form \eqref{aaa}-\eqref{aad}
and set the instanton number $k=1$, thus dropping
 the $i,j$ indices. Contrary to the ``$SU(2)$ as $Sp(1)$'' treatment
 of Ref.~\cite{MO-I}, now the ADHM constraints \eqref{fconea} and
 \eqref{fconeb} do not disappear in the one-instanton sector. Instead,
 Eq.~\eqref{fconeb} says that $a'_n$ is real,
\begin{equation}
a'_n \ \equiv \ -\com_n \ \in \ {\Bbb R}^4\ , 
\elabel{xcon}\end{equation}
after which Eq.~\eqref{fconea} collapses to
\begin{equation}\wbar^\dalpha_u \, w_{u \dbeta} \ = \ 
\rho^2\,\delta^\dalpha_{\ \dbeta}  \ .
\elabel{wcon}\end{equation}
The quantities $\rho$ and $\com_n$ will soon be identified with the instanton
scale size and space-time position, respectively.
It is convenient to put $w$ in the form:
\begin{equation}w_{u\aD}
\ = \ \rho\grp_{\sst [N] \times [N]} \ 
\begin{pmatrix}0_{\sst [N-2] \times [2]} \\ 
  1_{\sst [2]\times [2]}\end{pmatrix} \ , \qquad 
\grp \in {SU(N) \over SU(N-2)}\ .
\elabel{wcon2}
\end{equation}
Setting $\grp=1$ initially, we find for $\Delta$ and $f\,$:
\begin{equation}
\Delta_{\sst [N+2]\times [2]} \ = \ 
\begin{pmatrix} 0_{\sst [N-2] \times [2]} \\ 
  \rho \cdot 1_{\sst [2]\times [2]} \\
  y_{\sst [2]\times [2]}\end{pmatrix}\ ,\qquad
f  \ = \ {1 \over y^2 + \rho^2} \ , 
\elabel{dlsi}\end{equation}
with $y=  (x-\com)$. 
Equations \eqref{sga}-\eqref{prim} then amount to
\begin{equation}
V_{\sst [N]\times [N]}  \ = \ 
\begin{pmatrix} 1_{\sst [N-2]\times [N-2]} & 0 \\
0 & \left( {y^2 \over y^2 + \rho^2} \right)^{1/2} 
1_{\sst [2]\times [2]} \end{pmatrix}
\elabel{veq}\end{equation}
and
\begin{equation}U'_{\sst [2]\times [N]} \ = \ 
\begin{pmatrix} 0_{\sst [2]\times [N-2]} &\ , \ &
-\left( {\rho^2 \over y^2 (y^2+ \rho^2)} \right)^{1/2} 
y_{\sst [2]\times [2]}  \end{pmatrix}\ .
\elabel{ueq}\end{equation}
The gauge field then 
follows from Eq. \eqref{vdef}:
\begin{equation}
v_n \ = \ \begin{pmatrix} 0 & 0 \\ 0 & v_n^{\sst SU(2)} \end{pmatrix}
\ .
\elabel{sutin}\end{equation}
Here
$v_n^{\sst SU(2)}$ is the standard singular-gauge $SU(2)$ instanton \cite{tHooft}
with  space-time position $\com_n\,$, scale-size $\rho$, and
in a fixed ``reference'' iso-orientation:
\begin{equation}
 v_n^{\sst SU(2)} (x)\ = \ {\rho^2 \ \etabar^a_{nm}\ (x-\com)^m \ \tau^a   
\over (x-X)^2 \ ((x-\com)^2 + \rho^2)}
\ , 
\elabel{singin}\end{equation}
where $\etabar^a_{mn}$ is an 't Hooft eta symbol (see the Appendix).
For a general iso-orientation matrix $\grp$ we obtain instead
\begin{equation}
v_n \ = \ \grp \
\begin{pmatrix} 0 & 0 \\ 0 & v_n^{\sst SU(2)}\end{pmatrix}{\grp}^\dagger 
\ , \qquad 
\grp \in {SU(N) \over U (1) \times SU(N-2)}\ .
\elabel{inss}\end{equation}
We see that the instanton always lives in an $SU(2)$ subgroup of the $SU(N)$
gauge group. An explicit representation of this embedding is formed by
the three composite $SU(2)$ generators
\begin{equation}
\left(t^c\right)_{uv}\  =\ \rho^{-2}\,w^{}_{u\aD}
\left(\tau^c\right)^\aD_{\ \bD}\bar
w_{v}^\bD,\qquad c=1,2,3.
\elabel{embed}
\end{equation}

Next let us consider the general $k$-instanton configuration. In order
to make contact with familiar non-ADHM collective coordinates, it is
necessary to pass to the dilute instanton gas limit. In ADHM language
this corresponds to the limit
\begin{equation}
[a'_n\,,\,a'_m]\ \rightarrow\ 0 \quad \forall\ m,n\ .
\elabel{apcom}
\end{equation}
Note that this condition is invariant under the residual $U(k)$
symmetry \eqref{restw}. However, Eq.~\eqref{apcom} means that there
exists a preferred $U(k)$ transformation which simultaneously
diagonalizes the four $a'_n$ matrices:
\begin{equation}
a'_n\ =\ {\rm diag}\big((a'_n)_{11}^{},\ldots,(a'_n)_{kk}^{}\big)\ .
\end{equation}
With this choice, the individual collective coordinates of the $k$
far-separated instantons are the obvious analogs of the one-instanton
expressions \eqref{xcon}, \eqref{wcon} and \eqref{embed}:
\begin{subequations}
\begin{align}
\com_n^i\ &=\ -(a'_n)_{ii}^{}\, ,
\elabel{xconmulti}\\
\rho_i^2\ &=\ \hf\wbar^\dalpha_{iu} \, w^{}_{ui \dalpha} \, ,
\elabel{wconmulti}\\
\left(t^c_i\right)_{uv}\  &=\ \rho_i^{-2}\,w^{}_{ui\aD}
\left(\tau^c\right)^\aD_{\ \bD}\bar
w_{iv}^\bD\, ,
\elabel{embedmulti}
\end{align}
\end{subequations}
where the (unsummed) index $i=1,\ldots,k$ labels the individual instantons.

\subsection{Construction of the adjoint fermion zero modes}

In a supersymmetric theory the gauge field $v_m$ is accompanied 
by a gaugino $\lambda^A$, where the index $A$ runs over the number of
independent supersymmetries. In particular,  the $\N=4$ model has an
$SU(4)_R\cong SO(6)_R$ $R$-symmetry, and the $\lambda^A$ transform in
the fundamental representation of the $SU(4)_R$ or equivalently in the
spinor representation of the $SO(6)_R$.

In the leading semiclassical approximation, the $\lambda^A$ are
 replaced by the non-trivial solutions to the covariant Weyl equation
 in the ADHM background,
\begin{equation}\Dbarslash \lambda^A\  =\ 0\ ,
\elabel{nontriv}\end{equation}
where $\Dbarslash=\sigmabar^n\D_n$.
By the index theorem, the zero modes of $\Dbarslash$ comprise $2kN$
independent Grassmann degrees of freedom for each \susy.  As discussed
in \cite{NSVZ} in the one-instanton context, these zero modes can be thought
of as the superpartners of the instanton.  Explicit expressions for
the adjoint fermion zero modes in the ADHM background were first
obtained in \cite{CGTone}. In our notation they read (cf.~Eq.~(7.1) of
\cite{MO-I}):
\begin{equation}
(\lambda^A_\alpha)^{}_{uv} \ = \ 
\Ubar^\lambda_{u} \M^A_{\lambda i} f_{ij}\bbar^\rho_{\alpha j}U_{\rho v} \ - \ 
\Ubar^\lambda_{u} 
b_{\lambda i\alpha}\, f_{ij} \Mbar^{\rho A}_{j} U_{\rho v} \ . 
\elabel{lam}\end{equation}
Here $\M^A_{\lambda i}$ and $\Mbar^{\rho A}_j$ are constant $(N+2k)\times k$ 
and $k\times (N+2k)$ matrices of Grassmann collective coordinates;
they can be viewed as either  
two  real Grassmann matrices or as two complex 
Grassmann matrices which are Hermitian conjugates of one another.

In order to verify that the ansatz \eqref{lam} satisfies the Weyl equation
\eqref{nontriv}, we use the general differentiation formulae (see
Eqs.~\eqref{vdef}, \eqref{cmpl}, \eqref{del} and \eqref{fac} above)
\begin{equation}\partial_nf\ =\
-f\cdot\partial_n(\hf\Deltabar^\aD\Delta_\aD)\cdot f\ =\
\begin{cases}
-f\cdot\sigmabar_n^{\aD\alpha}\bbar_\alpha\Delta_\aD\cdot f & \\
-f\cdot\Deltabar^\aD b^\alpha\sigma_{n\alpha\aD}\cdot f & \end{cases}
\elabel{fderivs}\end{equation}
together with 
\begin{equation}\begin{split}
\D_n(\Ubar\J U)\ &\equiv\ \partial_n(\Ubar\J U)+[v_n,\Ubar\J U]\\
&=\ \Ubar\,\partial_n\J\,U-\bar Ub^\alpha\sigma_{n\alpha\aD}
f\bar\Delta^\aD\J U-\bar U\J\Delta_\aD f\bar\sigma_n^{\aD\alpha}\bar
b_\alpha U\, ,
\elabel{genderiv}\end{split}\end{equation}
valid for any $\J(x)$. From 
Eqs.~\eqref{lam}-\eqref{genderiv} we then calculate:
\begin{equation}\Dbarslash^{\dalpha\alpha}
\lambda^A_\alpha \ =\ 
2\Ubar b^\alpha f\big(\Deltabar^{\dalpha}\M^A+
\Mbar^A\Delta^\dalpha\big)f\bbar_\alpha U\ .
\elabel{zmid}\end{equation}
Hence the condition for a gaugino zero mode 
is the following two sets of
linear constraints on $\M^A$ and $\Mbar^A$ which ensure that the right-hand side
vanishes (expanding $\Delta(x)$ as $a+bx$) \cite{CGTone}:
\begin{subequations}
\begin{align}
\Mbar^{\lambda A}_{i} \, a_{\lambda j\dalpha}
 \ &= \ - \abar^\lambda_{i\dalpha} \, \M^A_{\lambda j}
\ ,\elabel{zmcona}\\
\Mbar^{\lambda A}_i \, b_{\lambda j}^\alpha \ &= \  \bbar_{i}^{\alpha\lambda}
 \, \M^A_{\lambda j}\ .  
\elabel{zmconb}\end{align}
\end{subequations}
 In a formal sense discussed
 in Ref.~\cite{KMS}, these fermionic constraints are the ``spin-$\hf$''
 superpartners of the original ``spin-1'' ADHM constraints
 \eqref{conea} and \eqref{coneb},
 respectively.  Note further that Eq.~\eqref{zmconb} is easily solved when
 $b$ is in the canonical form
\eqref{aaa}. 
With the ADHM index decomposition \eqref{rplam}, we set
\begin{equation}\M^A_{\lambda i} \ \equiv \ \M^A_{(u+l\beta)\, i} \ = \
\begin{pmatrix} \mu^A_{u i} \\  (\M^{\prime A}_\beta)_{li}\end{pmatrix}   \ ,\quad
\Mbar^{\lambda A}_i \ \equiv \ {\Mbar_i}^{u+l\beta,A} \ = \ 
\left( \mubar^A_{i u} \ ,\ (\Mbar^{\prime\beta A})_{il} \right)\ .
\elabel{mrep}\end{equation}
Equation \eqref{zmconb} then collapses to
\begin{equation}\Mbar^{\prime A}_\alpha\ = \
\M^{\prime A}_\alpha
\elabel{mctw}\end{equation}
which allows us to eliminate $\Mbar^{\prime A}$ in favor of
$\M^{\prime A}$.

\subsection{Classification and overlap formula for the fermion zero modes}

Let us classify these Weyl spinor zero modes. Two  can be immediately
distinguished, namely those proportional to the ADHM matrices $a$ and
$b$. We choose the following linear combinations
\begin{subequations}
\begin{equation}\M^A_{\lambda i}\ =\ 4b_{\lambda i}^\alpha\,\xi^A_\alpha\ ,\qquad
\Mbar^{\lambda A}_i\ =\ 4\bbar^\lambda_{\alpha i}\,\xi^{\alpha A}
\elabel{susymo}
\end{equation}
and
\begin{equation}\begin{split}
\M^A_{\lambda i}\ &=\ 4\,a_{\lambda i\dalpha}\etabar^{\dalpha
A}-4k^{-1}b^\alpha_{\lambda i}{\rm tr}_k(a'_{\alpha\aD})
\etabar^{\dalpha A}\\ \Mbar^{\lambda A}_i\ &=\
-4\abar^{\dalpha\lambda}_i\, \etabar^A_\dalpha+4k^{-1}\bar b_{\alpha
i}^{\lambda}{\rm tr}_k(\bar a^{\prime\aD\alpha})\etabar^A_\dalpha\ ,
\elabel{suconmo}\end{split}\end{equation}\end{subequations} 
where $\xi^A_\beta$ and $\etabar^{\dalpha A}$ are arbitrary spinor
parameters.  These are the so-called ``supersymmetric'' and
``superconformal'' zero modes,\footnote{Subtracting the trace terms in
\eqref{suconmo}, which amounts to 
 an admixture of the supersymmetric mode, corresponds to
performing the superconformal transformation about the center of the
multi-instanton, rather than about the arbitrary origin of
space-time.} respectively \cite{NSVZ}; they satisfy the fermionic
constraints \eqref{zmcona}-\eqref{zmconb} by virtue of the ADHM
constraints \eqref{conea}-\eqref{coneb}.

As for the remaining $8kN-16$ modes, it is simplest to focus first on
the one-instanton sector, $k=1$, with the instanton oriented as in
Eqs.~\eqref{dlsi}-\eqref{sutin}, and with $\grp$ originally set to unity. Apart
from the sixteen modes
\eqref{susymo}-\eqref{suconmo}, there are $8N-16$ additional fermionic
zero modes which are the superpartners to gauge orientations
\cite{Cordes}. They are constructed by setting $\M^{\prime A}=0$ and
also $\mu^A_u=0$ for $u=N-1$ or $N$, with arbitrary choices for
$\mu_u^A$ for $u\le N-2$; by inspection, these satisfy the constraints
\eqref{zmcona}-\eqref{zmconb}. Turning on the orientation matrix
$\grp$ as in Eq.~\eqref{inss} simply rotates these choices of $\M^A$ by
$\grp$. 

Next let us construct these modes for $k>1.$ Our strategy is first to
consider the bosonic gauge orientation modes, and then to act on them
with supersymmetry. By definition, these orientation modes must
preserve all gauge-invariant combinations of the bosonic collective
coordinates contained in the matrix $a$. The (global) gauge dependence
of a collective coordinate is carried by the index $u$ which
transforms in the fundamental representation of $SU(N)$; this index is
attached to the submatrix $w$ but not to the submatrix $a'$ (see
Eq.~\eqref{aaa}). 
A natural set of gauge-invariant collective coordinates (used
pervasively below) is obtained by
constructing bosonic bilinear variables $W$ in which $u$ is summed
over:
\begin{equation}\big(W_{\ \dbeta}^\dalpha\big)_{ij}=\bar w_{iu}^\dalpha
 \,w_{uj\dbeta}\ ,\quad
W^0={\rm tr}_2\,W,\quad W^c={\rm
tr}_2\,\tau^cW, \ \ c=1,2,3\ .
\elabel{bosbi}\end{equation}
By definition the infinitesimal gauge orientation modes $\delta w$ are
the ones which preserve all the $W$'s, i.e., which satisfy
\begin{equation}
\bar w^\aD_{iu}\delta w_{uj\bD}+\delta\bar w^\aD_{iu}w_{uj\bD}=0\, .
\elabel{stwbw}\end{equation}
Now we  consider the fermionic superpartners of these
modes. Under a supersymmetry transformation one simply has \cite{KMS}:
\begin{equation}
\delta w_{ui\aD}=i\bar\xi_{\aD A}\mu^A_{ui}\ , \qquad \delta
\bar w^\aD_{iu}=-i\bar\mu_{iu}^A\bar\xi^{\aD}_A\ .
\elabel{susytr}\end{equation}
Inserting Eq.~\eqref{susytr} into Eq.~\eqref{stwbw} produces the
gauge-invariant conditions
\begin{equation}
\bar\xi_{\bD A}\bar w^\aD_{iu}\mu_{uj}^A+\bar\xi^{\aD
}_A\bar\mu^A_{iu}w_{uj\bD}\ =\ 0\, 
\elabel{stwbe1}
\end{equation}
or equivalently,
\begin{equation}
 \bar w_{iu}^\aD\mu_{uj}^A=0 \quad\text{and}\quad 
\bar\mu_{iu}^Aw^{}_{uj\aD}=0\ .
\elabel{stwbe}\end{equation}
To satisfy these constraints, it is convenient to decompose $\mu^A$ as
follows:
\begin{equation}
\mu_{iu}^A=w_{uj\aD}(\zeta^{\aD
A})_{ji}+\nu_{iu}^A,\qquad
\bar\mu_{iu}^A=(\bar\zeta^{
A}_\aD)_{ij}\bar w_{uj}^\aD+\bar\nu_{iu}^A\ ,
\elabel{E44}\end{equation}
where $\nu^A$ lies in the orthogonal subspace to $w$:
\begin{equation}
\bar w_{iu}^\aD\nu_{uj}^A=0,\qquad \bar\nu_{iu}^Aw^{}_{uj\aD}=0\ .
\elabel{E45}\end{equation}
The superpartners of the bosonic coset coordinates are then precisely
the variables
$\{\nu^A,\bar\nu^A\}$.\footnote{It is worth
mentioning that although the coset coordinates correspond to bosonic zero
modes which are generated by Lagrangian symmetries, this is not true
of their fermionic partners.} 

Now let us count the number of fermion modes of these various
types. It is easily checked (see Eq.~\eqref{E46} below) that the
fermionic ADHM constraint \eqref{zmcona} only involves the
$\{\M^{\prime A},\zeta^A,\bar\zeta^A\}$ modes, and not the
$\{\nu^A,\nubar^A\}$ modes. Our convention will be to use this
constraint to eliminate the $\zetabar^A$ in favor of the
others. Counting real independent Grassmann degrees of freedom, we
thus find $8k^2$ of the $\M^{\prime A}$ modes (which include the eight exact
supersymmetric modes \eqref{susymo}), $8k^2$ of the $\zeta^A$ modes
(which,  linearly combined with the $\M^{\prime A}$ modes,
include the eight exact superconformal modes \eqref{suconmo}), and
$8kN-16k^2$ of the $\nu^A$ modes, totaling $8kN$. This is precisely as
expected: one initially has $8k(N+2k)$ real Grassmann parameters in
$\M^A$ and $\Mbar^A$, subject to $8k^2$ constraints from each of
Eqs.~\eqref{zmcona} and \eqref{zmconb}, for a net of $8kN$ independent
real gaugino zero modes.

Finally, we will need the expression for the overlap of two adjoint
fermion zero modes. Thanks to the ADHM formalism, the space-time
integration of the product of two classical fermion fields can be
equated to an ordinary trace over the product of the associated
collective coordinate matrices. The expression to be proved reads:
\begin{equation}
\int d^4x\,{\rm tr}_N\,\lambda_\alpha \zeta^\alpha\ =\ -{\pi^2\over2}
{\rm tr}_k\big[\bar{\cal M}(\Pinfty+1){\cal N}+
\bar{\cal N}(\Pinfty+1){\cal M}\big]\ .
\elabel{corrigan}
\end{equation}
Here 
\begin{equation}
\Pinfty\ =\ 
\underset{|x|\rightarrow\infty}{\rm lim}\ \P\ =\ 2-b\bbar\, ,
\elabel{Pinftydef}
\end{equation}
as per Eqs.~\eqref{cmpl} and \eqref{asymadhm},
and ${\cal M}$ and $\CN$ are the
collective coordinate matrices corresponding to $\lambda_\alpha$
 and $\zeta_\alpha$, respectively. 

The formula \eqref{corrigan}, attributed in
Ref.~\cite{OSB} to E. Corrigan \cite{Corrup}, is
proved in a similar way to the overlap formula for two bosonic vector
zero modes presented in Appendices~B-C of \cite{MO-I}. The strategy of the proof
is to show that the integrand is actually a total derivative,
\begin{equation}
{\rm tr}_N\,\lambda_\alpha \zeta^\alpha
\ =\ \eighth\partial_n\partial^n\,{\rm tr}_k\big[\bar\CM(\P+1)\CN f+
\bar\CN(\P+1)\CM f\big]\ ,
\elabel{express2}
\end{equation}
after which Eq.~\eqref{corrigan} follows from Gauss's law, together with the
asymptotic formulae of Sec.~II.3.
To verify this, let us first write out the
left-hand side of Eq.~\eqref{express2}:\footnote{Here, and in the
following, the trace on the right-hand side is either over instanton
or ADHM indices, depending on the context.}
\begin{equation}
{\rm tr}_N\,\lambda_\alpha \zeta^\alpha\ =\ 
{\rm tr}\big[\,(\bar\CN \P\CM+\bar\CM \P\CN)f\bar b_\alpha \P b^\alpha
f
+\bar\CM \P b^\alpha f\bar\CN \P b_\alpha f+\P\CM f\bar b_\alpha \P\CN
f\bar b^\alpha\,\big].
\elabel{lamzeta}
\end{equation}
We have used the cyclicity of the trace, together with the definition
\eqref{cmpl} for the projector $\P.$ Turning to the right-hand side of
Eq.~\eqref{express2}, one calculates:
\begin{equation}\begin{split}
\eighth\partial_n\partial^n\,{\rm tr}\big[\bar\CM(\P+1)\CN f\big]\
&=\ 
\fourth{\rm tr}\big[\,2\bar\CM\{\P,b^\alpha f\bar b_\alpha\}\CN f
-2\bar\CM\Delta_\aD f\bar b_\alpha \P b^\alpha f\bar\Delta^\aD\CN f
\\
&\qquad\qquad+2\bar\CM(\P+1)\CN f\bar b_\alpha \P b^\alpha f
-\bar \CM\Delta_\aD f \bar\sigma^{n\aD\alpha}\bar b_\alpha
\P\CN\partial_nf
\\&\qquad\qquad
-\bar\CM \P b^\alpha\sigma^n_{\alpha\aD}f\bar\Delta^\aD\CN\partial_nf\,
\big]
\\
&=\ 
\hf{\rm tr}\big[\,\CM f\bar b_\alpha \P b^\alpha\bar\CN(1-\P)
+\bar\CM(\P+1)\CN f\bar b_\alpha \P b^\alpha f
\\
&\qquad\qquad
-\CM f\bar b^\alpha \P\CN f\bar b_\alpha \P
-\bar\CM \P b_\alpha f\bar\CN \P b^\alpha f\,\big]
\\
&=\ 
\hf{\rm tr}_N\lambda_\alpha \zeta^\alpha
+\hf{\rm tr}\big[\,(\bar\CM\CN-\bar\CN\CM)f\bar b_\alpha \P b^\alpha
f\,\big]\ .
\elabel{mess}
\end{split}\end{equation}
Here the expressions on the right-hand sides follow from the
differentiation formulae
\begin{subequations}
\begin{align}
\partial_n\partial^n f\ &=\ 4 f\bar b_\alpha \P b^\alpha f\ ,\elabel{ddf}\\
\partial^n \P\ &=\ -\Delta_\aD f\bar\sigma^{n\aD\alpha}\bar b_\alpha
\P-\P b^\alpha\sigma^n_{\alpha\aD} f\bar\Delta^\aD\ ,\elabel{sdp}\\
\partial_n\partial^n\P\ &=\ 4\{\P,b^\alpha f\bar b_\alpha\}-4\Delta_\aD 
f\bar b_\alpha \P b^\alpha f\bar\Delta^\aD\ ,
\elabel{ddP}
\end{align}
\end{subequations}
together with the two alternate expressions in
 Eq.~\eqref{fderivs};
we have also invoked the relations \eqref{zmcona}, \eqref{zmconb} and
\eqref{cmpl} and, once again, cyclicity under the trace. From the
final rewrite in Eq.~\eqref{mess}, 
 the desired result \eqref{express2}
follows by inspection upon symmetrization in $\CM$ and $\CN$, \sl QED\rm.

\subsection{Construction of the adjoint Higgs bosons}

In addition to the gauge field $v_m$ and the four Weyl spinors
$\lambda^A_\alpha,$ the $\N=4$ vector supermultiplet contains three
complex scalars which likewise transform in the adjoint representation
of $SU(N)$. In $\N=2$ language, one of these three complex
scalars comes from the $\N=2$ gauge multiplet while the other two live
in a single adjoint matter hypermultiplet. It is convenient to
assemble these three complex scalars into an adjoint-valued
antisymmetric tensor field $A^{AB}(x)$, endowed with
the canonical kinetic energy
\begin{equation}\tfrac14\epsilon_{ABCD}\int d^4x\,\trN
\D_nA^{AB}\D_nA^{CD}\ .
\elabel{canKE}\end{equation}
The antisymmetric field is subject to
a specific reality condition:
\begin{equation}\hf\epsilon_{ABCD}A^{CD}\ =\ (A^{AB})^\dagger\, ,
\elabel{realcon}\end{equation}
where $\dagger$ acts only on gauge indices,
which implies that it transforms in the vector ${\bf 6}$
representation of $SO(6)_R$. We can transform to explicit vector
components via
\begin{equation}
A^{AB}=\frac1{\sqrt8}\bar\Sigma_a^{AB}A_a\ ,
\end{equation}
where the coefficients $\bar\Sigma^{AB}_a$ are defined in the Appendix.


In the leading semiclassical approximation, $A^{AB}(x)$ is replaced by
the solution of the classical Euler-Lagrange equation
\begin{equation}\D^2 A^{AB}\ =\ \sqrtwo
i\,[\,\lambda^A\,,\lambda^B\,]\ ,
\elabel{Higgseq}\end{equation}
where $\D^2$ is the covariant Klein-Gordon operator in the
multi-instanton  background, and $\lambda^A$ is the adjoint fermion zero mode
\eqref{lam}.
The construction of the solution to this equation for arbitrary Higgs
VEVs is one of the principal technical results of
Refs.~\cite{MO-I,KMS}. Here we will content ourselves to 
summarize the solution in the exact
conformal case of vanishing VEVs. One finds that $A^{AB}$ has the
additive form
\begin{equation}
i\,A^{AB}(x) \ =\
{1\over2\sqrtwo}\,\Ubar\,\big(\M^B f\Mbar^A
-\M^A f\Mbar^B\big)U\ +\  \Ubar \cdot
\begin{pmatrix}0_{\sst [N]\times [N]}^{} &0_{\sst[N]\times[2k]} \\
0_{\sst[2k]\times[N]} 
&\A^{AB}_{\sst [k]\times[k]}\otimes 1_{\sst[2]\times[2]}\end{pmatrix}
\cdot U \ .
\elabel{Aonedef}\end{equation}
The $k\times k$ anti-Hermitian collective coordinate matrix $\A^{AB}$
is defined as the solution to the inhomogeneous linear equation
\def\Atotdagger{{\cal A}_{\rm tot}^\dagger}
\begin{equation}\bigL\cdot\A^{AB}\ =\ \Lambda^{AB}\ .
\elabel{thirtysomething}\end{equation}
Here 
\def\Lambdabar{\bar\Lambda}
\begin{equation}\Lambda^{AB}\ =\ {1\over2\sqrtwo}\,
\big(\,\Mbar^A\M^B -\Mbar^B\M^A \,\big)
\elabel{newmatdef}\end{equation}
and $\bigL$ is a positive self-adjoint linear operator that maps the  space of
 $k\times k$ scalar-valued (anti-)Hermitian matrices onto
 itself.\footnote{Positivity of $\bigL$ follows from the Hermiticity
 of $a'_n$ and the fact that $W^0$ is the product of a (non-square)
 matrix with its Hermitian conjugate: $W^0=\bar w^\aD w_\aD$.} 
Explicitly, if $\Omega$ is such a matrix, then $\bigL$ is
 defined as \cite{MO-I}
\begin{equation}\begin{split}
\bigL\cdot\Omega\ &=\ 
\hf\{\,\Omega\,,\,W^0\,\}\,-\,\hf\trtwo\big(
[\,\abar'\,,\,\Omega\,]a'-\abar'[\,a'\,,\,\Omega\,]\big)\\
&=\ 
\hf\{\,\Omega\,,\,W^0\,\}\,+\,[a_n',[a_n',\Omega]]\ ,
\elabel{bigLreally}\end{split}\end{equation}
where $W^0$ is the Hermitian  $k\times k$ matrix 
\begin{equation}W^0_{ij}\ =\ \wbar^\dalpha_{i u} \  
w^{}_{u j\dalpha} \ , 
\quad  W^{0\dagger} \ = \ W^0
\elabel{Wdef}\end{equation}
which we introduced in Eq.~\eqref{bosbi}.
{}From Eqs.~\eqref{thirtysomething}-\eqref{Wdef} one sees that $\A^{AB}$ and
$\Lambda^{AB}$ transform in the adjoint representation of the residual
$U(k)$ \eqref{res} (i.e., like $a'$ and $\M^{\prime A}$).

Defined in this way, the Higgs field $A^{AB}(x)$ correctly
satisfies the equation \eqref{Higgseq}; see Sec.~7 of Ref.~\cite{MO-I} for calculational
details.
We also note that the constraints \eqref{conea}, \eqref{coneb},
\eqref{zmcona}, \eqref{zmconb}
and \eqref{thirtysomething} may be thought of as the
``spin-1,'' ``spin-$\hf$,''  and
``spin-0'' components of an $\N=4$ supermultiplet of constraints
\cite{MO-II,DKMn4}. 

\rsen\section{Construction of the Multi-Instanton Action}

Having constructed the gauge, gaugino, and scalar components of the
$\N=4$ superinstanton in Sec.~II, we
now derive the $k$-instanton action $\Skinst$. The expression we will
derive reads \cite{DKMn4}:
\begin{equation}
\Skinst\ =\ {8\pi^2k\over g^2}\ -ik\theta\ +\ S^k_{\rm quad}\ .
\elabel{Skinstdef}\end{equation}
Here $\Skquad$ is a particular fermion quadrilinear term, with one
fermion collective coordinate chosen from each of the four gaugino
sectors $A=1,2,3,4\,$:
 \begin{equation}\Skquad\ =\ 
{\pi^2\over
g^2}\,\epsilon_{ABCD}\,{\rm tr}_k\,\Lambda^{AB}
{\cal A}^{CD}\
=\ {\pi^2\over
g^2}\,\epsilon_{ABCD}\,{\rm tr}_k\,\Lambda^{AB}
\bigL^{-1}\Lambda^{CD}\  ,  
\elabel{Skquadef}\end{equation}
with $\bigL$ as in Eq.~\eqref{bigLreally}. 
As an important consistency check, note that this expression does not lift
any of the sixteen exact supersymmetric and superconformal zero modes
\eqref{susymo}-\eqref{suconmo}; indeed, thanks to the constraints 
\eqref{zmcona}-\eqref{zmconb} these do not appear in $\Lambda^{AB}$.
In contrast, the remaining $8kN-16$ fermionic modes do enter into
$\Skquad,$ and can therefore be lifted by bringing down powers of the
action from the exponent.

We know of three distinct derivations of
Eqs.~\eqref{Skinstdef}-\eqref{Skquadef}.  The first, given
in Ref.~\cite{DKMn4}, is somewhat indirect, as it uses the supersymmetry
properties of the collective coordinates; the second, presented below,
is more straightforward, albeit calculationally intensive. In
Ref.~\cite{DKMn4}, we considered the $\N=4$ model with $SU(2)$ gauge
group, on the Coulomb branch in which an adjoint Higgs VEV $\rmv$
breaks the gauge symmetry down to $U(1)$. Using methods from previous
work \cite{MO-I,MO-II} we calculated the fermion bilinear terms in the
$k$-instanton action; they are proportional to either $\rmv$ or
$\bar\rmv$.  Then we reasoned as follows. It is known that the supersymmetry
algebra can be implemented directly on the (unconstrained)
multi-instanton collective coordinates $\{a,\M^A\}$
\cite{MO-II,NSVZ}. Physical quantities such as the multi-instanton
collective coordinate action should be a supersymmetric invariant.  The fermion
bilinear terms calculated in Ref.~\cite{DKMn4} can be shown not to form
a supersymmetric invariant expression by themselves. However, the unique
addition of the quadrilinear term \eqref{Skquadef}, taken together
with the bilinear terms, does produce a supersymmetric invariant
expression. Furthermore, in the conformal limit $\rmv\rightarrow0,$
only the quadrilinear term survives; in this limit it becomes a supersymmetric invariant
by itself. Note that, without first introducing the VEV, we would not
have been able to fix the overall multiplicative constant in front of
$\Skquad$ using supersymmetry alone.
Note further that while the derivation detailed in
Ref.~\cite{DKMn4} was only for the gauge group $SU(2)$, with the
notational redefinitions of \cite{KMS} it immediately extends to
general $SU(N)$.
An alternative derivation of the action will be presented in Sec.~IV.2 below using
D-brane methods.

Here, instead, we will present a more straightforward derivation of
the quadrilinear action \eqref{Skquadef}. Our strategy is simply to
plug the semiclassical expressions \eqref{vdef}, \eqref{lam} and
\eqref{Aonedef} for the gauge, gaugino and scalar fields into the
$\N=4$ component Lagrangian, and to carry out the spatial
integrations. Such a direct approach might appear unpromising, since
in general the functional form of the gauge field cannot be written
down explicitly. (Remember that for $k>3$ the nonlinear ADHM
constraints \eqref{fconea} cannot be solved in closed form; see
Eqs.~\eqref{conea}-\eqref{conec} \it ff\rm.)  However, as in the $\N=2$ models
considered in \cite{MO-I,MO-II,AOYAMA}, it turns out that the integrations can
nevertheless be accomplished using Gauss's law, and that only the
asymptotic expressions \eqref{asymadhm} are actually needed. Here are
the details.

At leading semiclassical order, the relevant parts of the $\N=4$
component Lagrangian are the gauge, Higgs and Yukawa terms
\begin{equation}\begin{split}
S\ &=\ \Lgauge\ +\ \Lhiggs\ +\ \Lyuk
\\
&=\ 
-{1\over2g^2}\int d^4 x \ \tr_{N}\,v_{mn}^2\ +\
{i\theta\over16\pi^2}\int d^4 x \ \tr_{N}\,{v_{mn} \ }^*v^{mn} 
\\
&\qquad+\ 
{1\over4g^2}\epsilon_{ABCD}\int d^4x\,\trN
\D_nA^{AB}\D_nA^{CD}\ 
\ +\ {i\over\sqrtwo g^2}\,\epsilon_{ABCD}\int d^4x\,\trN
A^{AB}\,[\lambda^C,\lambda^D]\ .
\elabel{Ltot}\end{split}\end{equation} 
As discussed at the opening of Sec.~II, the other terms in the component
Lagrangian, involving the antifermions $\lambdabar,$ or the auxiliary
components $F$ and $D$, only turn on at higher order in $g^2$ and can
be neglected for present purposes.  
The first line in Eq.~\eqref{Ltot} yields the usual $k$-instanton
contribution
\begin{equation}\Lgauge\ =\ -2\pi ik\tau\ =\ {8\pi^2k\over g^2}\ -\
ik\theta\ .
\elabel{kaction}\end{equation}
With an integration by parts together with the scalar equation of motion
\eqref{Higgseq}, the last two terms may be rewritten as
\begin{equation}\Lhiggs+\Lyuk\ =\ g^{-2}\epsilon_{1BCD}\int d^4x\big(\,
\partial_n(\trN A^{1B}\D_n A^{CD})\ +\ i\sqrtwo\trN
\lambda^1[\lambda^B,A^{CD}]\,\big)\ .
\elabel{lastwo}\end{equation}
For definiteness we have singled out the supersymmetry index `1';
alternatively this and subsequent equations may be rewritten in a
manifestly $SU(4)_R$ invariant way. The first term here, being a
total derivative, may be converted to a surface integral on the sphere
at infinity. Since it is gauge-invariant we can evaluate it in any
convenient gauge. In particular, in the instanton singular gauge
defined in Sec.~II.3 above, Eqs.~\eqref{asymadhm} and \eqref{Aonedef}
imply that in the limit $|x|\rightarrow\infty$
\begin{equation}A^{AB}\ \sim\ {1\over x^2}\ ,\qquad \D_nA^{AB}\ \sim\
{1\over x^3}
\elabel{AABasym}\end{equation}
so that this surface contribution vanishes. 

We will now show that the commutator term in Eq.~\eqref{lastwo} can likewise
be rewritten as a surface term. The general construction parallels
that used in Ref.~\cite{AOYAMA} for deriving the fermion quadrilinear in the
simpler case of the $\N=2$ model coupled to fundamental hypermultiplets.
 The basic idea is to construct quantities
$\psibar^\dalpha$ and $\Xi_\alpha$ such that
\begin{equation}
\hf\epsilon_{1BCD}[\lambda^B_\alpha,A^{CD}]\ =\ 
\Dslash_{\alpha\dalpha}\psibar^\dalpha+\Xi_\alpha\ ,
\elabel{psibardef}\end{equation}
where $\Dslash_{\alpha\dalpha}=\sigma^n_{\alpha\dalpha}\D_n$, and
$\Xi_\alpha$ is itself an adjoint fermion zero mode of the form
\begin{equation}
(\Xi_\alpha)_{uv} \ = \ 
\Ubar^\lambda_{u} \N_{\lambda i} f_{ij}\bbar^\rho_{\alpha j}U_{\rho v} \ - \ 
\Ubar^\lambda_{u} 
b_{\lambda i\alpha}\, f_{ij} \Nbar^\rho_{j} U_{\rho v} \ . 
\elabel{lamxi}\end{equation}
The collective coordinate matrix $\N$ (which will be seen to depend in a
nontrivial way on the original collective coordinates $\{a,\M^A\}$) is
subject to the usual zero mode conditions from Sec.~II.5:
\begin{subequations}\begin{align}\Nbar^{\lambda}_{i} \, a_{\lambda j\dalpha}
 \ &= \ - \abar^\lambda_{i\dalpha} \, \N_{\lambda j}
\ ,\elabel{zmconxic}\\
\Nbar^{\lambda}_i \, b_{\lambda j}^\alpha \ &= \  \bbar_{i}^{\alpha\lambda}
 \, \N_{\lambda j}\ .
\elabel{zmconxid}\end{align}\end{subequations}
With the rewrite \eqref{psibardef}, the commutator term in Eq.~\eqref{lastwo}
becomes
\begin{equation}\begin{split}
2\sqrtwo i\int d^4x\,\trN\lambda^{1\alpha}(\Dslash_{\alpha\dalpha}
\psibar^\dalpha+\Xi_\alpha)\ &=\ 2\sqrtwo i\int
d^4x\,\partial_n(\trN\lambda^{1\alpha}\sigma^n_{\alpha\dalpha}
\psibar^\dalpha)\\ & +\ \sqrt2\,i\pi^2{\rm
tr}_k\big[\Mbar^{1}(\Pinfty+1)\N+ \Nbar(\Pinfty+1)\M^1\big]
\elabel{rewritecom}\end{split}
\end{equation} 
In the first term we have used the fact that $\Dbarslash\lambda^1=0$
to pull the derivative outside the trace, while in the second term we
have invoked the zero mode overlap formula \eqref{corrigan} (which
likewise follows from a surface integration). The reader can verify
below \it a posteriori\rm, using the asymptotics of Sec.~II.3, that
$\lambda^1\sim x^{-3}$ and $\psibar\sim x^{-2}$; hence the first term
on the right-hand side of Eq.~\eqref{rewritecom}, like its Higgs
counterpart in Eq.~\eqref{lastwo}, gives a vanishing contribution at
infinity and may be dropped. Instead, it is the second term which is
entirely responsible for the final answer \eqref{Skquadef}.

Let us return to Eq.~\eqref{psibardef}, and solve for
$\psibar^\dalpha$ and $\Xi_\alpha$. As usual in ADHM calculus, some
educated guesswork is required. To this end we expand the left-hand
side, using Eqs.~\eqref{lam} and \eqref{Aonedef}:
\begin{equation}\begin{split}
\hf\epsilon_{1BCD}[\lambda^B_\alpha,A^{CD}]\ &=\ 
\hf\epsilon_{1BCD}\,\Big(\,{1\over\sqrtwo}\,
\Ubar\M^Bf\bbar_\alpha\P\M^Df\Mbar^CU
\ -\ {1\over\sqrtwo}\,\Ubar b_\alpha f\Mbar^B\P\M^Df\Mbar^CU
\\+\ 
&\Ubar\M^Bf\bbar_\alpha\P\cdot
\begin{pmatrix}0 &0 \\
0 &\A^{CD}\end{pmatrix}
\cdot U
\ -\ 
\Ubar b_\alpha f\Mbar^B\P\cdot
\begin{pmatrix}0 &0 \\
0 &\A^{CD}\end{pmatrix}
\cdot U\,\Big)
\ -\ \hbox{H.c.}
\elabel{writeout}\end{split}\end{equation}
Here $\cal P$ is the projection operator defined in Eq.~\eqref{cmpl} above.
Now, for an arbitrary quantity of the type $\Ubar\J U,$ the chain-rule identity
\begin{equation}
\Dslash_{\alpha\dalpha}\big(\Ubar\J U)\ = \
\Ubar(\delslash_{\alpha\dalpha}\J)U+
2\Ubar b_\alpha f\Deltabar_\dalpha\J U+2\Ubar\J\Delta_\dalpha
f\bbar_\alpha U
\elabel{chainrule}\end{equation}
follows trivially from Eq.~\eqref{genderiv}.
 Since $\partial_n\Delta=b\sigma_n,$ 
a comparison of Eqs.~\eqref{writeout} and \eqref{chainrule} motivates the ansatz:
\def\one{{(1)}}
\def\two{{(2)}}
\def\three{{(3)}}
\begin{equation}\psibar_\dalpha\ =\ \psibar^\one_\dalpha+
\psibar^\two_\dalpha+\psibar^\three_\dalpha
\elabel{psiansatz}\end{equation}
where
\begin{subequations}\begin{align}
\psibar^\one_\dalpha\ &=\
-{1\over8\sqrtwo}\epsilon_{1BCD}
\Ubar\M^Bf\Deltabar_\dalpha \M^D f\Mbar^C U\ -\ \hbox{H.c.}\ ,\elabel{psionedef}\\
\psibar^\two_\dalpha\ &=\
-\quarter\epsilon_{1BCD}
\Ubar\M^Bf\Deltabar_\dalpha\cdot 
\begin{pmatrix}0 &0 \\
0 &\A^{CD}\end{pmatrix}\cdot U\ -\ \hbox{H.c.}\ ,
\elabel{psitwodef}\\
\intertext{and}
\psibar^\three_\dalpha\ &=\
\Ubar\cdot
\begin{pmatrix}0 &0 \\ 0 & \G_\dalpha\end{pmatrix}\cdot U\ ,\quad\bar\G_\dalpha=-\G_\dalpha
\ ,\quad
\partial_n\G_\dalpha=0\ .
\elabel{psithreedef}\end{align}
\end{subequations}
We expect $\psibar^\one_\dalpha$ to account (more or less) for the
first  line of Eq.~\eqref{writeout}, and $\psibar^\two_\dalpha$ to
account (more or less) for the next  line. The presence of
$\psibar^\three_\dalpha$, while less obviously motivated at this
stage, will be needed to ensure that the quantity $\Xi_\alpha$ defined in
Eq.~\eqref{psibardef} obeys the zero-mode constraints
\eqref{zmconxic}-\eqref{zmconxid}.

By explicit calculation using Eqs.~\eqref{fderivs}, \eqref{zmcona},
\eqref{zmconb}, \eqref{chainrule} and \eqref{psionedef}, one finds:
\begin{equation}\begin{split}
\Dslash_{\alpha\dalpha}\psibar^{\one\dalpha}\ =\ 
{1\over2\sqrtwo}\,\epsilon_{1BCD}\,\Big(\,&\Ubar\M^Bf\bbar_\alpha\P\M^D
f\Mbar^CU
\\-\ &\Ubar b_\alpha f\Mbar^B(\P-1)\M^Df\Mbar^CU\,\Big)
\ -\ \hbox{H.c.}
\elabel{writeone}\end{split}\end{equation}
Except for the ``$-1$'' in the final term, this  reproduces
the first  line of Eq.~\eqref{writeout}, as expected. Similarly one calculates:
\begin{equation}\begin{split}
\Dslash_{\alpha\dalpha}\psibar^{\two\dalpha} =
\hf\epsilon_{1BCD}\Big(&
\Ubar\M^Bf\bbar_\alpha(\P+1)\cdot
\begin{pmatrix}0 &0 \\
0 &\A^{CD}\end{pmatrix}
\cdot U
 -
\Ubar b_\alpha f\Mbar^B(\P-1)\cdot
\begin{pmatrix}0 &0 \\
0 &\A^{CD}\end{pmatrix}
\cdot U
\\
&-\ \Ubar b_\alpha f\cdot\trtwo\Deltabar
\begin{pmatrix}0 &0 \\
0 &\A^{CD}\end{pmatrix}\Delta\cdot f\Mbar^BU
\,\Big)
\ -\ \hbox{H.c.}
\\
= \hf\epsilon_{1BCD}\Big(&
\Ubar\M^Bf\bbar_\alpha\P\cdot
\begin{pmatrix}0 &0 \\
0 &\A^{CD}\end{pmatrix}
\cdot U
\ -\ 
\Ubar b_\alpha f\Mbar^B\P\cdot
\begin{pmatrix}0 &0 \\
0 &\A^{CD}\end{pmatrix}
\cdot U
\\
&+\ \Ubar b_\alpha f\Big(\Mbar^B\cdot
\begin{pmatrix}0 &0 \\
0 &\A^{CD}\end{pmatrix}-\A^{CD}\Mbar^B\Big)U
\\
&-\ {1\over\sqrtwo}\Ubar b_\alpha f\Mbar^B\M^D f\Mbar^C U
\,\Big)
\ -\ \hbox{H.c.}
\elabel{writetwo}\end{split}\end{equation}
so that the second line in Eq.~\eqref{writeout} is accounted for as well.
Here the final equality follows from the commutator identity 
\begin{equation}\begin{split}
\trtwo\Deltabar
\begin{pmatrix}0 &0 \\
0 &\A^{CD}\end{pmatrix}\Delta\ 
&=\ 
-\bigL\cdot\A^{CD}\ +\ \{\A^{CD}\,,\,f^{-1}\}
\\&=\ 
-{1\over2\sqrtwo}\,
\big(\,\Mbar^C\M^D -\Mbar^D\M^C \,\big)
\ +\ \{\A^{CD}\,,\,f^{-1}\}
\elabel{comid}\end{split}\end{equation}
implied by Eqs.~\eqref{del}-\eqref{fac} and
\eqref{thirtysomething}-\eqref{bigLreally}.  Finally,
\begin{equation}\begin{split}
\Dslash_{\alpha\dalpha}\psibar^{\three\dalpha}\ 
&=\
2\Ubar b_\alpha f\Deltabar_\dalpha\cdot
\begin{pmatrix}0 &0 \\ 0 &\G^\dalpha\end{pmatrix}\cdot U\ -\ \hbox{H.c.}
\\
&=\
2\Ubar b_\alpha f\Big(\abar_\dalpha\cdot
\begin{pmatrix}0 &0 \\ 0 &\G^\dalpha\end{pmatrix}-\G^\dalpha\abar_\dalpha\Big)U
\ -\ \hbox{H.c.}
\elabel{psithreeb}\end{split}\end{equation}
where to obtain the final equality we have decomposed
$\Deltabar=\abar+\xbar\bbar,$ commuted $\xbar\bbar$ through $\G,$ and
replaced it by $-\abar$ thanks to the orthogonality relation \eqref{uan}.

Next we sum the expressions \eqref{writeone}, \eqref{writetwo} and
\eqref{psithreeb}, and compare to the right-hand side of
Eq.~\eqref{writeout}. By inspection, we confirm the ans\"atze
\eqref{psibardef}-\eqref{lamxi}, where the Grassmann zero mode matrix
has the form
\begin{equation}\N\ =\ \hf\epsilon_{1BCD}\Big(\,
\begin{pmatrix}0&0\\0&\A^{CD}\end{pmatrix}\M^B\ -\ \M^B\A^{CD}\,\Big)
+2\begin{pmatrix}0&0\\0&\G^\dalpha\end{pmatrix}a_\dalpha-2a_\dalpha\G^\dalpha\ .
\elabel{calNdef}\end{equation}  
Up till this point we have yet to solve for $\G^\dalpha$. This is
accomplished by inserting $\N$ into the fermionic constraints
\eqref{zmconxic}-\eqref{zmconxid}.  One finds that
Eq.~\eqref{zmconxid} is satisfied automatically by the expression
\eqref{calNdef}.  In contrast, Eq.~\eqref{zmconxic} amounts to $2k^2$
independent real linear constraints, which is precisely the number
required to fix the anti-Hermitian $k\times k$ matrices $\G^\dalpha$
completely (the explicit form for $\G^\dalpha$ is not required).

 Finally we insert this expression for $\N$ into the final term in
Eq.~\eqref{rewritecom}. Thanks to the constraints
\eqref{zmcona}-\eqref{zmconb} on $\M^1,$ the last two terms in
Eq.~\eqref{calNdef} do not contribute to the inner product. It is easily
checked, using the self-adjointness of $\bigL$,
 that the first two terms in Eq.~\eqref{calNdef} precisely
reproduce the $SU(4)_R$ invariant quadrilinear \eqref{Skquadef}, as
claimed.

\rsen\section{The Multi-Instanton  Collective Coordinate  Integration Measure}

In this section we present an in-depth discussion of the
multi-instanton collective coordinate measure. We begin with the
so-called ``flat measure,'' in which the various bosonic and fermionic
collective coordinates described in Sec.~II are integrated over as
Cartesian variables; in the flat measure, the ADHM supermultiplet of
constraints is implemented simply as a product of
$\delta$-functions. Section IV.1 reviews the field-theory derivation
of the flat measure given in Refs.~\cite{measure1,DHKM,KMS} using
uniqueness arguments, whereas Sec.~IV.2 presents an independent
rederivation of the flat measure directly from superstring theory by
exploiting the world-volume description of D-instantons in the
presence of D$3$-branes.

However, the flat measure is not optimal for our present purposes, for
the following reason. Starting in Sec.~V we will explore how the ADHM
multi-instanton formalism combines in a natural way with the large-$N$
limit. Unfortunately, for large $N$, the number of collective
coordinates contained in the matrices $\{a,\M^A\}$ grows linearly with
$N$. In order to implement a traditional saddle-point approach, it is
desirable to change to a new set of collective coordinates whose
number stays fixed as $N$ grows. A strong clue for how to proceed
comes from anticipating Sec.~VI, where we finally calculate the
correlators $G_n$ ($n=16,$ 8 or 4) and related higher-partial-wave
correlators, and put Maldacena's conjecture to the test. A common
feature of these $n$-point functions is that the $n$ insertions are
all gauge-invariant composite operators. This suggests that we change
integration variables to a smaller set of gauge-invariant collective
coordinates. With this application in mind, Sec.~IV.3 is devoted to
the construction of the so-called ``gauge-invariant measure'' from the
flat measure. Specifically, this entails calculating the Jacobians
needed to switch from the original multiplet of ADHM variables, to the
gauge-invariant bosonic and fermionic collective coordinates which we
have already identified in Sec.~II.6. Pleasingly, it turns out that
after one integrates out the gauge coset, the number of remaining
gauge-invariant variables no longer grows with $N$, but only grows
with the topological number $k$.

\subsection{The ADHM multi-instanton measure}

We begin by reviewing the ``flat measure'' on the space of ADHM
 variables.  As the small-fluctuations determinants in a self-dual
 background cancel between the bosonic and fermionic sectors in a
 four-dimensional supersymmetric theory \cite{dadda}, the relevant
 measure is the one inherited from the Feynman path integral on
 changing variables from the fields to the collective coordinates
 which parametrize the $k$-instanton moduli space ${\rm M}^k_{\rm
 phys}$. In principle the super-Jacobian for this change of variables
 can be calculated by evaluating the normalization matrices of the
 appropriate bosonic and fermionic zero-modes.  In practice, this
 involves resolving the ADHM constraints and can only be accomplished
 explicitly for $k\leq 2$.

Instead, the heuristic strategy adopted in Refs.~\cite{measure1,DHKM,KMS}
 is to postulate a measure, and to verify \it a posteriori \rm that
 its form is fixed uniquely by various consistency requirements
enumerated below.  In particular, in the ${\cal N}=4$ supersymmetric
 theory, the measure has the following form \cite{DHKM,KMS}:
\begin{equation}\begin{split}
&\int \dmuphys\ =\ {2^{-k^2/2}(C''_1)^k\over{\rm Vol}\,U(k)}
\int d^{4k^2} a' \
d^{2kN} \wbar \ d^{2kN} w  
\prod_{A=1,2,3,4}
\ d^{2k^2} \M^{\prime A}  
\ d^{kN} \mubar^A \ d^{kN} \mu^A\\
&\times \prod_{B=2,3,4}
\ d^{k^2} \A^{1B}\
\prod_{r=1,\ldots,k^2}\Big[
\prod_{c=1,2,3}
\delta\big(\tfrac12{\rm
tr}_k\,T^r(\trtwo\, \tau^c \abar a)\big)
\prod_{A=1,2,3,4}
\prod_{\aD=1,2}\delta\left({\rm tr}_k\,T^r(\Mbar^A a_\aD + \abar_\aD \M^A
)\right)\\ &\qquad\qquad\qquad\qquad\times
\prod_{B=2,3,4}
\delta\big({\rm tr}_k\,T^r(\bigL\cdot\A^{1B} -
\Lambda^{1B})\big)\Big]\ .
\elabel{dmudef}\end{split}\end{equation}
In writing the above, we have used a normalization for the integrations that
is different, but more convenient for our purposes, from that in
\cite{DHKM}. This difference accounts 
for the overall factor of $2^{-k^2/2}$. The
integrals over the $k\times k$ matrices $a'_n$, $\CM^{\prime A}$ and
${\cal A}^{AB}$ are defined as the integral over the components with
respect to a Hermitian basis of $k\times  k$ matrices $T^r$ normalized
so that ${\rm tr}_k\,T^rT^s=\delta^{rs}$. These matrices also provide
explicit definitions of the $\delta$-function factors in the way indicated.
Encoded in this measure is information about the $SU(2)$ embeddings of
the instantons inside $SU(N)$;
it is pleasing that the ADHM parametrization replaces
the traditional trigonometric variables of the coset matrix $\grp$
describing this embedding, by Cartesian
variables endowed with a flat measure (apart from the
$\delta$-function insertions).

We can offer the following consistency checks on this measure:

(i) In the $1$-instanton sector, Eq.~\eqref{dmudef} reduces to
\begin{equation}\begin{split}
&\int\,d\mu^1_{\rm phys}\ \ =\ {C''_1 \over 2^{3/2} \pi} \int d^{4} a'
\ d^{2N} \wbar \ d^{2N} w \prod_{A=1,2,3,4} d^{2} \M^{\prime A}\ d^{N}
\mubar^A \ d^{N} \mu^A \prod_{B=2,3,4}d\A^{1B} \\&\times\
\prod_{c=1,2,3} \delta\big(\tfrac12\trtwo\,\tau^c \wbar w\big)
\prod_{A=1,2,3,4}\prod_{\aD=1,2} \delta\big(\mubar^A w_\aD + \wbar_\aD
\mu^A \big) \prod_{B=2,3,4}\delta \big( \bigL\cdot\A^{1B} -
\Lambda^{1B} \big)\ .
\elabel{onesimp}\end{split}\end{equation} 
The first $\delta$-function is then resolved as per
Eqs.~\eqref{wcon}-\eqref{wcon2}; also the third $\delta$-function is
trivial since in the one-instanton sector $\bigL\equiv2\rho^2$ as per
Eqs.~\eqref{bigLreally}, \eqref{Wdef} and \eqref{wcon}. In this way,
the bosonic part of Eq.~\eqref{onesimp} precisely reproduces the
standard one-instanton measure \cite{tHooft,Bernard}; in particular,
the position $X_n$, size $\rho$, and group iso-orientation $\grp$ of
the instanton follows from the identifications made in Sec.~II.4
above.  Likewise, the fermionic collective coordinates in
Eq.~\eqref{onesimp} can be identified with the supersymmetric and
superconformal modes \eqref{susymo}-\eqref{suconmo}, and with the
superpartners of the iso-orientations zero modes discussed in
Sec.~II.6. To summarize, in the one-instanton sector we have
\begin{equation}
\det_2 W=\rho^4\,,\qquad\det\BL\equiv\BL=2\rho^2\,
\qquad\zeta^{\aD A}=4\bar\eta^{\aD A}\,,\qquad{\cal M}^{\prime
A}_\alpha=4\xi_\alpha^A\ .
\elabel{1instid}\end{equation}
After integrating over the global iso-orientation of the instanton (as we do
for the multi-instanton measure in Sec.~IV.3), the one-instanton
measure is
\begin{equation}
\int d\mu_{\rm phys}^1
={2^{2N-69/2}\pi^{2N-2}C^{\prime\prime}_1\over(N-1)!(N-2)!}\int\rho^{4N-13}\,
d^4\com\, d\rho \prod_{A=1,2,3,4}d^2\xi^A\, d^2\bar\eta^A\,
d^{(N-2)}\nu^A\, d^{(N-2)}\bar\nu^A.
\end{equation}
Comparing this with the one-instanton Bernard measure \cite{Bernard}
suitably generalized to an ${\cal N}=4$ theory (see Eq.~(13) of
Ref.~\cite{DKMV}\footnote{We have multiplied
that result by $2^8$ due to the difference in
 conventions for integrating Weyl fermions.}), 
we extract the normalization factor
\begin{equation}
C^{\prime\prime}_1=2^{-2N+1/2}\pi^{-6N}g^{4N}
\end{equation}
needed below.

(ii) The $k$-instanton measure should be 
dimensionless, since the $\beta$-function vanishes in this $\N=4$ model.
And indeed this is satisfied by the expression \eqref{dmudef},
 since the engineering length dimensions are 
$[a]=[\Lambda^{1B}]=1$, $[\M]=1/2$, $[d \M]=-1/2$, and $[\A^{1B}]=-1.$

(iii) If one inserts into the $\N=4$
measure \eqref{dmudef} a mass term for (say) gauginos
$\M^3$ and $\M^4,$ of the form
\begin{equation}
\exp\,im\,\tr_N\big(\Mbar^4(\Pinfty+1)\M^3
+\Mbar^3(\Pinfty+1)\M^4\big)\, ,
\elabel{insertmass}\end{equation}
(see Eq.~\eqref{corrigan})
and passes to the decoupling limit $m\rightarrow\infty$  in the manner
described in Sec.~5 of Ref.~\cite{DHKM}, the measure
properly flows down to the $\N=2$ measure given in Ref.~\cite{KMS}.

(iv) The measure should preserve a net of $8kN$ unsaturated Grassmann
 integrations at the $k$-instanton level, in other words, $8kN$ exact
 fermion zero modes.  It is easy to see that this counting is obeyed
 by the right-hand side of Eq.~\eqref{dmudef}: $8k^2$ fermionic
 $\delta$-functions saturate $8k^2$ out of the $8k^2+8kN$ fermionic
 integrations over $\{\M^{\prime A}, \mu^A,\mubar^A\},$ leaving $8kN$ exact
 fermion zero modes.  While there is no index theorem in the $\N=4$
 model guaranteeing this number, due to the absence of a $U(1)$ anomaly, the
 fact that this theory can be deformed into an $\N=2$ model, where the
 anomaly exists, forces this na\"\i ve counting nonetheless.

(v) The  invariance of the measure \eqref{dmudef} under the residual $U(k)$
symmetry \eqref{res} is obvious.

(vi) Cluster decomposition in the dilute-gas limit  of large
space-time separation between instantons fixes the overall constant 
in the $k$-instanton measure \eqref{dmudef} in terms of the one instanton
factor $C''_1$. The derivation is analogous to that given in \cite{measure1,KMS}.

(vii) As with all physically relevant quantities, 
the $k$-instanton measure has to be a \susic\ invariant.
This important requirement can be directly checked by
performing the supersymmetry  transformations on the collective coordinates
detailed in Ref.~\cite{KMS}. The key point here is that the three
$\delta$-functions in Eq.~\eqref{dmudef} represent the spin-1, spin-$\hf$
and spin-0 ADHM constraints \eqref{conea}, \eqref{zmcona} and
\eqref{thirtysomething}, respectively, which
together form an $\N=4$ supermultiplet of constraints. Note that in
the final $\delta$-function in Eq.~\eqref{dmudef} we have singled out the
supersymmetry index `1', so that, na\"\i vely, the
invariance is only guaranteed for  supersymmetry transformations
$\xi_AQ^A+\bar\xi_A\bar Q^A$ with $A=1$. However, since the $\A^{1B}$
integration is trivially accomplished using
\begin{equation}
\int\prod_{B=2,3,4}d^{k^2}\A^{1B} \
\prod_{r=1,\ldots,k^2}
\prod_{B=2,3,4}
\delta\big({\rm tr}_k\,T^r(\bigL\cdot\A^{1B} - \Lambda^{1B})\big)
=\big({\det}_{k^2}\bigL\big)^{-3} 
\elabel{trivint}\end{equation}
which is independent of the `1' direction, the expression \eqref{dmudef} is
actually invariant under all four supersymmetries; i.e. it is
$SU(4)_R$ invariant. 

(viii) Finally we can make the following uniqueness argument \cite{DHKM}.
Since the $\delta$-functions
 in Eq.~\eqref{dmudef} are dictated by the ADHM formalism,
and since the resulting measure turns out to be
 a supersymmetric  invariant and also preserves the correct number of
fermion zero modes,
we claim that the ansatz \eqref{dmudef} is in fact unique.
To see why, let us consider including an additional
function of the collective coordinates, $F(a,\M^A),$ in the integrand
of Eq.~\eqref{dmudef}. To preserve supersymmetry, we  require that $F$
be a supersymmetry  invariant.  It is a fact that 
any non-constant function that is a supersymmetry  invariant
must contain fermion bilinear pieces (and possibly higher powers of
fermions as well). By the rules of Grassmann integration, such bilinears
would necessarily
lift some of the adjoint fermion zero modes contained in the collective
coordinate matrices $\M^A.$ But since
Eq.~\eqref{dmudef} contains precisely the right number
of unlifted fermion zero modes as per (iv),
this argument rules
out the existence of a non-constant function $F$. Moreover, any
constant $F$ would be absorbed into the overall
multiplicative factor, which is already fixed by cluster decomposition. 

In the following subsection, we shall give an alternative, more
direct,
derivation of the $\N=4$ measure \eqref{dmudef}, as well as the
quadrilinear term \eqref{Skquadef}, from string theory.
(Note that the $\N=2,$ $\N=1$ and (classical) $\N=0$ measures
presented in \cite{measure1,DHKM,KMS}  
may in turn be derived from the $\N=4$ measure, and thus ultimately from string
theory, by renormalization group
decoupling; see item (iii) above and Sec.~5 of \cite{DHKM}.)
But first, we should note that the results of the $\A^{1B}$
integration, given by Eq.~\eqref{trivint}, may be elegantly combined with the
quadrilinear action \eqref{Skquadef} derived in Sec.~III, via the introduction of
auxiliary Gaussian $k\times k$ matrices of variables $\chi_{AB}$ transforming in the vector
${\bf 6}$ of $SO(6)_R$: 
\begin{equation}
(\det_{k^2}\BL)^{-3}\exp\,-S_{\rm quad}^k\ =\ \pi^{-3k^2}\int
d^{6k^2}\chi\exp\big[-{\rm tr}_k\,\chi_a\BL
\chi_a+4\pi ig^{-1}{\rm
tr}_k\,\chi_{AB}\Lambda^{AB}\big]. 
\elabel{E53}\end{equation}
As in Sec.~II.7, the antisymmetric tensor $\chi_{AB}$ satisfies a
pseudo-reality condition
\begin{equation}
\tfrac12\epsilon^{ABCD}\chi_{CD}=\chi_{AB}^\dagger\ ,
\elabel{E54}\end{equation}
where $\dagger$ acts only on instanton indices and not on $SU(4)_R$ indices;
$\chi_{AB}$  can be written as an explicit $SO(6)_R$ vector $\chi_a,$
$a=1,\ldots, 6$, by using the
coefficients $\Sigma^a_{AB}$ defined in the Appendix:
\begin{equation}
\chi_{AB}={1\over\sqrt8}\Sigma^a_{AB}\chi_a\, .
\elabel{E54.1}\end{equation}
In the $SO(6)$ representation the six $k\times k$ matrices $\chi_a$
are Hermitian. This kind of
transformation is a well-known tool for analyzing the large-$N$ limit
of field theories with four-fermion interactions, like the Gross-Neveu
and Thirring models \cite{GN}. We shall find that this transformation
is absolutely crucial in our analysis, since it introduces a new set
of bosonic collective coordinates, namely $\chi_a$, which will have a
very important r\^ole to play in the relation of the Yang-Mills theory
to the superstring theory.

\subsection{D-instantons and the ADHM Measure}

In this section we will show that the ${\cal N}=4$ ADHM measure of
\cite{DHKM}, described in Sec.~IV.1, can actually be derived using
D-brane techniques.  We will consider ${\cal N}=4$ supersymmetric
Yang-Mills theory realized on the world-volume $N$ D3-branes in Type
IIB string theory. According to \cite{D1,D2}, D-instantons located on
the D3-branes are equivalent to Yang-Mills instantons in the
world-volume gauge theory.  In the following we determine the
collective coordinate measure for $k$ D-instantons,
$\dmuphys \exp-\Skinst,$
in the presence of
$N$ D3-branes. In a limit where the gauge theory on the D3-branes
decouples from gravity, we find that the D-instanton measure coincides
with our earlier results for the $k$-instanton ADHM measure.  In this
Section, unlike the preceding one, we will not keep track of the
overall normalization of the measure.

We begin by briefly reviewing some basic facts about D-branes in Type II 
string theory \cite{P}. A D$p$-brane is defined in the first instance as a
$p$-dimensional hyperplane in nine spatial dimensions on which 
open strings can end. The massless states in the open string 
spectrum give rise to massless fields which propagate on the $p+1$ 
dimensional world-volume of the D-brane. Specifically, 
the massless modes or collective coordinates of a single D$p$-brane come 
from dimensional reduction to $d=p+1$ of a 
$U(1)$ vector multiplet of ${\cal N}=1$ supersymmetry in $d=10$. 
In contrast, closed string modes propagate in the ten dimensional bulk. 
After dimensional reduction, the ten-dimensional gauge field $A_{\mu}$, 
$\mu=0,1,2,\ldots 9$, yields a $p+1$ dimensional gauge field and 
$9-p$ real scalars. 
The scalars specify the location of the D-brane in the $9-p$ dimensions 
transverse to its world-volume.  The $d=10$ multiplet also includes a 
Majorana-Weyl fermion, $\Psi$. The sixteen independent components of $\Psi$
correspond to sixteen fermion zero modes of the D-brane. This number 
reflects the fact that the D-brane is a BPS configuration which breaks 
half of the 32 supersymmetries of the Type II theory. 

When $k$ parallel D$p$-branes are located at different points in their 
common transverse dimensions their massless degrees of freedom 
are simply $k$ copies of the collective coordinates described 
above, which corresponds to gauge group $U(1)^{k}$. However, 
when two or more D-branes coincide, additional states corresponding
to open strings stretched between the two branes become massless 
leading to enhanced gauge symmetry
\cite{W135}. In the maximal case, where 
all $k$ D$p$-branes coincide, the unbroken gauge group is $U(k)$. 
The low-energy effective action for the 
world-volume theory can be obtained from 
dimensional reduction of ten dimensional super-Yang-Mills theory 
(SYM$_{10}$) with gauge group $U(k)$. The ten-dimensional action is, 
\begin{equation}
S^{(10)}_{k}=\frac{1}{g^{2}_{10}}\int\, d^{10}x\, {\rm tr}_{k}
\big(\tfrac{1}{2}F_{\mu\nu}^{2} + i\bar{\Psi}\Gamma_{\mu}{\cal D}_{\mu}
\Psi\big)\, ,
\elabel{10daction}
\end{equation}
where ${\cal D}_{\mu}\Psi = \partial_{\mu}\Psi-i[A_{\mu},\Psi]$ is 
the covariant derivative in the adjoint representation and 
$F_{\mu\nu}=\partial_{\mu}A_{\nu}-\partial_{\nu}A_{\mu}-i[A_{\mu},A_{\nu}]$ 
is the ten-dimensional field strength. 
The $\Gamma$-matrices obey the ten-dimensional Euclidean Clifford algebra,  
$\{\Gamma_{\mu},\Gamma_{\nu}\}=2\delta_{\mu\nu}$.  The particular 
representation of the $\Gamma_{\mu}$ used below is given explicitly in 
the Appendix. 

Dimensional reduction to $p+1$ dimensions proceeds by setting all 
spacetime derivatives in the reduced directions to zero. 
As in the case of a single D$p$-brane, the ten dimensional gauge field 
yields a $p+1$ dimensional gauge field and $9-p$ real scalars, 
all in the adjoint representation of the 
gauge group. Configurations with some or all of the D-branes separated 
in spacetime then correspond the Coulomb branch of the world-volume 
gauge theory. In terms of string theory parameters, the Yang-Mills coupling 
constant in $d=p+1$ is identified as $g^{2}_{p+1}=2(2\pi)^{p-2}e^{\phi}
\alpha^{\prime(p-3)/2}$.
This concludes the discussion of features which are common to all D$p$-branes. 
The two cases which interest us most here are 
the D-instanton ($p=-1$) and the $D3$ brane (along the way we will 
also meet the $p=5$ and $p=9$ cases). First of all we recall that in the 
case of $N$ parallel D3-branes the low-energy effective theory on 
the brane world-volume is ${\cal N}=4$ supersymmetric Yang-Mills theory 
with gauge group $U(N)$. This configuration is the starting point 
for Maldacena's analysis \cite{MAL}. The dimensionless four-dimensional 
gauge coupling is related to the string coupling as per Eq.~\eqref{corresp}.
 In general the four-dimensional 
fields which propagate on the brane also have couplings 
to the ten-dimensional graviton and the other closed string modes. 
To decouple the four-dimensional theory from gravity it is necessary to take 
the limit $\alpha'\rightarrow 0$. In particular we will take this limit 
with the string-coupling held fixed and small. This gives a weakly coupled 
gauge theory on the D3-brane. Note that this limit is quite different from the 
large-$N$ decoupling limit considered by Maldacena in which the D3-branes 
can reliably be described as classical solutions of Type IIB supergravity. 
As emphasized in previous sections, Maldacena's limit leads instead to 
the strong coupling limit ($g^{2}N\gg 1$) of the four-dimensional gauge 
theory.       

A D-instanton (or D$(-1)$-brane) corresponds to the extreme case where 
the dimensional reduction is complete and 
the world-volume is a single point in Euclidean spacetime. 
Correspondingly, rather 
than finite mass or tension, a single D-instanton has finite action 
$S_{\rm cl}=2\pi(e^{-\phi}-ic^{(0)})$ where $c^{(0)}$ is the 
Ramond-Ramond scalar of the IIB theory. As above, the collective coordinates 
of a charge-$k$ D-instanton correspond to a $U(k)$ vector multiplet of 
SYM$_{10}$. The diagonal components of the ten-dimensional 
gauge field $A_{\mu}$ specify the location of $k$ D-instantons in
${\Bbb R}^{10}$.   
As we dimensionally reduce to $d=0$, these degrees of freedom 
are $c$-numbers rather than fields. In addition to a constant part equal 
to $kS_{cl}$, the action of a charge-$k$ D-instanton also depends on the 
collective coordinates via the dimensional reduction of \eqref{10daction},  
\begin{equation}
S_{k}=-\frac{1}{2g^{2}_{0}} {\rm tr}_{k}\,[A_{\mu},A_{\nu}]^{2} + 
{1\over g^2_0}{\rm tr}_k\,\bar{\Psi}\Gamma_{\mu}[A_{\mu},\Psi],
\elabel{zerodaction}
\end{equation}
where $g_{0}\sim (\alpha')^{-2}$.

In addition to the manifest $SO(10)$ symmetry under ten-dimensional 
rotations, the action is trivially invariant under translations of the 
form $A_{\mu}\rightarrow A_{\mu} + x_{\mu}1_{\sst [k]\times[k]}$. 
Hence $k^{-1}{\rm tr}_{k}\,A_{\mu}$, which corresponds to the abelian factor of 
the $U(k)$ gauge group, is identified with the center-of-mass coordinate 
of the charge $k$ D-instanton. 
The fermionic symmetries of the collective coordinate action are, 
\begin{equation}\begin{split} 
\delta A_{\mu} &= i\bar{\eta}\Gamma_{\mu} \Psi\ , \notag \\
\delta \Psi\,\, &=  -[A_{\mu},A_{\nu}]\Gamma^{\mu\nu}{\eta} \, + \, 
1_{\sst [k]\times [k]}\,\varepsilon\ ,
\elabel{susy}
\end{split}\end{equation}
where $\Gamma_{\mu\nu}=i[\Gamma_{\mu},\Gamma_{\nu}]/4$. 
The sixteen components of the Majorana-Weyl  
$SO(10)$ spinor $\varepsilon$  
correspond to the sixteen zero modes of the D-instanton configuration 
generated by the action of the $D=10$ supercharges. Like the bosonic 
translation modes, these modes live in 
the abelian factor of the corresponding $U(k)$ field, 
$k^{-1}{\rm tr}_{k}\,\Psi$. 
In contrast, the Majorana-Weyl spinor $\eta$ parametrizes the sixteen supersymmetries 
left unbroken by the D-instanton. These are analogous to the collective 
coordinate supersymmetries of the ADHM formalism described in \cite{MO-II}.     

In ordinary field theory, instantons yield non-perturbative corrections 
to correlation functions via their saddle-point contribution to the 
Euclidean path integral. In the semiclassical limit, the 
path-integral in each topological sector reduces to an ordinary integral 
over the instanton moduli space. The extent to 
which similar ideas apply to D-instantons is less clear, in part because 
string theory lacks a second-quantized formalism analogous to the path 
integral. Despite this, there is considerable evidence that 
D-instanton contributions to string-theory amplitudes also reduce 
to integrals over collective coordinates at weak 
string coupling \cite{GG1}. In this case the relevant collective coordinates 
are the components the ten-dimensional $U(k)$ gauge field 
$A_{\mu}$ and their superpartners $\Psi$. According to Green and 
Gutperle \cite{GG1,GG2,GG3,GG4}, the charge-$k$ D-instanton contributions 
to the low-energy correlators of the IIB theory are governed by the 
partition function, 
\begin{equation}
{\cal Z}_{k} = \frac{1}{{\rm Vol}\,U(k)}\,
\int_{U(k)}\,d^{10}A\,d^{16}\Psi\, \exp-S_{k}\ .
\elabel{ukpfn}
\end{equation}
This partition function can be thought of as the collective coordinate 
integration measure for $k$ D-instantons. In particular, the leading 
semiclassical contribution of $k$ D-instantons to the  
correlators of the low-energy supergravity fields is obtained by
inserting the classical value of each field in the integrand of 
\eqref{ukpfn}. Because of the symmetries described above, the 
collective coordinate action $S_{k}$ does not depend on the 
$U(1)$ components of the fields $A_{\mu}$ and $\Psi$. 
Hence, to obtain a non-zero answer, 
the inserted fields must include at least sixteen fermions to 
saturate the  corresponding Grassmann integrations. As in field theory 
instanton calculations, the resulting amplitudes can be interpreted in 
terms of an instanton-induced vertex in the low-energy effective action.     
The spacetime position of the D-instanton, $x_{\mu}$, is interpreted as the 
location of the vertex. In particular, the work of Green and Gutperle 
\cite{GG1} has 
focused on a term of the form $R^{4}$ in the IIB effective action 
(here $R$ is the ten-dimensional curvature tensor) and its supersymmetric 
completion. 

So far, we have only considered D-instantons in the IIB theory in a flat 
ten dimensional background and in the absence of other branes. In order 
to make contact with four-dimensional gauge theory 
we need to understand how these 
ideas apply to D-instantons in the presence of D3-branes. In particular 
we wish to determine how the D-instanton measure \eqref{ukpfn} 
is modified by the introduction of $N$ parallel D3-branes.  Conversely, 
in the absence of D-instantons, the theory 
on the four-dimensional world volume of the D3-branes is ${\cal N}=4$ $U(N)$
super Yang-Mills theory. Hence a related question is how the D-instantons 
appear from the point of view of the four-dimensional theory on the D3 
branes.  In fact, the brane configuration considered here is a special 
case of a system which has been studied intensively in the past. 
The general case involves a configuration of $k$ D$p$-branes  
in the presence of $N$ $D(p+4)$-branes, with all branes parallel. 
As we will review below, in each of these cases the lower-dimensional brane 
corresponds to a Yang-Mills instanton in the world-volume gauge theory of the 
higher \cite{D2}. 
We begin by reviewing the maximal case $p=5$, which was first 
considered (in the context of Type I string theory) by Witten \cite{W1}. 
The cases with $p<5$ then follow straightforwardly by dimensional
reduction.   

We start by considering a theory of $k$ parallel D5-branes in isolation. 
As above, the world-volume theory is obtained by dimensional reduction 
of ten dimensional ${\cal N}=1$  super-Yang-Mills theory with gauge group 
$U(k)$. The resulting theory in six dimensions has two Weyl supercharges 
of opposite chirality and is conventionally denoted as ${\cal N}=(1,1)$ 
SYM.\footnote{Some convenient facts about six-dimensional supersymmetry 
are reviewed in \cite{CHS} (See page 67 in particular).} After dimensional 
reduction, the $SO(10)$ Lorentz group of the Euclidean theory in ten 
dimensions is broken to $H=SO(6)\times SO(4)$. The $SO(6)$ factor is 
the Lorentz group of the six dimensional theory while the $SO(4)$ is 
an $R$-symmetry. The ten dimensional gauge field $A_{\mu}$ splits up into 
an adjoint scalar $a'$ in the vector representation of $SO(4)$ and 
a six dimensional gauge field $\chi_{a}$ with $a=1,\ldots,6$. 
Explicitly we set $A_{\mu}=(a'_{n}, \chi_{a})$ where $n=0,1,2,3$ is an 
$SO(4)$ vector index. 

In order to describe the fermion content of the theory we will consider the 
covering group of $H$, $\bar{H}=SU(4)\times SU(2)_{L}\times SU(2)_{R}$. 
We introduce indices $A=1,2,3,4$ and $\alpha,\aD=1,2$ corresponding to
each factor. As mentioned above, a 
ten dimensional Majorana-Weyl spinor can be decomposed into two Weyl spinors of 
opposite chirality in six dimensions. The corresponding representation 
of $SO(10)$ decomposes as,  
\begin{equation} 
{\bf 16}\rightarrow 
({\bf 4},{\bf 2},{\bf 1})\oplus ({\bf \bar{4}},{\bf 1},{\bf 2})
\elabel{repn}
\end{equation}
under $\bar{H}$. An explicit decomposition of the ten-dimensional 
spinor $\Psi$ in terms of the six dimensional spinors, 
${\cal M}^{\prime A}_{\alpha}$  and $\lambda_{\dot{\alpha}A}$ is given in 
the Appendix. 
As in the previous sections of this 
paper, it will often be convenient to rewrite the $SO(6)$ vector $\chi_{a}$ 
as a (quasi-real) antisymmetric tensor $\chi_{AB}$ using
Eq.~\eqref{E54.1}, where $\chi_{AB}$ is subject to the reality
condition \eqref{E54}.

The fields $(\chi_a, {\cal M}^{\prime A}_\alpha, \lambda^\aD_A, a'_n)$  form a vector 
multiplet of ${\cal N}=(1,1)$ supersymmetry in six dimensions. 
In terms of an ${\cal N}=(0,1)$ subalgebra, the ${\cal N}=(1,1)$ 
vector multiplet splits up into an ${\cal N}=(0,1)$ vector multiplet 
containing $\chi_a$ 
and $\lambda^\aD_A$ and an adjoint hypermultiplet containing $a'_n$ and 
${\cal M}^{\prime A}_\alpha$. The action of the ${\cal N}=(1,1)$ theory is   
\begin{equation}
S^{(6)}=\frac{1}{g_{6}^{2}}S_{\rm gauge}+S_{\rm matter}^{({\rm a})}\ ,
\elabel{de6}
\end{equation}
where
\begin{equation}
S_{\rm gauge}=\int\, d^{6}x \,{\rm tr}_{k}\big(\tfrac{1}{2}F^{2}_{ab}
-\sqrt{2}\pi\lambda_{\dot{\alpha}A}\left(\bar{\Sigma}_{a}^{AB}
{\cal D}_{a}\right)\lambda^{\dot{\alpha}}_B-\tfrac{1}{2}D^{2}_{mn}\big)\ ,
\elabel{sgauge}
\end{equation}
and
\begin{equation}
S_{\rm matter}^{\rm (a)}  =  \int\, d^{6}x \,{\rm tr}_{k}
\big(\left({\cal D}_{a}a'_{n}\right)^{2}
-\sqrt{2}\pi {\cal M}^{\prime\alpha A}\left(\Sigma^{a}_{AB}
{\cal D}_{a}\right){\cal M}^{\prime B}_{\alpha} 
+ i\pi 
[a'_{\alpha\dot{\alpha}},{\cal M}^{\prime\alpha A}]
\lambda^{\dot{\alpha}}_{A} +iD_{mn}[a'_{m},a'_{n}]\big)\ .
\elabel{sa}
\end{equation}
In the above, we have rescaled the fields so that the six-dimensional 
coupling constant, $g_{6}^{2}\sim \alpha'$, only appears in front of 
the action of the ${\cal N}=(1,1)$ vector multiplet. The reason for the 
unconventional normalization of the fermion kinetic terms will become 
apparent below. For later convenience 
we have also introduced a real anti-self dual auxiliary field for the vector 
multiplet, $D_{mn}=-(^{*}D_{mn})$. $D$ transforms in the adjoint representation 
of $SU(2)_{R}$, which can be made explicit by writing 
$D_{mn}=-D^{c}\bar{\eta}^{c}_{mn}$, where the eta symbol is defined in
the Appendix. 

Following \cite{D2}, the next step is to introduce $N$ D9-branes of the 
Type IIB theory whose world-volume fills the ten dimensional 
spacetime.\footnote{In fact a IIB background with non-vanishing D9-brane charge 
suffers from inconsistencies at the quantum level. 
This is not relevant here because the D9-branes in 
question are just a starting point for a classical dimensional 
reduction.} These are analogous to the $N$ D3-branes in the 
$p=-1$ case on which we will eventually focus. The world volume 
theory of the D9-branes in isolation (i.e. in the absence of the D5-branes) 
is simply ten dimensional $U(N)$ supersymmetric Yang-Mills. As explained 
by Douglas \cite{D2}, the effective action for this system contains a 
coupling between the field strength $V_{mn}$ of the 
world-volume gauge field and the  
the rank six antisymmetric tensor field $C_{\mu\nu\cdots\rho}$ 
which comes from the Ramond-Ramond sector of the Type IIB theory. The 
latter field is dual to the three-form field strength which appears in 
the Type IIB supergravity action. This coupling has the form,  
\begin{equation}
\int\, C\wedge V\wedge V,
\elabel{douglas}
\end{equation} 
where $C$ and $V$ are written as a six-form and a 
two-form respectively and the integration is over the ten-dimensional world 
volume of the D9-branes. 
The same six-form field also couples minimally 
to the Ramond-Ramond charge carried by D5-branes.  Hence a configuration 
of the $U(N)$ gauge fields with non-zero second Chern class, $V\wedge V$, acts 
as a source for D5-brane charge. More concretely, if the D9-brane 
gauge field is chosen to be independent of six of the world-volume 
dimensions and an ordinary Yang-Mills instanton is embedded in the 
remaining four dimensions, then the resulting configuration has the same 
charge-density as a single D5-brane. Both objects are also BPS saturated and 
therefore they also have the same tension. Further confirmation of the 
identification of D5-branes on a D9-brane as instantons was found in 
\cite{D2} where the gauge-field background due to a Type I D5-brane 
was shown to be self-dual via its coupling to the world-volume of a D1-brane 
probe.     

 As described above, D5-branes appear as BPS saturated instanton 
configurations on the D9-brane which break half 
of the supersymmetries of the world-volume theory. Conversely, 
the presence of D9-branes also break half 
the supersymmetries of the D5-brane world-volume theory described by 
the action \eqref{de6}. Specifically, the ${\cal N}=(1,1)$ supersymmetry  
of the six-dimensional theory is broken down to the ${\cal N}=(0,1)$ 
subalgebra described above equation \eqref{de6}. To explain how this happens 
we recall that open strings stretched between branes give rise to fields 
which propagate on the D-brane world-volume. So far we have only included 
the adjoint representation fields which arise from strings stretching between 
pairs of D5-branes. As our configuration now includes both D5- and 
D9-branes there is the additional possibility of states corresponding to 
strings with one end on each of the two different types of brane.  
As the D5-brane and D9-brane ends of the string carry $U(k)$ and $U(N)$ 
Chan-Paton indices respectively, the resulting states transform in the 
$(\Bk,\BN)$ representation of $U(k)\times U(N)$. 

In fact the additional degrees of 
freedom comprise $kN$ hypermultiplets of ${\cal N}=(0,1)$ supersymmetry 
in six dimensions \cite{W1}. 
As these hypermultiplets cannot be combined to form 
multiplets of ${\cal N}=(1,1)$ supersymmetry, the residual supersymmetry 
of the six dimensional theory is ${\cal N}=(0,1)$ as claimed above. 
Each hypermultiplet contains two complex scalars 
$w_{ui\dot{\alpha}}$. Here, as previously, $i$ and $u$ are fundamental representation 
indices of $U(k)$ and $U(N)$ respectively. The fact that hypermultiplet 
scalars transform as doublets of the $SU(2)$ $R$-symmetry is familiar 
from ${\cal N}=2$ theories in four dimensions. Each hypermultiplet also 
contains a pair of complex Weyl spinors, $\mu_{ui}^{A}$ and  
$\bar{\mu}_{iu}^{A}$. The six-dimensional action for the hypermultiplets 
is, 
\begin{equation}\begin{split}
S_{\rm matter}^{\rm (f)} & =  \int\, d^{6}x \,{\rm tr}_k
\big({\cal D}^{F}_{a}\bar{w}_{u}^{\dot{\alpha}}{\cal D}^{F}_{a}
w_{u \dot{\alpha}}
-2\sqrt{2}\pi \bar{\mu}_{u}^{A}\left(\Sigma^{a}_{AB}
{\cal D}^{F}_{a}\right)\mu^{B}_{u} \\ 
&   \qquad\qquad\qquad\qquad\ +i\pi\left(\bar{\mu}^{A}_{u}w_{u\dot{\alpha}}+
\bar{w}_{u\dot{\alpha}}\mu^{A}_{u}\right)\lambda^{\dot{\alpha}}_{A}
+D^{c}\bar{w}_{u\dot{\alpha}}
(\tau^{c})^{\dot{\alpha}}_{\ \dot{\beta}}w^{\dot{\beta}}_{u}\big)\ .
\elabel{sf}
\end{split}\end{equation}
The scalar and fermion kinetic terms in the above 
action include the  fundamental representation 
covariant derivative, ${\cal D}^{F}_{a}w=\partial_{a}w-iw\chi_a$. 
The remaining two terms in \eqref{sf} are fundamental representation versions 
of the Yukawa coupling and D-term which appear in \eqref{sa}. The complete 
action of the $d=6$ theory is $g_6^{-2}S_{\rm gauge}+S^{\rm (a)}_{\rm matter}+
S^{\rm (f)}_{\rm matter}$. 

The ${\cal N}=(0,1)$ supersymmetry transformations for 
this action can be deduced from the 
supersymmetry transformations of ten dimensional Yang-Mills theory 
in Eq.~\eqref{10daction}. The latter are
\begin{equation}\begin{split}
\delta A_{\mu}  & =  i\bar{\eta}\Gamma_{\mu}\Psi\ ,\\ 
\delta \Psi \, & =  -iF_{\mu\nu}\Gamma^{\mu\nu}\eta\ ,  
\elabel{10dsusy}\end{split}\end{equation}
The ${\cal N}=(0,1)$ supersymmetry of the six-dimensional action 
\eqref{sgauge}, \eqref{sa} and \eqref{sf} is then  
obtained as the subalgebra of the ten-dimensional ${\cal N}=1$ algebra 
obtained by choosing, in the notation of the Appendix, 
\begin{equation}
\eta=\frac{1}{\sqrt{2\pi}}\,  
\begin{pmatrix} 1 \\ 0\end{pmatrix}\,\otimes\, 
\begin{pmatrix} 0  \\ 
\bar{\xi}^\aD_{A} \end{pmatrix}\, .
\elabel{cov2}
\end{equation} 
This yields the transformations, 
\begin{equation}\begin{split}
\delta a'_{n} =  \frac{1}{2}\bar{\xi}_{\dot{\alpha}A} 
\bar{\sigma}^{\dot{\alpha}\alpha}_{n} {\cal M}^{\prime A}_{\alpha},&\qquad
 \delta{\cal M}^{\prime A}_{\alpha} =  \frac{2\sqrt{2}}{\pi}
({\cal D}_{a}a'_{n})\bar{\Sigma}_{a}^{AB}
\sigma^{n}_{\alpha\dot{\alpha}} \bar{\xi}^\aD_{B},\\
\delta \chi_{a} =  i\sqrt{2}\bar{\xi}_{\dot{\alpha}A} 
\bar{\Sigma}_{a}^{AB} \lambda^{\dot{\alpha}}_B,&\qquad 
\delta \lambda^\aD_{A} =\frac{1}{\pi}F^{ab} 
\bar{\Sigma}_{\phantom{ab}A}^{ab\ B}\bar{\xi}^\aD_{B} -\frac{i}{\pi}[a'_{m},a'_{n}] 
\bar{\sigma}^{mn\aD}_{\phantom{mn\bD}\bD}\bar{\xi}^\bD_{A}, \\ 
\delta w_{\dot{\alpha}u} =  \bar{\xi}_{\dot{\alpha}A} \mu_{u}^A,&\qquad 
\delta \mu^{A}_{u} =  \frac{2\sqrt{2}}{\pi}\bar{\Sigma}^{AB}_{a}  ({\cal D}^{F}_{a} 
w_{\dot{\alpha}u})\bar{\xi}_{B}^{\dot{\alpha}}.
\elabel{susy3}
\end{split}\end{equation}
After dimensional reduction to $d=0$ and elimination of $\chi_a$ and 
$\lambda^\aD_{A}$ by their equations of motion, 
these transformation rules for the remaining variables 
$a'_{n}$, ${\cal M}^{\prime A}_\alpha$, 
$w_{u\dot{\alpha}}$ and $\mu_{u}^{A}$ agree 
with the collective coordinate supersymmetry transformations Eqs.~(13a) and 
(13b) of \cite{DHKM} (suitably generalized to gauge group $SU(N)$). 

The various fields of the six dimensional theory and their transformation 
properties under $U(k)\times U(N)\times \bar{H}$ are tabulated below:

\begin{center}
\begin{tabular}{cccccc}
\hline
\Rowspace & $U(k)$ &  $U(N) $ & $ SU(4) $ & $ SU(2)_{L} $ & $ SU(2)_{R} $ \\
\Rowspace $\chi$ & {\bf adj} & ${\bf 1}$ & ${\bf 6}$ & ${\bf 1}$ & ${\bf 1}$ \\
\Rowspace $\lambda$ & {\bf adj} & ${\bf 1}$ & ${\bf \bar{4}}$ & ${\bf 1}$ & ${\bf 2}$ \\
\Rowspace $D$ & {\bf adj} & ${\bf 1}$ & ${\bf 1}$ & ${\bf 1}$ & ${\bf 3}$ \\
\Rowspace $a'$ & {\bf adj} & ${\bf 1}$ & ${\bf 1}$ & ${\bf 2}$ & ${\bf 2}$ \\
\Rowspace ${\cal M}'$ & {\bf adj} & ${\bf 1}$ & ${\bf 4}$ & ${\bf 2}$ & ${\bf 1}$ \\
\Rowspace $w$ & $\Bk$ & $\BN$ &  ${\bf 1}$ & ${\bf 1}$ & ${\bf 2}$ \\
\Rowspace $\bar w$ & $\bar\Bk$ & $\bar\BN$ & ${\bf 1}$ & ${\bf 1}$ & ${\bf 2}$ \\
\Rowspace$ \mu$ & $\Bk$ & $\BN$ & ${\bf 4}$ & ${\bf 1}$ & ${\bf 1}$ \\
\Rowspace $\bar{\mu}$ & $\bar\Bk$ & $\bar\BN$ & ${\bf 4}$ & ${\bf 1}$ & ${\bf 1}$ \\
\hline 
\end{tabular} 

\vspace{0.5cm}
Table 1: Transformation properties of the fields
\end{center} 

The reader will have noticed that we have chosen our notation so 
that each six-dimensional field has a counterpart, denoted by 
the same letter, in our discussion of the ADHM construction in Section III. 
The only exceptions are the fields $\lambda$ and $D$ whose significance 
will be explained below. Also the group indices on each $d=6$ field are 
in one-to-one correspondence with those on the ADHM variable of 
the same name. The physical reason for this correspondence is 
simple: the six dimensional fields are the collective 
coordinates of the D5-branes. As the D5-branes are equivalent to the 
Yang-Mills instantons, the vacuum moduli space of the $U(k)$ gauge theory 
on the D5-branes should coincide with the $k$-instanton moduli space 
described by the ADHM construction. As the only scalar fields in the 
six-dimensional theory lie in ${\cal N}=(0,1)$ hypermultiplets, the relevant 
vacuum moduli space is conventionally referred to as a Higgs branch.  
Precisely how the proposed equivalence arises was explained in \cite{W1} 
where it was shown that the D-term vacuum conditions which follow from the 
six-dimensional action coincide with the non-linear constraint 
equations of the ADHM construction. This and several new aspects of this 
correspondence will be demonstrated explicitly below. 

Starting from the configuration of D5 and D9-branes described above, 
the general case of parallel D$p$- and D$(p+4)$-branes with $p<5$ can be 
obtained by dimensional reduction on the brane world-volumes. In particular 
the case of D0- and D4-branes in the Type IIA theory has been studied 
extensively because of its application as a lightcone matrix model of the 
$(2,0)$ theory in six dimensions \cite{A}. 
In this case we will perform one further 
dimensional reduction and study $k$ D-instantons in the presence of 
$N$ D3-branes. Following the above discussion, this corresponds  
to $k$ Yang-Mills instantons in the four-dimensional ${\cal N}=4$ 
theory. The symmetry group $U(k)\times U(N)\times\bar{H}$ of the six 
dimensional system now has a simple interpretation in terms of the 
four-dimensional theory: $U(k)$ is the internal symmetry of the ADHM 
construction, $U(N)$ is the gauge group, $SU(4)$ is the $R$-symmetry 
group of the ${\cal N}=4$ supersymmetry algebra and $SO(4)=SU(2)_{L}\times SU(2)_{R}$ 
is the four-dimensional Lorentz group. In six dimensions the  
scalar fields were all in ${\cal N}=(0,1)$ hypermultiplets, and hence the 
resulting vacuum moduli space was a Higgs branch. After dimensional reduction 
to $d=0$, the six-dimensional gauge fields $\chi_a$ become scalars 
which can acquire VEVs. The resulting Coulomb branch corresponds to 
motion of the D-instantons in the six dimensions transverse to the 
D3-branes. We will see below that $\chi_a$ is just the auxiliary field 
introduced at the end of Sec.~IV.1 to bilinearize the 
four-fermion term in the instanton action.  

We will now write down the 
collective-coordinate measure which determines the leading semiclassical 
contribution of $k$ D-instantons to correlation functions of the low-energy 
fields of the IIB theory in the presence of $N$ D3-branes. From 
the above discussion, the appropriate generalization of \eqref{ukpfn}, 
is obtained by dimensionally reducing the partition function 
of the six-dimensional theory with action 
$g_6^{-2}S_{\rm gauge}+S^{\rm (a)}_{\rm matter}+
S^{\rm (f)}_{\rm matter}$ defined in \eqref{sgauge}, \eqref{sa} and \eqref{sf} 
to zero dimensions. Thus we have,  
\begin{equation}
{\cal Z}_{k,N} = \frac{1}{{\rm Vol}\,U(k)}\,
\int d^{6}\chi\, d^{8}\lambda\, d^{3}D\,d^{4}a'\,d^{8}{\cal M}'\, 
d^{2}w\,d^2\bar w\, d^{4}\mu\, d^{4}\bar{\mu}\, 
e^{-S_{k,N}}\ , 
\elabel{final1}
\end{equation}
where $S_{k,N}=g_0^{-2}S_{G} + S_{K}+S_{D}$ with,  
\begin{subequations}
\begin{align}
S_{G} & = {\rm tr}_{k}\big(-[\chi_a,\chi_b]^2+\sqrt{2}i\pi 
\lambda_{\dot{\alpha}A}[\chi_{AB}^\dagger,\lambda_B^{\dot{\alpha}}] 
+2D^{c}D^{c}\big)\, ,\elabel{p=-1actiona} \\
S_{K} & =  -{\rm tr}_{k}\big([\chi_a,a'_{n}]^2 
+\chi_a\bar{w}^\aD_{u}
w_{u\dot{\alpha}}\chi_a + \sqrt{2}i\pi 
{\cal M}^{\prime \alpha A}[\chi_{AB},
{\cal M}^{\prime B}_{\alpha}]+2\sqrt{2} i \pi
\bar{\mu}_{u}^{A}\chi_{AB}\mu^{B}_{u}\big)\ ,\elabel{p=-1actionb} \\
S_{D} & =  i  \pi{\rm tr}_k\big(
[a'_{\alpha\dot{\alpha}},{\cal M}^{\prime\alpha A}]\lambda^{\dot{\alpha}}_{A}
+\bar{\mu}^{A}_{u}w_{u\dot{\alpha}}\lambda^{\dot{\alpha}}_{A}+
\bar{w}_{u\dot{\alpha}}\mu^{A}_{u}
\lambda^{\dot{\alpha}}_{A} + D^{c}\left(W^{c}-i [a'_{n},a'_{m}]
\bar{\eta}^{c}_{nm}\right) \big)\ .
\elabel{p=-1actionc}
\end{align}\end{subequations}
Note that $S_{G}$ arises 
from dimensional reduction of $S_{\rm gauge}$ and $S_{K}+S_{D}$ comes
from dimensional reduction of $S_{\rm matter}^{\rm (a)}+S_{\rm
matter}^{\rm (f)}$. 
Specifically $S_{K}$ contains the six-dimensional gauge couplings of the 
hypermultiplets, while $S_{D}$ contains the Yukawa couplings and D-terms.  

Semiclassical correlation functions of the light fields can be 
calculated by replacing each field with its value in the D-instanton 
background and performing the collective coordinate integrations 
with measure \eqref{final1}. In the case of the low-energy gauge fields on 
the D3-brane, the relevant classical configuration is simply the charge-$k$ 
Yang-Mills instanton specified by the ADHM data which appears explicitly 
in the action \eqref{p=-1actiona}-\eqref{p=-1actionc}. Note that the measure \eqref{final1} 
depends explicitly on the string length-scale through the 
zero-dimensional coupling $g^{2}_{0}\sim (\alpha')^{-2}$ which appears 
in the action. As a consequence correlation functions which 
include fields inserted at distinct spacetime points $x_{i}$ and $x_{j}$ 
will have a non-trivial expansion in powers of 
$\sqrt{\alpha'}/ |x_{i}-x_{j}|$. In order to decouple the world-volume 
gauge theory from gravity must take the limit $\alpha'\rightarrow 0$. 
Hence we take the strong coupling limit $g^{2}_{0}\rightarrow \infty$, 
in the exponent of \eqref{final1}.\footnote{See page 4 of \cite{A} 
for the corresponding discussion in the $p=0$ case.}   
Our final answer for the modified D-instanton measure is 
therefore, 
\begin{equation}
{\cal Z}_{k,N}  =  \frac{1}{{\rm Vol}\,U(k)}\,
\int d^{6}\chi d^{8}\lambda d^{3}D\,d^{4}a'\,d^{8}{\cal M}'\, 
d^{4}w\, d^{4}\mu\, d^{4}\bar{\mu}\, 
\exp\left(-S_{K}-S_{D}\right) \ .
\elabel{final}
\end{equation}
As mentioned above, we have not kept track of the overall 
normalization of this expression which includes a numerical 
constant that depends on $k$ and $N$ as well as some power of the 
four-dimensional gauge coupling.  

We can now make contact with our results for the ADHM measure in
Sec.~IV.1.\footnote{The difference between the gauge groups $U(N)$ and
$SU(N)$ is irrelevant in this context because the instantons always
live in the nonabelian factor.}  In particular, the equation of motion
for $D$ is precisely the non-linear ADHM constraint \eqref{fconea}.
Similarly the equation of motion for $\lambda$ is the fermionic
constraint \eqref{zmcona}. Integration over these variables yields the
first two sets of $\delta$-functions in the measure formula
\eqref{dmudef}. Further, $S_{K}$ can be compactly rewritten in the
notation of Sec.~IV.1 as,
\begin{equation}
S_{K}={\rm tr}_k\,\chi_a{\bf L}\chi_a - 4\pi i 
{\rm tr}_{k}\,\chi_{AB}\Lambda^{AB}\ .
\end{equation} 
This is equal to (minus) the exponent appearing in equation
\eqref{E53}. On integrating out the gauge field 
$\chi_a$, the instanton action reduces to the fermion quadrilinear term 
\eqref{Skquadef}. We have therefore reproduced our result for the ADHM measure, 
up to an overall normalization constant. 
In addition the action $S_{K}+S_{D}$ is invariant under eight supercharges 
which are inherited from the ${\cal N}=(0,1)$ theory in six dimensions.      
The resulting supersymmetry transformations are identical to the collective 
coordinate supersymmetries of the ${\cal N}=4$ ADHM measure given in 
\cite{DHKM} (suitably generalized for gauge group $U(N)$ 
as in \cite{KMS}). There are also eight additional fermionic symmetries which 
only appear after taking the decoupling limit $\alpha'\rightarrow 0$.  
These correspond to the half of the 
superconformal transformations of the ${\cal N}=4$ theory which leave   
the instanton invariant.   

It is interesting to compare the above discussion of D-instantons in the 
presence of D3-branes with the corresponding results for the 
D$p$/D($p+4$) system with $p>0$. 
In the $p=0$ case considered in \cite{A}, we have D0-branes which 
correspond to solitons in the 
$4+1$ dimensional gauge theory on the D4-brane world-volume. These solitons 
are just four-dimensional Yang-Mills instantons thought of as 
static finite-energy configurations in five dimensions.  At weak coupling, we 
expect the dynamics of these solitons to be correctly 
described by supersymmetric quantum mechanics 
on the instanton moduli space. Hence, it is not surprising 
that precisely this quantum mechanical system is obtained in \cite{A} 
by dimensionally reducing the six-dimensional ${\cal N}=(0,1)$ theory 
described above down to a single time dimension. 
On adding another world-volume dimension, we obtain the D1/D5 system 
discussed by Witten in \cite{W2}. The D1-branes are now strings in 
a six dimensional Yang-Mills theory and, in the decoupling limit, 
their world-sheet dynamics 
is described by a two-dimensional ${\cal N}=(4,4)$ non-linear 
$\sigma$-model with the ADHM moduli space as the target manifold. 
This $\sigma$-model has kinetic terms for the coordinates on the target 
and their superpartners and as usual the supersymmetric completion of 
the action involves a four-fermion term which couples to the Riemann tensor 
of the target. If we reduce this action back to $d=1$ by discarding 
spatial derivatives we obtain the quantum 
mechanics of \cite{A}.   If we then reduce to $d=0$ by discarding 
time derivatives also, the only term of the $\sigma$-model action 
which survives is the four-fermion term. This is the 
origin of the fermion quadrilinear term \eqref{Skquadef} in the 
collective coordinate action. 

\subsection{The gauge-invariant measure}

In Secs.~IV.1-IV.2 we have presented  field-theory and
string-theory derivations of the flat measure,
Eq.~\eqref{dmudef}. However, as stated earlier, this form of the
measure is inadequate to our present purposes, for two reasons:

(i) The number of bosonic and fermionic variables grows with $N$. 
Such a growth is inconvenient in light of our stated
aim, in Sec.~V below, to treat the large-$N$ limit in a traditional
saddle-point approximation.

(ii) The collective coordinates $w$ and $\mu^A$ carry a $u$ index
which transforms in the fundamental representation of the (global) $SU(N)$
gauge group. In other words, they do not define a gauge-invariant set
of variables. This is inefficient, given that in Sec.~VI below, we
intend to evaluate only correlators of gauge-invariant operators.

Accordingly, our present task is to change variables from $w$ and
$\mu^A$ to the gauge-invariant set of collective coordinates already
introduced in Sec.~II.6. After integrating out the degrees of freedom
corresponding to the gauge coset\footnote{Actually
physical gauge transformations involve modding out by an additional
$U(1)$ corresponding to overall phase rotations since these
transformations lie in the auxiliary group $U(k)$. However,  we shall loosely
refer to \eqref{gcoset} as the ``gauge coset''.}
\begin{equation}
{SU(N)\over SU(N-2k)}\ ,
\elabel{gcoset}
\end{equation}
 we will find that the number of remaining integration
variables no longer grows with $N$, but only with $k$. The main
calculational aim of this subsection is to derive the bosonic and
fermionic Jacobians associated with these changes of variables. The
reader who is uninterested in the details of our algebra can skip
directly to our final result for the gauge-invariant measure given in
Eq.~\eqref{gaugeinvmeas} below. Note that this expression 
incorporates the auxiliary Gaussian variables $\chi_a$ 
introduced in Eqs.~\eqref{E53}-\eqref{E54.1}.

We begin the Jacobian calculation in the bosonic sector where we need
to change variables from the collective coordinates
$w_{ui\dot\alpha}$, whose $u$ index transforms in the fundamental
representation of the gauge group $SU(N)$, to the gauge-invariant
bilinear variables $(W^{\dot\alpha}_{\ \dot\beta})^{}_{ij}$ defined in
Eq.~\eqref{Wdef}. 

Let us consider the more general mathematical problem of an $N\times
K$ dimensional complex matrix $v^l_u$ where $u=1,\ldots,N$ and $l=1,\ldots,K,$
and let us always assume $N\ge K$ (which is appropriate for the large-$N$
limit to follow).  In the case at hand, $l$ stands for the composite
index $(i,\dot\alpha)$ and so
\begin{equation}
K\ =\ 2k\ ,\quad N\ge2k\ ,
 \elabel{Kdef}\end{equation}
 where $k$ is the topological
number as always.
  Furthermore we assume that these
variables (read: the $w$'s) are endowed with a flat integration
measure
\begin{equation}
{\mathcal D}v\ =\ 
\prod_{u=1}^N\prod_{l=1}^K\,dv_u^l\,d v_u^{l*}\ .
\elabel{measreal}\end{equation}

A suitable $SU(N)$ gauge transformation $\grp$ puts the matrix $v_u^l$ into
upper-triangular form:
\def\xmax{x_{\scriptscriptstyle\rm max}}
\def\hf{\textstyle{{1\over2}}}
\def\Ovec{{\vec0}}
\begin{equation}
\begin{pmatrix}v_1^1&\cdots&v_1^K\\ \vdots&\ddots&\vdots\\  
v_N^1&\cdots&v_N^K\end{pmatrix}
\ =\ \grp(\omega_x)\cdot\begin{pmatrix}\xi_1^1&\xi_1^2&\cdots&\xi_1^K\\ 
0&\xi_2^2&\cdots&\xi_2^K\\ 
{}&0&\ddots&\vdots\\ 
\vdots&{}&\ddots&\xi_K^K\\ 
{}&{}&{}&0\\  {}&{}&{}&\vdots\\  0&0&\cdots&0\end{pmatrix}\ .
\elabel{uptri}\end{equation}  
The $\xi_k^l$ are complex except for the diagonal elements $\xi_l^l$ which we can choose
to be real. The group parameters $\omega_x$ 
coordinatize the coset $SU(N)/SU(N-K)$ which
acts non-trivially on the $\xi$ matrix; the
index $x$ runs over $1,\ldots,\xmax$ where 
\begin{equation}
\xmax\ =\ 2KN-K^2\ .
\elabel{xmaxdef}\end{equation}
Obviously the number of
independent real parameters, namely $2KN,$ is the same on both
sides of Eq.~\eqref{uptri}. 

While \it a priori \rm the $\xi_l^k$ are only defined for $l\le k\le
K,$ it is convenient to extend them to $k<l\le K$ as well, by defining
$\xi_l^k=(\xi_k^l)^*$.  In terms of these extended variables we
define the gauge-invariant bilinears $W_{kl}$
via the matrix equation
\begin{equation}
W\ =\ v^\dagger\,v\ =\ 
\begin{pmatrix}\xi_1^1&0&\cdots&0\\ 
\xi_2^1&\xi_2^2&{}&\vdots\\ 
\vdots&\vdots&\ddots&0\\ 
\xi_K^1&\xi_K^2&\cdots&\xi_K^K\end{pmatrix}
\begin{pmatrix}\xi^1_1&\xi^2_1&\cdots&\xi^K_1\\ 
0&\xi^2_2&\cdots&\xi^K_2\\ 
\vdots&{}&\ddots&\vdots\\ 
0&\cdots&0&\xi_K^K\end{pmatrix}.
\elabel{Wequals}\end{equation}
Note that there are as many real degrees of freedom in the $\{W_{kl}\}$ as
in the $\{\xi_k^l\}.$ From Eq.~\eqref{Wequals} it follows that
\begin{equation}
\det_KW\ =\ \Big(\prod_{l=1}^K\xi_l^l\Big)^2\ .
\elabel{detWdef}\end{equation}

We wish to calculate the Jacobian induced by changing
integration variables from the original collective
coordinates $v_u^l,$ to the gauge-invariant bilinears $W_{kl}$
together with the coset parameters $\omega_x$. It is useful to pass
 through an intermediate change of variables involving the $\xi_k^l$
 rather than the $W_{kl}.$ The desired
 Jacobian can then be expressed as a quotient:
\begin{equation}
{\partial\big(\{v_u^l,v_u^{l*}\}
\big)\over\partial\big(\{W_{kl},\omega_x\}\big)}
\ =\ 
{\partial\big(\{v_u^l,v_u^{l*}
\}\big)\over\partial\big(\{\xi_l^k,\omega_x\}\big)}
\bigg/
{\partial\big(\{W_{kl}\}\big)\over\partial\big(\{\xi_l^k\}\big)}\ .
\elabel{quotientjac}
\end{equation}
What we aim to show is that this Jacobian actually factors cleanly into a
function of the gauge-invariant parameters $W_{kl},$ times a function
of the coset variables $\omega_x$; integrating over the gauge coset
then turns the latter function into some overall normalization
constant that we will evaluate explicitly.

To this end, let us analyze, in turn, the denominator and the numerator on the
right-hand side of Eq.~\eqref{quotientjac}. The denominator is readily
evaluated from Eq.~\eqref{Wequals} by induction in $K$.
For $K=1$ one has simply $\partial W_{11}/\partial\xi^1_1=2\xi_1^1.$
{}From Eq.~\eqref{Wequals} we can also easily relate the Jacobian for $K$ to
that for $K-1$; one finds
\begin{equation}\begin{split}
{\partial\big(\{W_{kl}\}\big)\over\partial\big(\{\xi_l^k\}\big)}\bigg|_K^{}
\ &=\ 
{\partial W_{KK}\over\partial\xi_K^K}
\Big(\prod_{l=1}^{K-1}{\partial W_{lK}\over\partial\xi_l^K}
{\partial W_{Kl}\over\partial\xi_K^l}\Big)
{\partial\big(\{W_{kl}\}\big)\over\partial\big(\{\xi_l^k\}\big)}\bigg|_{K-1}^{}
\\ &=\
2\xi^K_K\Big(\prod_{l=1}^{K-1}\big(\xi_l^l\big)^2
\,\Big)
{\partial\big(\{W_{kl}\}\big)\over\partial\big(\{\xi_l^k\}\big)}\bigg|_{K-1}^{}\ .
\elabel{inductcomp}
\end{split}\end{equation}
It follows by induction that 
\begin{equation}
{\partial\big(\{W_{kl}\}\big)\over\partial\big(\{\xi_l^k\}\big)}
\ =\ 2^K\,\xi^K_K\,(\xi^{K-1}_{K-1})^3\cdots(\xi^1_1)^{2K-1}\ .
\elabel{denomanswer}\end{equation}

Next, we evaluate the numerator on the right-hand side of 
Eq.~\eqref{quotientjac}. At this stage, it is useful to make an explicit choice of
coset coordinates $\omega_x$. We choose them to be simply the
independent entries of the first $K$ orthonormal columns $u^l$ of $\Upsilon$:
\begin{equation}
u^l_u=\Upsilon_{ul}\ ,\quad u^{l\dagger}\cdot u^{l'}=\delta^{ll'}\ .
\end{equation}
The $N$-vectors $u^l$ then provide the well-known parametrization of the
coset as a product of spheres \cite{GILMORE}. To see this, $u^1$ is a unit
vector in an $N$-dimensional complex space and consequently parametrizes
$S^{2N-1}$. The second vector $u^2$ is a similar unit vector, but one which is
orthogonal to $u^1$, and consequently parameterizes
$S^{2N-3}$. Continuing this chain of argument, we see that the
vectors $\{u^l\}$ parametrize the product of spheres
\begin{equation}
{SU(N)\over SU(N-K)}\simeq S^{2N-1}\times S^{2N-3}\times\cdots\times S^{2N-2K+1}\ .
\end{equation} 

{}From \eqref{uptri} we can read off the expansion of the vectors $v^l$
in terms of this basis:
\begin{equation}
v^l=\sum_{k=1}^l\xi^l_ku^k\ .
\end{equation}
The numerator on the right-hand side of Eq.~\eqref{quotientjac} can be
determined through the following iterative process. First we start
with $v^1=\xi_1^1u^1$, whose measure can be written in polar
coordinates as
\begin{equation}
\prod_{u=1}^N dv_u^1dv_u^{1*}=2^N(\xi_1^1)^{2N-1}\,d\xi_1^1\,d^{2N-1}\hat\Omega_1\ .
\end{equation}
Here $d^{2N-1}\hat\Omega_1$ is the usual measure for the solid angles on
$S^{2N-1}$ parametrized by $u^1$. Continuing the process on the next vector
$v^2=\xi_1^2u^1+\xi_2^2u^2$, we have
\begin{equation}
\prod_{u=1}^N dv_u^2dv_u^{2*}=d\xi_1^2d\xi_1^{2*}
\cdot2^{N-1}(\xi_2^2)^{2N-3}\,d\xi_2^2\,d^{2N-3}\hat\Omega_2\ .
\end{equation}
In general
\begin{equation}
\prod_{u=1}^N dv_u^ldv_u^{l*}=\prod_{k=1}^{l-1}d\xi_k^ld\xi_k^{l*}
\cdot2^{N-l-1}(\xi_l^l)^{2N-2l+1}\,d\xi_l^l\,d^{2N-2l+1}\hat\Omega_l\ .
\end{equation}
where $\hat\Omega_l$ is parametrized by $u^l$. Hence
\begin{equation}\begin{split}
{\mathcal D}v=&
2^{NK-K(K-1)/2}(\xi_1^1)^{2N-1}(\xi_2^2)^{2N-3}\cdots(\xi_K^K)^{2N-2K+1}
d\xi_1^1\cdots d\xi_K^K\,\prod_{k<l}d\xi^l_k\,d\xi^{l*}_k\\
&\qquad\qquad\qquad\times d^{2N-1}\hat\Omega_1\,d^{2N-3}\hat\Omega_2\cdots
d^{2N-2K+1}\hat\Omega_K\ .
\elabel{numanss}\end{split}\end{equation}

Using Eqs.~\eqref{detWdef}, \eqref{quotientjac}, \eqref{denomanswer}
and \eqref{numanss}, we therefore obtain
\begin{equation}
{\mathcal D}v=
2^{NK-K(K+1)/2}(\det_{K}
W)^{N-K}\,d^{K^2}W\,d^{2N-1}\hat\Omega_1\,d^{2N-3}
\hat\Omega_2\cdots d^{2N-2K+1}\hat\Omega_K\ .
\end{equation}
Thus, when this measure is used to integrate only
gauge-invariant quantities, as we shall do in Sec.~VI, then only the
$\det\, W$ piece survives; the remaining terms are integrated over
the gauge coset to give an overall normalization constant reflecting
the volume of the coset \eqref{gcoset}. Re-introducing $k=K/2$ and
the vectors $w$, we state our final result as
\begin{equation} 
\int_{\rm gauge\atop coset}d^{2kN}w\, d^{2kN}\bar w=c_{k,N}\left({\rm
det}_{2k}W\right)^{N-2k} \, d^{k^2}W^0\prod_{c=1,2,3}d^{k^2}W^c\, .
\elabel{E36.1}\end{equation} 
The integral over the $2k\times 2k$
matrix $W$ has been written as four separate integrals over the
$k\times k$ matrices $W^0$ and $W^c$, defined in \eqref{bosbi}, with
respect to the basis $T^r$, defined below
\eqref{dmudef}.\footnote{This accounts for an additional factor of
$2^{-2k^2}$.} The normalization constant is precisely
\begin{equation}\begin{split}
c_{k,N}\ &= 2^{2kN-4k^2-k}\,{\rm Vol}\,S^{2N-1}\cdot{\rm
Vol}\,S^{2N-3}\cdots{\rm Vol}\,S^{2N-4k+1}\\&=\
{2^{2kN-4k^2+k}\pi^{2kN-2k^2+k}\over\prod_{i=1}^{2k}(N-i)!}\ .
\elabel{E37.1}\end{split}\end{equation} 

In changing variables from the  $w$'s to the 
 $W$'s, one must also, of course,
specify the integration domain for the latter. Since these
variables are the inner products of vectors they will be constrained
by various triangle inequalities. Fortunately, these technicalities
will not be important for us,  the reason being
 that the saddle-point values of the $W$'s obtained in Sec.~5.2
 satisfy all such triangle inequalities by inspection.

We now turn to the fermionic sector. As explained in Sec.~II.6, the
fermionic equivalent of integrating out the gauge coset is to
integrate out the Grassmann variables $\nu^A$ and $\bar\nu^A$ defined
in Eqs.~\eqref{E44}-\eqref{E45}. Since these coordinates are
orthogonal to the $\bar w$ and $w$ vectors, respectively, it is easy
to see from \eqref{zmcona} that they do not appear in the fermionic
ADHM constraints; however, they do appear in the fermion quadrilinear
term \eqref{Skquadef}.  The Jacobian for the change of variables from
the original fermionic coordinates $\{\mu^A,\bar\mu^A\}$ 
to the variables
of Sec.~II.6, namely $\{\zeta^A,\bar\zeta^A,\nu^A,\bar\nu^A\}$, is, for each
value of $A$,\footnote{We define the integrals over the $k\times k$
matrices $\zeta^{\aD A}$ and $\bar\zeta^{\aD A}$ with respect to the
Hermitian basis $T^r$, as for the bosonic variables $a'_n$ and $W^0$,
and the fermionic variables ${\cal M}^{\prime A}_\alpha$, above.}
\begin{equation}
{\partial\big(\{\mu^A,\bar\mu^A\}\big)\over
\partial\big(\{\zeta^A,\bar\zeta^A,\nu^A,\bar\nu^A\}\big)}
=\left(\det_{2k}\, W\right)^{-k}\ .
\elabel{jacznu}\end{equation}

In order to integrate out the $\nu^A$ and $\bar\nu^A$ variables it is
useful to split the fermion bilinear \eqref{newmatdef} that couples to $\chi_{AB}$ in
\eqref{E53} as
\begin{equation}
\Lambda^{AB}=\hat\Lambda^{AB}+\tilde\Lambda^{AB},
\elabel{E50}\end{equation}
where the first term has components just depending on $\{\nu^A,\bar\nu^A\}$:
\begin{equation}
\hat\Lambda^{AB}_{ij}={1\over2\sqrt2}\left(\bar\nu_{iu}^A\nu_{uj}^B-
\bar\nu_{iu}^B\nu_{uj}^A\right),
\elabel{E51}\end{equation}
and the second term depends on the remaining variables
\begin{equation}
\tilde\Lambda^{AB}={1\over2\sqrt2}\big(\ 
\bar\zeta^A_\aD W^\aD_{\ \bD}\zeta^{\bD B}-\bar\zeta^B_\aD W^\aD_{\
\bD}\zeta^{\bD A}
+\{{\cal M}^{\prime\alpha A},{\cal M}^{\prime B}_\alpha\} \big)\ .
\elabel{E52}\end{equation}
We can now explicitly integrate out the $\nu^A$'s and $\bar\nu^A$'s:
\begin{equation}
\int\prod_{A=1,2,3,4}d^{k(N-2k)}\nu^A\, d^{k(N-2k)}\bar\nu^A\,
\exp\big[\sqrt8\pi
ig^{-1}{\rm tr}_k
\,\chi_{AB}\bar\nu^A\nu^B\big]=\left(8\pi^2\over
g^2\right)^{2k(N-2k)} \left({\rm det}_{ 4k}\chi\right)^{N-2k}\ ,
\elabel{intoutnu}\end{equation}
where the determinant is of $(\chi_{AB})_{ij}$ viewed as a
$4k\times 4k$ matrix.

Actually, going to a gauge invariant measure carries with it
a very significant
advantage: we can integrate out the ADHM $\delta$-functions explicitly. In
the bosonic sector, one notices that the ADHM constraints are written
as 
\begin{equation}
0=W^c+[\,a'_n\,,\,a'_m\,]\,\trtwo\,\tau^c\sigmabar^{nm}= 
 W^c - i [\,a'_n\,,\,a'_m\,]\,\etabar^c_{nm}\ .
\elabel{E27}\end{equation}
in terms of the gauge invariant coordinates. Notice that they are {\it
linear\/} in $W^c$ and consequently the $W^c$ integrals simply remove
the bosonic ADHM $\delta$-functions in \eqref{dmudef} (giving rise to the
numerical factor of $2^{3k^2}$ from the $\tfrac12$'s in the arguments of
the $\delta$-functions). There is a similar happy story in the fermionic
sector: in terms of the new variables the fermionic ADHM constraints
\eqref{zmcona} are
\begin{equation}
\bar\zeta_{\bD}^AW_{\ \,\aD}^{\bD}+W_{\aD\bD}\zeta^{\bD A}+
\big[{\cal M}^{\prime\alpha A},a'_{\alpha\aD}\big]=0.
\elabel{E46}\end{equation}
These equations can be used to eliminate the $8k^2$ variables $\bar\zeta_\aD^A$
The relevant integral is simply 
\begin{equation}
\int d^{2k^2}\sigma^A\,
\prod_{r=1,\ldots,k^2}\prod_{\aD=1,2}\delta
\Big({\rm tr}_k\,T^r(\bar\zeta_{\bD}^AW_{\ \,\aD}^{\bD}+W_{\aD\bD}
\zeta^{\bD A}+
\big[{\cal M}^{\prime\alpha A},a'_{\alpha\aD}\big])\Big)
=\left(\det_{2k}\, W\right)^{k}.
\end{equation}
Notice that the factor on the right-hand side conveniently cancels the
Jacobian of \eqref{jacznu}.

Putting everything together we now have a much simpler form for the
measure for integrating gauge invariant quantities:
\begin{equation}\begin{split}
&\int\dmuphys\,e^{-\Skinst}\ =\ 
{g^{8k^2}2^{4kN-19k^2/2+k/2}e^{2\pi ik\tau}c_{k,N}
\over\pi^{2kN+11k^2}\,{\rm Vol}\,\U(k)}\int d^{k^2}W^0\,d^{4k^2}a'\,
d^{6k^2}\chi\,\prod_{A=1,2,3,4}d^{2k^2}\M^{\prime A}\,
d^{2k^2}\zeta^A
\\
&\qquad\qquad\times\ \left({\det^{}_{2k} W}{\det^{}_{4k}\chi}\right)^{N-2k}\,
\exp\big[4\pi i g^{-1}\,{\rm tr}_k\,\chi_{AB}\tilde \Lambda^{AB}
\ -\ {\rm tr}_k\,\chi_a\BL\chi_a\big]\, .
\elabel{gaugeinvmeas}
\end{split}\end{equation}
In particular, since the number of integration variables no longer
grows with $N$ (in contradistinction to the flat measure), this
expression is immediately amenable to a large-$N$ saddle-point analysis, to which we
now turn.

\rsen\section{The Large-$N$ Limit in a Saddle-Point Approximation} 	

\subsection{The one-instanton measure revisited, and the emergence of
the $S^5$}

As a prelude to  the technically  demanding case of
multi-instantons, it is instructive to revisit the one-instanton
sector, and in particular, to derive the result \eqref{CNdef} cited
in Sec.~I.3 above \cite{DKMV}. In this simple setting, we will
formulate a large-$N$ saddle-point method which naturally extends to
$k>1,$ and we will already see the emergence of the $S^5$ factor expected from
the AdS/CFT correspondence.

Consider the gauge invariant measure \eqref{gaugeinvmeas} for
$k=1$. In this case we can greatly simplify the expression: in
addition to the relations \eqref{1instid} one has
$\tilde\Lambda^{AB}=0$,
and also \cite{DKMV}
\begin{equation}
\det_4\,\chi\ =\
\tfrac1{64}\big(\epsilon^{ABCD}\chi_{AB}\chi_{CD}\big)^2\ \equiv\
\tfrac1{64}\sum_{a=1}^6(\chi_a)^2\ .
\elabel{chidef1}
\end{equation}
The first equality in Eq.~\eqref{chidef1} is a well-known
factorization property of $4\times4$ antisymmetric matrices, while the
second equality involves the $SO(6)$ rewrite \eqref{E54.1}.
Introducing six-dimensional polar coordinates $\chi_a=\{r,\sfc\}$, one
then finds for the measure:
\begin{equation}
\int d\mu_{\rm phys}^1\,e^{-S_{\rm inst}^1}\ =\
{g^8e^{2\pi i\tau} \over2^{31}\pi^{13}(N-1)!(N-2)!}
\int d^4\com\,d\rho\,d^5\sfc\,\rho^{4N-7}\,I_N
\prod_{A=1,2,3,4}
d^2\xi^A\, d^2\bar\eta^A\ .
\elabel{OneIM}
\end{equation}
Here $I_N$ denotes the $r$ integration, which it is instructive to
separate out:
\begin{equation}
I_N=\int_0^\infty\, dr\,r^{4N-3}\,e^{-2\rho^2r^2}=
\tfrac12(2\rho^2)^{-2N+1}\int_0^\infty\,dx\,x^{2N-2}\,e^{-x}
=
\tfrac12(2\rho^2)^{1-2N}(2N-2)!\ .
\elabel{INdef1}
\end{equation}
{}From Eqs.~\eqref{OneIM}-\eqref{INdef1} one sees that the $X_n$ and $\rho$
integrations assemble into the scale-invariant $AdS_5$ volume form
$d^4X\,d\rho\,\rho^{-5}$. Moreover we can perform the integration over
$\sfc$, yielding ${\rm Vol}\,S^5=\pi^3$, to recapture the numerical
factor $C_N$ quoted in Eq.~\eqref{CNdef} (where the remaining powers
of two come from the factors of 96 associated to each of the operator
insertions).

Equations \eqref{OneIM}-\eqref{INdef1} are the principal result of
\cite{DKMV}, and, we stress, are exact for all $N$. However, this
exactness can be traced to the factorization property \eqref{chidef1},
which is special to the one-instanton sector. In order to generalize
the above to $k>1$, at least in the large-$N$ limit, it is important
to reproduce these results in an alternative way, using saddle-point
methods. To this end, note that the integral $I_N$ is nothing but a
Gamma function.  For large $N$ it is well approximated by Stirling's
formula, which, the reader will recall, is derived as follows.  First
one rescales $r\rightarrow\sqrt Nr$, or equivalently
\begin{equation} 
\chi_a\rightarrow\sqrt N\chi_a \ , \elabel{E56}\end{equation} 
so that
$N$ factors out of the exponent.  The integral then becomes
\begin{equation}
I_N=N^{2N-1}
\int_0^\infty\, dr\,r^{-3}\,e^{2N(\log r^2-\rho^2r^2)}\ ,
\end{equation}
which is in a form amenable to a standard saddle-point evaluation. 
The saddle-point is at $r=\rho^{-1}$ and, to leading order, a Gaussian
integral around the solution gives
\begin{equation}
\lim_{N\rightarrow\infty}\,I_N\ =\
\rho^{2-4N}N^{2N-1}e^{-2N}\sqrt{\pi\over N}\ , 
\end{equation}
which is valid up to $1/N$ corrections.

While this is all very elementary, we have uncovered something truly
surprising. A single Yang-Mills instanton in an ${\cal N}=4$
supersymmetric gauge theory is effectively parametrized at large $N$
by its position $X_n$, scale size $\rho,$ and a point $\hat\Omega$ on
$S^5$ (as well as the sixteen fermionic collective coordinates). Notice
that this parametrization {\it only\/} emerges in the large-$N$ limit
where the solution of the saddle-point equations identifies the radius
of the sphere with the inverse of the instanton scale size; i.e.~the
radius of the sphere with the radial variable of $AdS_5$. 
These parameters are precisely the coordinates required to specify the position of 
an object in the ten-dimensional space $AdS_5\times S^5$,
where $\rho^{-1}$ is a radial coordinate on the $AdS_5$. Given
Maldacena's conjectured correspondence between correlation functions
in the supergravity theory in ten dimensions and the four-dimensional gauge
theory on its boundary, this leads us to identify a large-$N$
Yang-Mills instanton with a (singly-charged) D-instanton. In this
identification, the sixteen supersymmetric and superconformal zero-modes of
the Yang-Mills instanton $\{\xi_\alpha^A,\bar\eta^{\aD A}\}$ are
identified with the sixteen supersymmetries of the Type IIB string theory
broken by the D-instanton denoted $\varepsilon$ in Eq.~\eqref{int}.

\subsection{Na\"\i vely the $k$-instanton moduli space contains $k$
copies of $AdS_5\times S^5$}

We now turn to the case of general topological number $k$. For
$k>1$ one quickly abandons any hope of performing the integrations in
Eq.~\eqref{gaugeinvmeas} exactly. Instead, let us extend the large-$N$
saddle-point treatment developed in Sec.~V.1.  As in the one-instanton
sector, we first perform the rescaling \eqref{E56}.  The
gauge-invariant measure \eqref{gaugeinvmeas} then becomes:
\begin{equation}\begin{split}
\int\dmuphys\,e^{-\Skinst}\ =\ & {g^{8k^2}N^{k^2}e^{2\pi ik\tau}
\over2^{27k^2/2-k/2}\,\pi^{13k^2}\,{\rm Vol}\,\U(k)}\int
d^{k^2}W^0\,d^{4k^2}a'\,
d^{6k^2}\chi\,\prod_{A=1,2,3,4}d^{2k^2}\M^{\prime A}\, d^{2k^2}\zeta^A
\\ &\times\ \left({\det^{}_{2k} W}{\det^{}_{4k}\chi}\right)^{-2k}\,
\exp\big[4\pi i g^{-1}\sqrt{N}\,{\rm tr}_k\,\chi_{AB}\tilde
\Lambda^{AB} \ -\ NS_{\rm eff}\big]\ .  \elabel{E57.2}
\end{split}\end{equation}
Here, in anticipation of the large-$N$ limit, we have already applied
Stirling's formula to the numerical prefactor \eqref{E37.1}.  We have also
introduced the ``effective $k$-instanton action'' $S_{\rm eff}$ in
which all the terms which scale with $N$ are collected:\footnote{As previously,
we translate back and forth as convenient between the
antisymmetric tensor representation $\chi_{AB}$ and the $SO(6)_R$
vector representation $\chi_a$, $a=1,\ldots,6$ \eqref{E54.1}.}
\begin{equation}S_{\rm
eff}=-2k(1+3\log2)-\log{\rm det}_{2k}W-\log{\rm det}_{4k}\chi +{\rm
tr}_k\,\chi_a\BL\chi_a\, . \elabel{E57}\end{equation} This expression
involves the $11k^2$ bosonic variables comprising the eleven
independent $k\times k$ Hermitian matrices $W^0$, $a'_n$ and
$\chi_a$. The remaining components $W^c$, $c=1,2,3,$ are eliminated
in favor of the $a'_n$ via the ADHM constraint \eqref{E27}.  The
action is invariant under the $\U(k)$ symmetry \eqref{restw} which
acts by adjoint action on all the variables.

With $N$ factored out of the exponent, the measure is in a form which
is amenable to a saddle-point treatment as $N\rightarrow\infty$. The
coupled saddle-point equations read:
\begin{subequations}
\begin{align}
\epsilon^{ABCD}\left(\BL\cdot\chi_{AB}\right) \chi_{CE}\ &=\
\tfrac12\delta^D_E \,1_{\sst [k]\times[k]}\ ,\elabel{E58}\\
\chi_a\chi_a\ &=\ \tfrac12(W^{-1})^0\ ,\elabel{E59}\\
[\chi_a,[\chi_a,a'_n]]\ &=\ i\bar\eta^c_{nm}[a'_m,(W^{-1})^c]\ .
\elabel{E60}
\end{align}
\end{subequations}
These are obtained by varying $S_{\rm eff}$ with respect to the matrix
elements of $\chi$, $W^0$ and $a'_n$, respectively, and rewriting
``log$\,$det'' as ``tr$\,$log.''  We have employed the $k\times k$
matrices
\begin{equation}
(W^{-1})^0=
{\rm tr}_2\,W^{-1},\qquad
(W^{-1})^c=
{\rm tr}_2\,\tau^cW^{-1}.
\elabel{E60.1}\end{equation}

The general solution to these coupled saddle-point equations is easily
found. It has the simple property that all the quantities are diagonal
in instanton indices,
\begin{subequations}\begin{align}
W^0\ &=\ {\rm diag}\big(2\rho_1^2,\ldots,2\rho_k^2\big)\ ,
\elabel{E61a}\\ \chi_a\ &=\ {\rm
diag}\big(\rho_1^{-1}\sfc^1_a,\ldots,\rho_k^{-1}\sfc^k_{a}\big)\
,\elabel{E61b} \\ a'_n\ &=\ {\rm
diag}\big(-\com^1_n,\ldots,-\com^k_n\big)\ ,
\elabel{E61c}\end{align}\end{subequations} 
up to a common adjoint
action by the $\U(k)/U(1)^k_{\scriptscriptstyle\rm diag}$ residual
symmetry. For each value of the instanton index $i=1,\ldots,k$, the
six-dimensional unit vector $\sfc_a^{i}$ parametrizes an independent
point on $S^5$,
\begin{equation}
\sfc^i_{a}\sfc^i_{a}=1\qquad(\text{no sum on }i)\, ,
\elabel{E62}\end{equation}
where the radius of the $i^{\rm th}$ $S^5$ factor is $\rho_i^{-1}$.

 A simple picture of this leading-order saddle-point solution emerges:
it can be thought of as $k$ independent copies of the one-instanton
saddle-point solution described in Sec.~V.1, where the $i^{\rm th}$
instanton is parametrized by a point $(X_n^i,\rho_i,\hat\Omega_a^i)$
on $AdS_5\times S^5$. Additional insight into this solution emerges
from considering the $SU(2)$ generators $(t^c_i)^{}_{uv}$
describing the embedding of the $i^{\rm th}$ instanton inside
$SU(N)$. From Eqs.~\eqref{embedmulti}, \eqref{bosbi} and \eqref{E61a}
one derives the commutation relations
\begin{equation}
[\,t^a_i\,,\,t^b_j\,]\ =\ 2i\delta_{ij}\,\epsilon_{abc}\,t^c_i
\elabel{relcomm}
\end{equation}
so that at the saddle-point, thanks to the Kronecker-$\delta,$ the $k$
individual instantons lie in $k$ mutually commuting $SU(2)$
subgroups. Actually this feature follows from large-$N$ statistics
alone,\footnote{Consider the analogous problem of $k$ randomly
oriented vectors in ${\Bbb R}^N$ in the limit $N\rightarrow\infty;$
clearly the dot products of these vectors tend to zero simply due to
statistics.}  and has nothing to do with either the existence of
supersymmetry or with the details of the ADHM construction;
nevertheless the fact that Eq.~\eqref{relcomm} emerges from the
saddle-point equations is a reassuring check on our formalism.  In
this one particular respect, the solution \eqref{E61a}-\eqref{E61c}
can be said to be dilute-gas-like. Another important property is that
the effective action \eqref{E57} evaluated on these saddle-point solutions is
zero; hence there is no exponential dependence on $N$ in the final
result. Finally we should make the technical point that, thanks to
the diagonal structure of these solutions, they are automatically consistent
with the triangle inequalities on the boson bilinear $W$ discussed in
the paragraph following Eq.~\eqref{E37.1}; hence we never need to
specify more explicitly the integration limits on the $W$ variables.

We are now squarely faced with the puzzle described in
Sec.~I: the fact that we have obtained $k$ copies of
$AdS_5\times S^5$ ostensibly contradicts the supergravity side of the AdS/CFT
correspondence, where only one copy of this moduli space appears
at each topological level $k$ (e.g., see the supergravity expression
\eqref{grnfcn} for the correlator $G_{16}$). In the following
subsection we show that this $k$-fold degeneracy is lifted when one
considers small fluctuations in the neighborhood of the generic solution
\eqref{E61a}-\eqref{E61c}, and that the
Yang-Mills multi-instanton moduli space indeed collapses to a single
copy, precisely as dictated by Maldacena's conjecture.

\subsection{The exact $k$-instanton moduli space collapses to one copy
of $AdS_5\times S^5$}

   Let us repeat some general lore about saddle-point 
methods. To evaluate a multi-dimensional integral in saddle-point
approximation, it never suffices merely to identify the space of
saddle-point solution(s). One must also evaluate the leading
non-vanishing small-fluctuations integral in the background of the
solution(s).  Typically this small-fluctuations integral is Gaussian,
and produces a prefactor proportional to $(\det\,{\rm M})^{-1/2}$
where M is the quadratic form. If M has null-vector directions, one
needs to go beyond quadratic order in small fluctuations, specifically
up to the order where the degeneracy in these directions is
lifted. Of course, the degeneracy may persist to
all orders reflecting 
the existence of a moduli-space of saddle-point solutions. One must then 
introduce collective coordinates which parametrize the 
flat directions of the leading-order action\footnote{In the following, 
"leading-order action" denotes those terms in the exponent of the integrand 
which are of order $N$.} and restrict the small-fluctuations integration to 
directions orthogonal to the moduli. 
However, the flat 
directions of the leading-order action can still be 
lifted at higher order in $1/N$. For example, 
this happens when the small fluctuations determinant described 
above depends explicitly on the moduli. 
On exponentiating the determinant we find a potential term 
of order $N^{0}$ which corrects the leading-order action and 
lifts the flat directions. 
This effect is particularly 
important when the space of saddle-point solutions is non-compact 
as the potential term can render the otherwise 
divergent integral over the collective coordinates finite. Finally, 
there may be flat directions which remain unlifted to all orders in $1/N$ 
(typically because they are protected by a symmetry). In this case the 
corresponding collective coordinates parametrize a moduli-space of 
exactly degenerate stationary-points of the effective 
action to all orders in $1/N$.   

Now let us apply these general considerations to the particular space of
saddle-point solutions obtained in Sec.~V.2 above. At the
$k$-instanton level, to leading order in $1/N$, this solution space has
the geometry of $(AdS_5\times S^5)^k$. However, a small-fluctuations
analysis in the background of a generic solution point confirms that
most of this degeneracy is actually lifted at the next order by 
the effect described above. Specifically, the flat directions corresponding 
to the {\em relative} positions of the $k$ instantons on 
$AdS_{5}\times S^{5}$ are lifted at order $N^{0}$. 
Thus we find that the exact, unlifted, bosonic moduli
space consists of only one copy of $AdS_5\times S^5$ which corresponds 
to the center-of-mass position of the instantons. Furthermore, we
will see that the large-$N$ saddle-point expansion can be reformulated 
as an expansion in fluctuations around the exact saddle-points 
parametrized by these moduli. In particular, the leading order 
in this expansion can be elegantly reinterpreted as a partition function of a
standard supersymmetric $SU(k)$ gauge theory, and in this way, reduced
to previously known results. We will see that this matches precisely
the description of D-instantons in string theory.

In order to verify these claims, let us first perform a
small-fluctuations expansion in the neighborhood of a generic solution
point on  $(AdS_5\times S^5)^k$. By ``generic'' we mean that for all
$i$ and $j,$
\begin{equation}
\com_n^i\neq \com_n^j\ ,\quad\sfc_a^i\neq \sfc_a^j\
,\quad\rho_i\neq\rho_j\ .  \elabel{genericase}\end{equation} 
We will
truncate the expansion at Gaussian order, and restrict ourselves for
now to the bosonic degrees of freedom. Let us introduce $k\times k$
fluctuating fields $\{\,\delta W^0_{ij}\,,\,\delta\chi_{aij}\,,\,\delta a^{\prime
}_{nij}\,\}$, in the background of the solution
\eqref{E61a}-\eqref{E61c}, by defining:
\begin{subequations}\begin{align}
W^0\ &=\ 
{\rm diag}\big(2\rho_1^2,\ldots,2\rho_k^2\big)+\delta W^0
\ , \elabel{E61aa}\\ \chi_a\ &=\ {\rm
diag}\big(\rho_1^{-1}\sfc^1_a,\ldots,\rho_k^{-1}\sfc^k_{a}\big)+\delta\chi_a\
,\elabel{E61bb} \\ a'_n\ &=\ {\rm
diag}\big(-\com^1_n,\ldots,-\com^k_n\big)+\delta a'_n\ .
\elabel{E61cc}\end{align}\end{subequations} 
Inserting these expressions into $S_{\rm eff}$, one finds that
the Gaussian integration is  governed by the effective quadratic action
\begin{equation}
S^{(2)}=\sum_{1\le i,j\le k}\Big(v_{ij}\cdot
v_{ij}\left\vert\sigma'_{ij}\cdot\sigma'_{ij}\right\vert- \left\vert
v_{ij}\cdot \sigma'_{ij}\right\vert^2+\left\vert
u_{ij}\cdot\sigma_{ij}\right\vert^2\Big).
\elabel{quadfluct}\end{equation} 
Here $\sigma$ and $\sigma'$ denote the
matrix-valued 11-vectors of fluctuating fields\footnote{Inner products
on such 11-vectors are defined as $(v^1,v^2,v^3)\cdot(u^1,u^2,u^3)=
v^1_au^1_a+v^2_nu^2_n+v^3u^3$.}
\begin{equation}
\sigma_{ij}=\big(\rho_i\rho_j\delta\chi_{aij}\,,\,\delta a^{\prime}_{nij}\,,\,
(2\rho_i\rho_j)^{-1}\delta W^0_{ij}\big)\ ,\qquad
\sigma'_{ij}=\big(\rho_i\rho_j\delta\chi_{aij}\,,\,\delta
a^{\prime}_{nij}\,,\,0\big)\ ,
\end{equation}
while $v_{ij}$ and $u_{ij}$ are matrix-valued 11-vectors formed from
parameters of the saddle-point solution:
\begin{equation}
v_{ij}=\big(\rho_i^{-1}\sfc_a^i-\rho_j^{-1}\sfc_a^j\,,\,
(\rho_i\rho_j)^{-1}(\com_n^i-\com_n^j)\,,\,0\big)\ ,\qquad
u_{ij}=\big(\rho_i^{-1}\sfc_a^i+\rho_j^{-1}\sfc_a^j\,,\,0\,,\,1\big).
\end{equation}

By inspection, the quadratic form \eqref{quadfluct} is positive
semi-definite.  The null-vectors can be classified as
follows. For the diagonal elements, the variations 
\begin{equation}
\sigma_{ii}=\big(\rho_i^2\delta\chi_{aii}\,,\,\delta
a^{\prime}_{nii}\,,\,-2\rho_i\sfc^i\cdot\delta\chi_{ii}\big)\ ,
\end{equation}
have zero eigenvalue. These correspond to motion along the
$10k$-dimensional solution manifold \eqref{E61a}-\eqref{E61c}
parametrized by $\{\com_n^i,\sfc_a^i,\rho_i\}$. As for the
off-diagonal elements, the only null-vectors are the $k(k-1)$
independent variations $\sigma_{ij}$ in the directions
\begin{equation}
\big(\rho_j\sfc_a^i-\rho_i\sfc_a^j\,,\,-\com_n^i+
\com_n^j\,,\,
\rho_i/\rho_j-\rho_j/\rho_i\big)\ ,\qquad i\neq j\, ,
\end{equation}
with no sum on $i$ and $j$.  These are precisely the variations
induced by the adjoint action of $\U(k)/U(1)^k_{\scriptscriptstyle\rm
diag}$ on the solution.  In
summary, at a generic point \eqref{genericase}, all the Gaussian
null-vectors correspond to directions which are already explicitly
taken into account by the moduli for $(AdS_5\times S^5)^k\times
U(k)/U(1)^k_{\scriptscriptstyle\rm diag}$.  Conversely, since there
are no additional null-vectors,
we conclude that at a generic solution point this moduli space is
complete (at least locally).

However the generic case \eqref{genericase} is not the full
story. In particular, whenever a vector $v_{ij}$ associated to a pair
of instantons is zero, additional flat directions appear at quadratic
order. The condition $v_{ij}=0$ requires that the two instantons live
at the same point in $AdS_5\times S^5$, i.e., $\com_n^i=\com_n^j$,
$\rho_i=\rho_j$, and $\sfc_a^i=\sfc_a^j$.  At this special point, all
the variations $\sigma_{ij}$ which are orthogonal to the vector
$u_{ij}$ are flat to quadratic order.  Correspondingly, the
determinantal prefactor $(\det{\rm M})^{-1/2}$ blows up. Additional
zeroes of this determinant appear whenever the positions of additional
instantons coalesce, leading to a greater divergence. 
As above, if one 
were to exponentiate this determinantal prefactor, one would naturally 
obtain a correction to the leading-order action of order $N^{0}$. 
In the present case this leads to an attractive singular potential 
which partially lifts the degeneracy in the moduli space, and draws
the $k$ distinct instantons to a {\it common center\/} in $AdS_5\times
S^5\,$! Loosely speaking, the attractive interaction results in a
$k$-instanton ``bound state''. However, the divergent nature 
of potential indicates that the above 
analysis in terms of quadratic fluctuations around the generic 
saddle-point is not adequate.  In particular, 
the extra zero modes appearing when two instantons approach each other 
mean that, in these regions of the moduli space, 
we must go to higher order in the fluctuation expansion.  

To perform a consistent saddle-point analysis, we should now expand to 
next-to-leading order around the generic solution and 
augment the quadratic action $S^{(2)}$ with a term $S^{(4)}$ quartic in 
$\delta W^{0}$, $\delta\chi_{a}$ and $\delta a'_{n}$. 
After integrating out these fluctuations, we should finally obtain a
convergent integral over those moduli of the generic saddle-point 
solution which describe the relative positions of the $k$ instantons on 
$AdS_{5}\times S^{5}$. In fact we will 
meet an important simplification which means that we will not 
need to perform these steps in full. 
We will find that 
the leading order in $1/N$ can actually be obtained by a much simpler 
proceedure: rather than expanding in fluctuations 
around the generic solution we will instead expand around the maximally 
degenerate solution identified above.
This amounts to treating the $k-1$ 
relative positions on $AdS_{5}\times S^{5}$ 
as fluctuations around the maximally-degenerate solution 
rather than moduli of the generic solution. 
One way of seeing that the two proceedures are equivalent at 
leading order in $1/N$ is to note that $S^{(2)}$, which by definition 
is quadratic in fluctuations around the generic solution, also happens 
to depend only quadratically on the moduli. This means that the 
exact expression \eqref{quadfluct} for $S^{(2)}$ can also be viewed 
as a term of quartic order in an expansion around 
the maximally-degenerate solution where the additional moduli are now
thought of as fluctuations. Thus the leading-order term in our 
new expansion of the action automatically contains the leading-order 
term in the old one. However, the leading term in the 
new expansion also contains additional quartic terms which lift the extra 
zero modes of the degenerate configurations thereby remove the divergent 
behaviour discovered above. 
In the following we shall find, up to an overall 
integral over one copy of $AdS_5\times S^5$, 
a finite well-defined answer for the leading term in the 
large-$N$ expansion of the ADHM measure.  

As discussed above we will now begin our small fluctuations 
analysis afresh, this time in the background of
the \it maximally degenerate \rm saddle-point solution
\begin{equation}
W^0=2\rho^2\,1_{\sst[k]\times[k]},\qquad
\chi_a=\rho^{-1}\sfc_a\,1_{\sst[k]\times[k]},\qquad
a'_n=-\com_n\,1_{\sst[k]\times[k]}\ ,
\elabel{specsol}
\end{equation}
which corresponds to $k$
instantons living at a common point $\{\com_n,\sfc_a,\rho\}$ in
$AdS_5\times S^5.$ (From the ADHM constraint \eqref{E27} it follows
that the remaining components of $W$ vanish: $W^c=0$ for $c=1,2,3.$)
This degenerate solution, unlike Eqs.~\eqref{E61a}-\eqref{E61c}, is
invariant under the residual $U(k)$. With the instantons sitting on
top of one another, it looks like the complete opposite of the dilute
instanton gas limit; however the instantons still live in $k$ mutually
commuting $SU(2)$ subgroups of $SU(N)$ as per Eq.~\eqref{relcomm},
which is a dilute-gas-like feature.  Notice that, just as in the one
instanton sector, the radius of the $S^5$ is fixed to be $\rho^{-1}$
at the saddle-point.

In order to expand about this special solution, one first needs to
factor out the exact moduli corresponding to the common position of
the $k$-instanton ``bound state'' on
$AdS_5\times S^5$. This is done in the following way.
For each $k\times k$ matrix, we  introduce a basis of traceless
Hermitian matrices $\hat T^r$, $r=1,\ldots,k^2-1$, normalized by ${\rm
tr}_k\,\hat T^r\hat T^s=\delta^{rs}$.  For each $k\times k$ matrix $v$
we separate out the ``scalar'' component $v_0$ by taking
\begin{equation}
v=v_01_{\sst [k]\times[k]}+\hat v^r\hat T^r\, .
\end{equation}
The change of variables from the $T^r$ basis used in \eqref{dmudef} $\{1_{\sst
[k]\times[k]},\hat T^r\}$ involves a Jacobian
\begin{equation}
d^{k^2}v=k^{\pm1/2}dv_0\,d^{k^2-1}\hat v,
\end{equation}
where $\pm1$ refers to bosons and fermions, respectively. For the
moment we continue to focus on the bosonic variables, which are
decomposed as follows:
\begin{subequations}
\begin{align}
a'_n&=-X_n1_{\sst [k]\times[k]}+\hat a'_n\ ,\elabel{asplitdef}\\
\chi_a&=\rho^{-1}\sfc_a1_{\sst [k]\times[k]}+\hat\chi_a\ .\elabel{chisplitdef}
\end{align}
\end{subequations}
By definition the traceless matrix variables $\hat a'_n$ and
$\hat\chi_a$ are the fluctuating fields (there is no need to write
$\delta\hat a'_n$ or $\delta\hat\chi_a$). Note that, in terms of 
the generic solution considered above, 
the diagonal entries of the matrices $\hat a'_{n}$ and $\hat{\chi}_{a}$ 
correspond to the moduli $u$ and $v$ 
while the off-diagonal elements correspond to fluctuations.  

Inserting Eqs.~\eqref{asplitdef}-\eqref{chisplitdef} into
Eq.~\eqref{E57} and Taylor expanding is a tedious though
straightforward exercise. As anticipated above, we will now expand 
to fourth order in the fluctuating fields around the solution parametrized 
by the ten exact $AdS_5\times S^5$ moduli. 
The expansion of the determinant terms in \eqref{E57} is facilitated
by first writing ``$\log\det$'' as ``${\rm tr}\log$'' and then
expanding the logarithm:
\begin{equation}\begin{split}
{\rm tr}_{2k}\log W\ =\ &2k\log\rho^2+{1\over\rho^2}{\rm tr}_k(\delta
W^0)-{1\over4\rho^4}{\rm tr}_k(\delta
W^0)^2
+{1\over12\rho^6}{\rm tr}_k(\delta W^0)^3\\&\qquad\qquad\qquad-{1\over32\rho^8}{\rm
tr}_k(\delta W^0)^4
+{1\over2\rho^4}{\rm
tr}_k\,\big[\hat a'_n,\hat a'_m\big]^2+\cdots\ ,
\elabel{expdw}\end{split}
\end{equation}
and
\begin{equation}
{\rm tr}_{4k}\log\chi
=-2k\log(8\rho^2)-2^5\rho^2{\rm
tr}_{4k}\,\big(\sfc^*\hat\chi\big)^2
+{2^9\rho^3\over3}{\rm tr}_{4k}\,\big(\sfc^*\hat\chi\big)^3-
2^{10}\rho^4{\rm tr}_{4k}\,\big(\sfc^*\hat\chi\big)^4+\cdots\ .
\elabel{expdchi}
\end{equation}
In these expansions we have  dropped fifth- and
higher-order terms in the fluctuating fields. 
Here we are anticipating the fact that these terms are not
needed to regularize the small-fluctuations integration. 
On obtaining a finite answer from the leading-order terms we may immediately 
rescale the integration variables in a standard way which shows that 
the fluctuations around the maximally degenerate saddle point are of 
order $N^{-1/4}$. 
The higher order terms in the exponent therefore yield 
subleading contributions in the large-$N$ expansion. In particular, this 
is true for the diagonal components of $\hat{a}'_{n}$ and 
$\hat{\chi}_{a}$ which correspond to the moduli of the generic saddle-point 
solution discussed earlier in this Section. This shows that our large-$N$ 
expansion around the maximally degenerate saddle-point is self-consistent. 
In Eq.~\eqref{expdw} we
have used Eq.~\eqref{E27}, while in Eq.~\eqref{expdchi} and in
subsequent equations we move back and forth as convenient between the
6-vector and the antisymmetric tensor representations of $\sfc$ and
$\chi$ using Eq.~\eqref{E54.1}. In particular, the $SO(6)$
orthonormality condition $\sfc\cdot\sfc=1$ becomes, in $4\times4$
matrix language,
\begin{equation}
\sfc\sfc^*=-\tfrac181_{\sst[4]\times[4]}\ ,\quad\hbox{or}\quad
\sfc^{-1}=-8\sfc^*\ ,
\elabel{resu2}
\end{equation}
which has been implemented in Eq.~\eqref{expdchi}.

Next we need a systematic method for re-expressing the traces over
 $4k\times4k$ matrices in Eq.~\eqref{expdchi} as traces over $k\times k$
 matrices. We will exploit the following ``moves'':\footnote{In the
 following, we should emphasize that $\dagger$ only acts on instanton
 indices, as per the reality condition \eqref{E54}, and {\it not\/} on
 $SU(4)$ matrix indices.}
\begin{subequations}
\begin{align}
\sfc^*\hat\chi&=-\hat\chi^\dagger\sfc-\tfrac14(\sfc\cdot\hat\chi)
1_{\sst[4]\times[4]}\ ,\elabel{resu1}\\ 
{\rm tr}_4\,E^\dagger F&={\rm tr}_4\,F^\dagger E=-\tfrac12(E\cdot F)\ ,\elabel{resu3}
\\ {\rm tr}_4\,E^\dagger FG^\dagger H&=\tfrac1{16}\big(E_aF_aG_bH_b-E_aF_bG_aH_b
+E_aF_bG_bH_a\big)\
.\elabel{resu4}
\end{align}
\end{subequations}
On the left-hand sides of Eq.~\eqref{resu3}-\eqref{resu4}, the
$4k\times4k$ matrices $\{E,F,G,H\}$ are assumed to be antisymmetric in
$SU(4)_R$ indices and subject to the usual conditions
\eqref{E54}-\eqref{E54.1}; the identity \eqref{resu1} follows from a
double application of Eq.~\eqref{E54}.  Using
Eqs.~\eqref{resu2}-\eqref{resu4} in an iterative fashion, it is then
easy to derive the following trace identities:
\begin{subequations}
\begin{align}
{\rm tr}_{4k}\,\big(\sfc^*\hat\chi\big)^2\ &=\
-{\rm
tr}_{4k}\,\hat\chi^\dagger\sfc\sfc^*\hat\chi-\tfrac14{\rm
tr}_k\big(\sfc\cdot\hat\chi\,
{\rm tr}_4(\sfc^*\hat\chi)\big)\notag\\
&=
{1\over2^3}{\rm tr}_k\,
\big(\sfc\cdot\hat\chi\big)^2-{1\over2^4}
{\rm tr}_k\,\hat\chi\cdot\hat\chi\, ,\elabel{ident1}\\
{\rm tr}_{4k}\,\big(\sfc^*\hat\chi\big)^3\ &=\ 
\tfrac18{\rm
tr}_{4k}(\hat\chi^\dagger\hat\chi\sfc^*\hat\chi)+\tfrac1{64}{\rm
tr}_k\,(\sfc\cdot\hat\chi)^2
\hat\chi\cdot\hat\chi-\tfrac1{32}{\rm
tr}_k\,(\sfc\cdot\hat\chi)^3\notag\\
&=\
-{1\over2^5}{\rm tr}_k\,
\big(\sfc\cdot\hat\chi\big)^3
+{3\over2^7}{\rm tr}_k\,\hat\chi\cdot\hat\chi\, \sfc\cdot
\hat\chi\, ,\elabel{ident2}\\
{\rm tr}_{4k}\,\big(\sfc^*\hat\chi\big)^4\ &=\ 
{\rm tr}_{4k}\big(
\tfrac1{64}\hat\chi^\dagger\hat\chi\hat\chi^\dagger\hat\chi-\tfrac1{32}
\hat\chi^\dagger\hat\chi(\sfc\cdot\hat\chi)\sfc^*\hat\chi
\notag\\ &\qquad\qquad\qquad\qquad\qquad
-\tfrac1{32}(\sfc\cdot\hat\chi)\sfc^*\hat\chi\hat\chi^\dagger\hat\chi
+\tfrac1{16}(\sfc\cdot\hat\chi)\sfc^*\hat\chi(\sfc\cdot\hat\chi)\sfc^*
\hat\chi\big)\notag\\
&=\
{1\over2^7}{\rm tr}_k\,\big(\sfc\cdot\hat\chi\big)^4-
{1\over2^7}{\rm tr}_k\,\big(\sfc\cdot\hat\chi\big)^2
\hat\chi\cdot\hat\chi
+{1\over2^{9}}{\rm tr}_k\,\big(\hat\chi\cdot\hat\chi\big)^2
-{1\over2^{10}}{\rm tr}_k\,\hat\chi_a\hat\chi_b
\hat\chi_a\hat\chi_b\, .\elabel{ident3}
\end{align}
\end{subequations}
As before, on the left-hand side of these formulae, $4k\times4k$
 matrix multiplication is implied, whereas on the right-hand side, all
 $SO(6)$ indices are saturated in standard 6-vector inner products,
 leaving the traces over $k\times k$ matrices.
%
%
%
%
%
%
%

{}From Eqs.~\eqref{E57}, \eqref{expdw}, \eqref{expdchi}, and
\eqref{ident1}-\eqref{ident3}, one obtains for the bosonic effective
action:
\begin{equation}
S_\rmb\ =\ S^{(2)}+S^{(3)}+S^{(4)}
\elabel{sumSb}
\end{equation}
where the quadratic, cubic and quartic actions are now given entirely
as $k$-dimensional (rather than $2k$ or $4k$-dimensional) traces:
\begin{subequations}
\begin{align}
S^{(2)}\ &=\ {\rm tr}_k\,\varphi^2\  ,\qquad 
\varphi=2\rho \sfc\cdot\hat\chi+{1\over2\rho^2}\delta W^0\, ,\elabel{S2def}
\\
S^{(3)}\ &=\ -{1\over12\rho^6}{\rm tr}_k\,(\delta W^0)^3+4\rho^3
{\rm tr}_k\,\sfc\cdot\hat\chi\,\hat\chi\cdot\hat\chi-{16\rho^3\over3}
{\rm tr}_k\,(\sfc\cdot\hat\chi)^3+{\rm tr}_k\,\delta
W^0\hat\chi\cdot\hat\chi
\notag\\
 &=\ 2\rho^2\,{\rm tr}_k\,\varphi\big(\hat\chi\cdot\hat\chi
-4(\sfc\cdot\hat\chi)^2\big)+\cdots\ ,\elabel{S3con}
\\
S^{(4)}\ &=\ 
-{1\over2\rho^4}{\rm
tr}_k\,\big[\hat a'_n,\hat a'_m\big]^2+
{1\over32\rho^8}{\rm
tr}_k(\delta W^0)^4-
{\rm tr}_k\,\big[\hat\chi_a,\hat a'_n\big]\big[\hat\chi_a
,\hat a'_n\big]
+8\rho^4{\rm tr}_k\,\big(\sfc\cdot\hat\chi\big)^4\notag\\ &\qquad\qquad\qquad-
8\rho^4{\rm tr}_k\,\big(\sfc\cdot\hat\chi\big)^2
\hat\chi\cdot\hat\chi
+2\rho^4{\rm tr}_k\,\big(\hat\chi\cdot\hat\chi\big)^2
-\rho^4{\rm tr}_k\,\hat\chi_a\hat\chi_b\hat\chi_a\hat\chi_b\notag
\\
 &=\ 
-{1\over2\rho^4}{\rm
tr}_k\,\big[\hat a'_n,\hat a'_m\big]^2
-8\rho^4{\rm tr}_k\,\big(\sfc\cdot\hat\chi\big)^2
\hat\chi\cdot\hat\chi
+2\rho^4{\rm tr}_k\,\big(\hat\chi\cdot\hat\chi\big)^2\notag\\
&\qquad\qquad\qquad+16\rho^4{\rm tr}_k\,\big(\sfc\cdot\hat\chi\big)^4-
\rho^4{\rm tr}_k\,\hat\chi_a\hat\chi_b\hat\chi_a\hat\chi_b
-{\rm tr}_k\,\big[\hat\chi_a,\hat a'_n\big]\big[\hat\chi_a
,\hat a'_n\big]+\cdots\ .\elabel{S4con}
\end{align}
\end{subequations}
Notice that only $k^2$ fluctuations, denoted $\varphi,$ are actually lifted at
quadratic order. This, in turn, implies that certain terms in
$S^{(3)}$ and $S^{(4)}$ are subleading, and can be omitted. Specifically,
the omitted terms in the final rewrites in Eqs.~\eqref{S3con}-\eqref{S4con}
contain, respectively, two or more, and one or more, factors of
the quadratically lifted $\varphi$ modes, and consequently are suppressed in
large $N$ (as a simple rescaling argument again confirms).

Now let us perform the elementary Gaussian integration over the
$\varphi$'s. Changing integration variables in Eq.~\eqref{E57.2} from
$d^{k^2}W^0$ to $d^{k^2}\varphi$ using Eq.~\eqref{S2def}, one finds:
\begin{equation}
\int d^{k^2}W^0\,e^{-N(S^{(2)}+S^{(3)})}\ =\ \Big({4\pi\rho^4\over
N}\Big)^{k^2/2}\, e^{-NS^{(4)\prime}}
\elabel{S4'def}
\end{equation}
where the new induced quartic coupling reads
\begin{equation}
S^{(4)\prime}\ =\ 
-\rho^4{\rm tr}_k\,
\big(\hat\chi\cdot\hat\chi-4(\sfc\cdot\hat\chi)^2\big)^2\ .
\elabel{S4'def1}
\end{equation}
Combining $S^{(4)\prime}$ with the original quartic coupling
\eqref{S4con} gives for the
effective bosonic small-fluctuations action:
\begin{equation} S_{ \rm b}=
-{1\over2}{\rm
tr}_k\left(\rho^{-4}[\hat a'_n,\hat a'_m]^2+2[\hat\chi_a,\hat a'_n]^2
+\rho^4[\hat\chi_a,\hat\chi_b]^2\right)\, .
\elabel{Sbanswer}
\end{equation}
Remarkably, all dependence on the unit vector $\sfc_a$ has canceled out.

Notice that apart from the absence of derivative terms, the expression
\eqref{Sbanswer} looks like a Yang-Mills field strength for
the gauge group $SU(k)$! We can make this explicit by introducing a
ten-dimensional vector field $A_\mu$,
\begin{equation}
A_\mu=N^{1/4}\big(\rho^{-1}\hat a'_n,\rho\hat\chi_a\big)\ ,\quad
\mu=0,\ldots,9\ ,
\elabel{amudef}\end{equation}
in terms of which
\begin{equation}
NS_\rmb(A_\mu)\ =\ -{1\over2}{\rm tr}_k\,\left[A_\mu,A_\nu\right]^2 \ .
\elabel{actymb}
\end{equation}
We recognize this as precisely the action of ten-dimensional $\SU(k)$
gauge theory, reduced to $0+0$ dimensions, i.e., with all derivatives
set to zero.

Now let us turn to the fermions. Since $\N=4$ supersymmetry in four
dimensions descends from $\N=1$ supersymmetry in ten dimensions, and
since all our saddle-point manipulations commute with supersymmetry,
we expect to find the $\N=1$ supersymmetric completion of the ten-dimensional
dimensionally-reduced action \eqref{actymb}, namely
\begin{equation}
NS_\rmf(A_\mu,\Psi)={\rm
tr}_k\,\bar\Psi\Gamma_\mu\left[A_\mu,\Psi\right]\ ,
\elabel{actymf}
\end{equation}
where $\Psi$ is a ten-dimensional Majorana-Weyl spinor, and
$\Gamma_\mu$ is an element of the ten-dimensional Clifford algebra. To
see how this comes about, we first separate out from the fermionic
collective coordinates the exact zero modes, in analogy
to Eqs.~\eqref{asplitdef}-\eqref{chisplitdef}: 
\begin{subequations}
\begin{align}
\M^{\prime A}_\alpha&=4\xi^A_\alpha1_{\sst
[k]\times[k]}+4\hat a'_{\alpha\aD}\bar\eta^{\aD A}+\hat\M^{\prime A}_\alpha\ ,\elabel{sosm}\\
\zeta^{\aD A}&=4\bar\eta^{\aD A}1_{\sst [k]\times[k]}+\hat\zeta^{\aD
A}\ .
\end{align}
\end{subequations}
Here  $\xi^A_\alpha$ and $\bar\eta^{\aD A}$ are the 
supersymmetric and superconformal fermion
modes \eqref{susymo}-\eqref{suconmo}.
Expanding the fermion coupling in the exponent of \eqref{E57.2} around the
special solution \eqref{specsol} and using the relations \eqref{E27}
 and \eqref{S2def}, we find 
\begin{equation}\begin{split}
 NS_{\rm f}=i
\Big({8\pi^2 N\over g^2}\Big)^{1/2}\,
{\rm tr}_k\Big[&
\big(\varphi
-2\rho(\sfc\cdot\hat\chi)\big)\rho\sfc_{AB}
\hat \zeta^{\aD A}\hat \zeta_\aD^B+\rho^{-1}\sfc_{AB}\big[\hat a'_{\alpha\aD},
\hat {\cal M}^{\prime\alpha A}\big]\hat \zeta^{\aD B}\\
&\qquad\qquad\qquad+\hat\chi_{AB}\big(\rho^2\hat 
\zeta^{\aD A}\hat \zeta_\aD^B+\hat {\cal M}^{\prime
\alpha A}\hat {\cal M}^{\prime B}_\alpha\big)\Big] \ .
\elabel{fermc}\end{split}\end{equation}
If we now define the
 $d=10$ Majorana-Weyl fermion field $\Psi$
\begin{equation}
\Psi=\sqrt{\pi\over2g}\,N^{1/8}\big(\rho^{-1/2}
\hat{\cal M}^{\prime A}_\alpha\,,\,
\rho^{1/2}\hat\zeta^{\aD A}\big)\ ,
\elabel{psidef}
\end{equation}
and the $\Gamma_\mu$ matrices according to 
Eq.~\eqref{rotca} below, we do in fact
recover the simple form \eqref{actymf}.  In moving from
Eq.~\eqref{fermc} to Eq.~\eqref{actymf} we have dropped the term
depending on $\varphi$; since $\varphi$ is a quadratically lifted bosonic
mode its contribution is suppressed in large $N$ compared to the other
couplings in Eq.~\eqref{fermc}, as a simple rescaling argument
confirms.  Note that unlike  the bosonic sector, the $\sfc_a$
dependence of the fermionic action does not actually disappear; as
discussed in the Appendix, it is simply subsumed into the representation
of the ten-dimensional Clifford algebra.

Finally our effective measure for the $k$ instantons has the form
\begin{equation}
\int\dmuphys\,e^{-\Skinst}\underset{N\rightarrow\infty}=
{\sqrt Ng^8e^{2\pi ik\tau}\over
k^32^{17k^2/2-k/2+25}\pi^{9k^2/2+9}
}\int\rho^{-5}d\rho\,d^4\com\, d^5\sfc\prod_{A=1,2,3,4}d^2\xi^A 
d^2\bar\eta^A\cdot\hat{\cal Z}_k\ ,
\elabel{hello}\end{equation}
where $\hat{\cal Z}_k$ is the partition function of an ${\cal N}=1$
supersymmetric $\SU(k)$ gauge theory in ten dimensions dimensionally reduced
to zero dimensions:
\begin{equation}\begin{split}
\hat{\cal Z}_k\ &=\ {1\over{\rm Vol}\,SU(k)}\int_{SU(k)}\, 
d^{10}A\, d^{16}\Psi\,e^{-S(A_\mu,\Psi)}\ ,\\
S(A_\mu,\Psi)\ &=\ N(S_{\rm b}+S_{\rm f})\ =\ 
-{1\over2}{\rm tr}_k\,\left[A_\mu,A_\nu\right]^2 
+{\rm
tr}_k\,\bar\Psi\Gamma_\mu\left[A_\mu,\Psi\right]\ .
\elabel{sukpart}
\end{split}
\end{equation}
Notice that the rest of the measure, up to numerical factors, is
independent of the instanton number $k$.
When integrating expressions which are independent of the $\SU(k)$
degrees-of-freedom, $\hat{\cal Z}_k$ is simply an overall constant
factor. A calculation of Ref.~\cite{MNS} revealed that $\hat{\cal
Z}_k$ is proportional to $\sum_{d|k}d^{-2}$, a sum over the positive integer
divisors $d$ of $k$. However, the constant of proportionality was
fixed definitively in Ref.~\cite{KNS} to give\footnote{In comparing to
Ref.~\cite{KNS}, it is important to note that our convention
for the normalization of the generators is  ${\rm tr}_k\hat
T^r\hat T^s=\delta^{rs}$, rather than $\tfrac12\delta^{rs}$ in
 Ref.~\cite{KNS}.}
\begin{equation}
\hat{\cal Z}_k=2^{17k^2/2-k/2-8}\pi^{9k^2/2-9/2}k^{-1/2}
\sum_{d\vert k}{1\over d^2}\ .
\elabel{parte}\end{equation}
In evaluating
$\hat{\cal Z}_k$, we have used 
\begin{equation}
{\rm Vol}\,\big(\SU(k)\big)\ 
=\ {2^{k-1}\pi^{k(k+1)/2-1}\over\prod_{i=1}^{k-1}i!}\ .
\end{equation}

In summary, on gauge invariant and $\SU(k)$ singlet operators, our
effective large-$N$ collective coordinate measure has the following
simple form:
\begin{equation}
\int\dmuphys\,e^{-\Skinst}\ \underset{N\rightarrow\infty}{=}\ {\sqrt Ng^8\over
2^{33}\pi^{27/2}}\,k^{-7/2}e^{2\pi ik
\tau}\sum_{d\vert k}{1\over d^2}
\int\,
{d^4\com\,d\rho\over\rho^5}\, d^5\sfc\prod_{A=1,2,3,4}d^2\xi^A 
d^2\bar\eta^A\ .
\elabel{endexp}
\end{equation}
As expected from the AdS/CFT correspondence (albeit
counter-intuitive from the point of view of the field theory), only
one copy of the $AdS_5\times S^5$ moduli space appears at any $k$.

\subsection{Comments on the ten-dimensional $SU(k)$ partition function}

It is worthwhile making some remarks about the $SU(k)$ partition
function $\hat{\cal Z}_k$ defined in Eq.~\eqref{sukpart}. To begin
with, it may surprise the reader that this integral even exists!  The
following na\"\i ve argument suggests that it diverges \cite{KNS}:
consider the basis where $A_1$ (say) is diagonal; then fluctuations in
the directions where the other $A_\mu$ are also diagonal are
un-suppressed.  However, this argument does not take into account that
for sufficiently large $k$ and/or $D$ ($D$ being the dimension of
space-time), these fluctuations are of measure zero. The question of
convergence has only recently been definitively settled in
Refs.~\cite{KNS}. For the purely bosonic version of the integral, it
converges in the cases $\{D=3,k\ge4\}$, $\{D=4,k\ge3\}$ and
$\{D\ge5,k\ge2\}$, and diverges otherwise. In the supersymmetric
version the convergence is even better; in fact no divergent cases are
believed to exist.

Second, we invite the reader to compare our gauge invariant measure
\eqref{hello} with the measure for D-instanton in flat space, the
$U(k)$ partition function \eqref{ukpfn}. Clearly the $U(1)$ part of
this partition function, describing the center-of-mass coordinates of
the D-instanton configuration has been generalized to the
(supersymmetrized) volume measure on $AdS_5\times S^5$, as in
\eqref{int}; however, the $SU(k)$ part of the partition function is
identical in both cases. Hence, we find a very attractive matching of
large-$N$ Yang-Mills instantons with string theory D-instantons. In
particular, on the D-instanton side, the intuition of \cite{GG1,GG2,GG3,GG4},
suggests that, as far as the contribution to the correlation functions
$G_n$, the D-instantons contribution can be thought of as being due to
a charge $k$ D-instanton bound state, is matched by our large-$N$
instanton analysis and our identification of the moduli space as one
copy of $AdS_5\times S^5$. 

Third, it is interesting that the large-$N$ description of $k$
Yang-Mills instantons, in an ${\cal N}=4$ supersymmetric $SU(N)$ gauge
theory, is described by a $SU(k)$ gauge theory. This kind of
``duality'' between gauge group and ADHM auxiliary group has been
noted previously in a string theory context \cite{W1,D1,D2}, as is
apparent from our discussion in Sec.~IV.2. For the other classical
groups, it is a well-known feature of the ADHM construction that 
the duality exchanges groups, i.e. $O(N)\leftrightarrow Sp(k)$ and
$Sp(N)\leftrightarrow O(k)$ \cite{OSB}, although it turns out that in each
of these cases large-$N$ instantons are described by a unitary gauge theory.

Finally, we should comment on the fact that $\hat{\cal Z}_k$ is
proportional to the non-integer expression $\sum_{d|k}\,d^{-2},$
rather than an integer as would normally be expected from the
Gauss-Bonnet-Chern (GBC) theorem. The normal expectation goes as follows.
We start by recalling  that the coefficient of the four-fermion term in the 
instanton action can be identified with the 
Riemann tensor of the instanton moduli 
space; this is familiar from supersymmetric quantum mechanics
\cite{AG} and the analysis of the 
three-dimensional theory with sixteen supercharges given in \cite{DKM3D}. 
To formally evaluate the $k$-instanton contribution to a correlator
such as $G_{16},$
one must bring down powers of the quadrilinear term to saturate 
the extra Grassmann integrations which are left over after the 
explicit fermion insertions are accounted for. The resulting bosonic integrand 
involves a power of the Riemann tensor with various indices 
contracted and is just the Gaussian curvature of the moduli space. 
The relevant moduli space here is 
actually a relative moduli space
where eight center-of-mass degrees of freedom, which are the superpartners 
of the sixteen exact zero modes, have been modded out. 
This space, which we will denote $\tilde{\cal M}_{k,N}$  
is obtained by taking a quotient of the full moduli space 
of $k$ ADHM instantons for gauge group $U(N)$, by translations, 
dilatations and global $SU(2)$ gauge rotations.

If the instanton moduli space were 
both compact and smooth the integral of the Gaussian curvature 
would simply be equal to 
its Euler character by the GBC theorem. 
In this well-behaved case the Euler character is 
also equal to the index of the Laplacian, or in physical terms, the Witten 
index of supersymmetric quantum mechanics defined 
on the manifold in question.  
In reality the ADHM moduli space is non-compact and 
has singular points where instantons coincide or shrink to zero size. 
The three-dimensional case considered in \cite{DKM3D} was somewhat better 
behaved as the moduli space of three-dimensional instantons 
(which are BPS monopoles), while still non-compact, is smooth.  
In these cases, one can still formally define a Witten index which now 
counts the number of {\em normalizable} zero energy states (weighted 
by fermion number). However, this index is in general no longer 
equal to the GBC integral which arises in the instanton 
calculation. Instead, the GBC integral is what is known as the principal 
or bulk contribution to the index which differs from the index itself by 
a surface term. 

Taking various normalizations into account, the 
result of Sec.~V.3 can be interpreted as saying that the 
principal contribution to the generalized index 
of supersymmetric quantum mechanics on 
$\tilde{\cal M}_{k,N}$ is equal to $\sum_{d|k}\,d^{-2}$ in the large-$N$ 
limit. 
Interestingly, the principal contribution to the corresponding index 
in M(atrix) quantum mechanics, which is actually just the $N=0$ case of the 
configuration considered in this section, is also known to be equal to 
$\sum_{d|k}\,d^{-2}$ \cite{MNS}. 
In both these cases 
the surface terms must certainly be non-zero as the principal 
contribution is fractional. This is to be contrasted with the 
three-dimensional example discussed in \cite{DKM3D},
where it was argued that the corresponding 
surface terms vanish (this vanishing was demonstrated explicitly 
in the two-instanton case). 
Finally, we note that the occurrence of integrals 
which formally give the Euler character of instanton moduli space 
suggests an interesting connection with the partition function of 
the twisted ${\cal N}=4$ theory evaluated in \cite{VW}.     

\rsen\section{Large-$N$ Instanton Correlation Functions}

Finally, we can use our gauge-invariant large-$N$ measure to calculate
correlation functions $\langle{\cal O}_1(x_1)\cdots{\cal
O}_n(x_n)\rangle$ of gauge-invariant composite operators.  Before we
consider specific $n$-point functions, let us make the following general
comments. At leading order in $N$, we can replace each operator
insertion with its classical $k$-instanton saddle-point value. Since the
saddle-point solution of Sec.~V.3 is relatively simple, this
observation greatly streamlines the form of the operator
insertions. We will restrict our attention to operators $\CO(x)$
consisting of a single trace on the gauge group index of a product
of adjoint scalars, fermions and field strengths. Each of these three adjoint
quantities is of the type $\Ubar\BX U$ where $\BX$ is some
matrix of ADHM variables; consequently $\CO$ has the generic form
\begin{equation}
{\cal O}(x)={\rm tr}_N\big[\bar U\BX_1U\bar U\BX_2U\cdots\bar U\BX_pU\big]=
{\rm tr}_{N+2k}\,\big[\P\BX_1\P\BX_2\P \cdots\P\BX_p\big]\ ,
\elabel{insop}\end{equation}
where $\P=U\bar U$ is the projection operator  \eqref{cmpl}.
It is easily checked that
at the saddle-point, the bosonic ADHM quantities $f$, $a'_n$, $\bigL$
and $\P$ collapse to
\begin{subequations}
\begin{align}
f&\rightarrow{}{1\over y^2+\rho^2}1_{\sst[k]\times[k]}\ , \quad
a'_n\rightarrow{}-X_n1_{\sst[k]\times[k]}\  ,\quad
\BL\rightarrow{}2\rho^2\ ,\elabel{spff}
\\
\P&\rightarrow{}1_{\sst[N+2k]\times[N+2k]}-{1\over y^2+\rho^2}\begin{pmatrix}
w_\aD\bar w^\aD & w_\aD y^{\aD\alpha} \\ y_{\alpha\aD}\bar w^\aD & y^2
1_{\sst[2k]\times[2k]} \end{pmatrix}\ ,\elabel{spuub}
\end{align}\end{subequations}
where $y=x-X$ as before. Deviations from these saddle-point values 
are suppressed by powers of $N^{-1/4}$ as follows from the rescaling
\eqref{amudef}.

The analogous replacement prescription for the fermionic ADHM
quantities is, in general, somewhat trickier. While these, too, are
amenable to a saddle-point analysis as described in Sec.~VI.2 below,
they are also subject to the stringent selection rules of Grassmann
integration.  In Sec.~VI.1 we will focus on the relatively
straightforward correlators $G_{16},$ $G_8$ and $G_4$ for which a
large-$N$ evaluation of the fermionic quantities is actually moot;
this is because all the Grassmann variables in the insertions must be
replaced by the sixteen exact supersymmetric and superconformal zero
modes in order to obtain a nonzero result. In contrast, in Sec.~VI.2
we will discuss a tower of higher partial-wave operators in which, in
addition to these sixteen exact modes, there are left-over Grassmann
collective coordinates which must be carefully analyzed in large $N$.

\subsection{Multi-instanton contributions to the correlators $G_n$}

In this subsection we analyze the three gauge-invariant chiral correlators
$G_n$, $n=16,$ 8 or 4, defined by \cite{BGKR}:
\begin{subequations}\begin{align}
G_{16}\ &=\
\langle\,\Lambda_{\alpha_1}^{1}(x_1)\cdots\Lambda_{\alpha_{16}}^{4}
(x_{16})\rangle\ ,\quad
 \Lambda_\alpha^A\ =\ g^{-2}\sigma^{mn}{}_\alpha^{\
\beta}\,{\rm tr}_N\,
v_{mn}\,\lambda_\beta^A\ ,
\elabel{e4}
\\
G_{8}\ &=\ \langle\,{\cal E}^{A_1B_1}(x_1)\cdots {\cal E}^{A_{8}B_{8}}
(x_{8})\rangle\ ,\quad
 {\cal E}^{AB} \ =\ g^{-2}\,{\rm tr}_N\, \big(\lambda^A \lambda^B +
t^{(AB)+}_{[abc]} A^a A^b A^c
 \big)\ ,
\elabel{e6}
\\
G_{4}\ &=\ \langle\,{\cal Q}^{a_1b_1}(x_1)\cdots {\cal Q}^{a_4b_4}
(x_{4})\rangle\ ,\quad
{\cal Q}^{ab} \ 
=\ g^{-2}\,{\rm tr}_N\,
\big(A^a A^b - \tfrac16\delta^{ab}A^c A^c \big)
\ ,
\elabel{e5}
\end{align}\end{subequations}
where $t$ in Eq.~\eqref{e6} is a numerical tensor.
We focus first on $G_{16}$, which we previously addressed in Secs.~I.2-I.3. 
 The component fields that make up the
composite operator $\Lambda^A_\alpha$ have the saddle-point form
\begin{equation}
v_{mn}(x)\ \rightarrow\ {4\over y^2+\rho^2}\,\bar U\cdot\begin{pmatrix}
0_{\sst[N]\times[N]} & 0_{\sst[N]\times[2k]}\\ 
0_{\sst[2k]\times[N]} & \sigma_{mn\beta}^{\phantom{mn}\
\gamma}1_{\sst[k]\times[k]} 
\end{pmatrix}\cdot U
\elabel{lofs}\end{equation} 
and\footnote{Here, and in the following,
the Weyl indices $\beta$ and $\gamma$ are contracted with $\bar U$ and
$U$, respectively.}
\begin{equation}
\lambda^A_\alpha(x)\ \rightarrow\ 
{1\over y^2+\rho^2}\,\bar U\cdot\begin{pmatrix}
0_{\sst[N]\times[N]} & 
4w_\aD\bar\eta^{\aD A}\delta_\alpha^{\ \gamma}\\
4\epsilon_{\alpha\beta}\bar\eta_\aD^A\bar w^\aD &
4\xi^A_\beta\delta_\alpha^{\ \gamma}1_{\sst[k]\times[k]}
-4\epsilon_{\alpha\beta}\xi^{\gamma A}1_{\sst[k]\times[k]}
\end{pmatrix}\cdot U\ +\ \cdots\   \elabel{logaugino}\end{equation}
as follows from Eqs.~\eqref{sdu} and \eqref{lam}.
Here, as always, the Grassmann parameters $\xi^A_\alpha$ and
$\etabar^{\aD A}$ measure the
strength of the sixteen exact supersymmetric and superconformal
zero modes; for a nonzero result, each of the sixteen $\Lambda^A_\alpha$ insertions must
saturate a distinct such mode. The omitted terms in
Eq.~\eqref{logaugino} stand for the remaining, lifted, fermion modes
which can therefore be dropped. Putting Eqs.~\eqref{logaugino}, \eqref{lofs} and
\eqref{spuub} together, gives us the leading order behavior of the
composite operator $\Lambda^A_\alpha\,$:
\begin{equation}
\Lambda^A_\alpha(x)=-{96k\over
g^2}\big(\xi^{A}_\alpha-\sigma^n_{\alpha\aD}\etabar^{\aD A}
\cdot(x-\com)_n\big){\rho^4\over(\rho^2+(x-X)^2)^4}\ ,\elabel{lamins}
\end{equation}
where the other fermion modes are neglected. Recalling the definition
\eqref{prop} of the supergravity propagator
$K_\Delta$, we see that this is precisely the
one-instanton expression \eqref{insertid} apart from the overall
factor of $k$, which comes trivially from tracing over a $k\times k$
unit matrix:
\begin{equation}
{\Lambda}^A_\alpha\Big\vert_{k\hbox{-}\text{inst}}\ =\
k\cdot{\Lambda}^A_\alpha\Big\vert_{1\hbox{-}\text{inst}}\ .
\elabel{trivially}
\end{equation}
Thus the sixteen insertions account for a factor $k^{16}$ relative to
the one-instanton calculation reviewed in Sec.~I.3.

Thanks to Eqs.~\eqref{endexp} and \eqref{lamins}, 
our final answer for the large-$N$ $k$-instanton contribution to $G_{16}$, from
the Yang-Mills side of the correspondence, therefore reads:
\begin{equation}\begin{split}
G_{16}(x_1,\ldots,x_{16})\,{\Big|}_{k\hbox{-}\rm inst}\ &=\
\langle\Lambda_{\alpha_1}^1(x_1)\cdots
\Lambda_{\alpha_{16}}^4(x_{16})\rangle
\,{\Big|}_{k\hbox{-}\rm inst}\\
&=\,{96^{16}\sqrt N\over2^{33}\pi^{27/2}g^{24}}k^{25/2}e^{2\pi
ik\tau}\sum_{d\vert k}{1\over d^2}
\int {d^4\com d\rho\over\rho^5}\,\prod_{A=1,2,3,4}d^2\xi^A\,d^2\etabar^A\,\\
&\times\ \big(\xi^1_{\alpha_1}-\sigma^n_{\alpha_1\aD}\etabar^{\aD 1}
\cdot(x_1-\com)_n\big)K_4(\com,\rho;x_1,0)
\\
\times\cdots&\times\ 
\big(\xi^4_{\alpha_{16}}-\sigma^n_{\alpha_{16}\aD}\etabar^{\aD 4}
\cdot(x_{16}-\com)_n\big)K_4(\com,\rho;x_{16},0)
\elabel{finalans}\end{split}
\end{equation}
To leading semi-classical order, this is in perfect agreement with the corresponding supergravity
expression defined by Eqs.~\eqref{foksh}-\eqref{grnfcn}. 
In particular, note that the factor of
$k^{16}$ from Eq.~\eqref{trivially}, taken together with the factors
of $k$ in Eq.~\eqref{endexp},  precisely  reproduces the leading
semiclassical approximation \eqref{foksh} to the modular form
$f_{16}(\tau,\bar\tau).$ Moreover, thanks to the proportionality
\eqref{trivially}, we have finally answered the puzzle posed in
Sec.~I.3; namely, we have seen how the $k$-instanton field strength
can look like a supergravity propagator for $k>1.$ For this to happen,
the dominant  $k$-instanton saddle-point configuration needed to look 
like $k$ instantons not only living
 in mutually commuting $SU(2)$'s, but also sharing
a common size and 4-position, precisely as we found in Sec.~V.3.

Next we consider $G_8$ and $G_4.$ 
In order to calculate the corresponding insertions ${\cal E}^{AB}$ and ${\cal
 Q}^{ab}$ one needs to utilize the ADHM expression for the scalar
 fields $A^{AB}(x)$ as given in Sec.~II.7.
For present purposes we need to saturate only
the supersymmetric and the superconformal fermion zero modes by the
insertions. Neglecting the other modes, one easily finds:
\begin{equation} iA^{AB} \ \rightarrow\ \sqrt{2}
\big(\xi^{\alpha A}-\etabar_{\aD}^A \sigmabar_k^{\aD \alpha} \cdot
(x-\com)^k\big) \sigma^{mn}{}_\alpha^{\
\beta}\big(\xi_\beta^B-\etabar^{\bD B}\sigma^l_{\beta\bD} \cdot
(x-\com)_l\big) \, v_{mn}\ , \elabel{elast}\end{equation} where
$v_{mn},$ in turn, is replaced by the saddle-point expression
\eqref{lofs}.  The computation of $G_4$ and $G_8$ proceeds similarly
to the computation of $G_{16}$. Note that only the
$\lambda^A\lambda^B$ term in ${\cal E}^{AB}$ is needed in the leading
semiclassical evaluation of $G_8$, as the $A^aA^bA^c$ term is
suppressed by a power of the coupling $g$ (as a careful rescaling of
the component fields confirms).  In particular, the $k$-instanton
insertions are again trivially related to their one-instanton
counterparts:
\begin{equation}
{\cal E}\Big\vert_{k\hbox{-}\text{inst}}\ =\
k\cdot{\cal E}\Big\vert_{1\hbox{-}\text{inst}}
\ ,\qquad
{\cal Q}\Big\vert_{k\hbox{-}\text{inst}}\ =\
k\cdot{\cal Q}\Big\vert_{1\hbox{-}\text{inst}}\ .
\elabel{trivially1}
\end{equation}
 Once again, thanks to these proportionality relations, the space-time
 dependence of the correlators at the $k$-instanton level will be
 identical to the one-instanton dependence. Furthermore, the relations
 \eqref{trivially1} account for a factor of $k^8$ and $k^4$ in $G_8$
 and $G_4,$ which conspires with the expression \eqref{endexp} to
 produce the leading semiclassical approximation \eqref{foksh} to the
 expected modular forms $f_8(\tau,\bar\tau)$ and $f_4(\tau,\bar\tau)$,
 respectively, as in Eq.~\eqref{yetae}.  It follows that, just as with
 $G_{16}$, the large-$N$ Yang-Mills results for $G_8$ and $G_4$
are in perfect
 agreement with the supergravity expectations \cite{BGKR}, order by
 order in the instanton expansion. For $G_4$, there is an identical
contribution from anti-instantons at the same order in $g^2$, as noted
previously.

\subsection{Kaluza-Klein modes on $S^5$ and Yang-Mills correlators}

Next let us consider the general problem of $n$-point functions in which the
insertions contain more than sixteen Grassmann modes. Of course, sixteen
of these must be used to saturate the exact supersymmetric and superconformal
modes, as in Sec.~VI.1. The question becomes how, in large $N$, to select
the remaining, lifted, fermion zero modes.

To answer this question, we need to analyze the effective $N$
dependence of these modes.  In analogy with the bosonic quantities
\eqref{spff}-\eqref{spuub}, it is useful to think of the unbroken
$\xi^A$ and $\etabar^A$ modes themselves as arising from a
saddle-point evaluation:
\begin{equation}
\M_\alpha^{\prime A}\rightarrow4\xi_\alpha^A1_{\sst[k]\times[k]}\, ,\qquad
\zeta^{\aD A}\rightarrow4\bar\eta^{\aD A}1_{\sst[k]\times[k]}\ .
\elabel{spvamz}
\end{equation}
Indeed, the rescaling \eqref{psidef} implies that the remaining,
 fluctuating, modes in $\M^{\prime A}$ and $\zeta^A$ (which were
 denoted $\hat\M^{\prime A}$ and $\hat\zeta^A$ in Sec.~V.3) are
 subleading compared to $\xi^A$ and $\etabar^A$ by a factor of
 $N^{-1/8}.$ There remain the modes $\nu^A$ and $\bar\nu^A$, which are
 distinct from the others in that they carry an $SU(N)$ index
 $u$. From their coupling to the $\chi_a$ field in
 Eq.~\eqref{intoutnu}, together with the rescaling \eqref{E56}, one
 sees that each $\bar\nu^A\nu^B$ pair in an insertion, for a fixed,
 unsummed value of the index $u$, costs a factor of $N^{1/2}$;
 however, summing on $u$ (as required by gauge invariance) then turns
 this $N^{1/2}$ suppression into an $N^{1/2}$ enhancement. In other
 words, $\nu^A$ and $\bar\nu^A$ factors in the insertions should each
 be thought of as being enhanced by $N^{1/4}$.  The large-$N$ rule of
 thumb for choosing fermionic collective coordinates
 in gauge-invariant correlators is now clear:
 as many of the modes as possible should be $\nu^A$ and $\bar\nu^A$
 modes, subject to sixteen of the modes being $\xi^A$ and $\etabar^A$
 modes to saturate these Grassmann integrations.

With this rule of thumb in hand, let us examine a specific set of
correlators of special interest to the AdS/CFT correspondence.
%
%
In Sec.~VI.1, we were concerned with correlation functions
of operators which on the supergravity side involved no dependence on
the position on $S^5$. In this subsection, we will show that the
dependence of operators on the position on $S^5$ is encoded on the
Yang-Mills side in the dependence of the operators on
the variables $\nu^A$ and $\bar\nu^A$. As just alluded to,
our discussion will further
elucidate the mysterious r\^ole played by the auxiliary variables $\chi_a$.
In general, operators depend upon the Grassmann 
variables $\nu^A$ and $\bar\nu^A$
which (we have seen) dominate in large $N$;
however, gauge invariant operators can only depend on the gauge
invariant $k\times k$ matrix combinations $\bar\nu^A\nu^B$. We now
prove that, to leading order in $N$, such a combination, in the final
integral with respect to the measure \eqref{endexp} is replaced by
\begin{equation}
\bar\nu^A\nu^B\rightarrow {\sqrt2g\rho N^{1/2}\over\pi
i}\epsilon^{ABCD}\sfc_{CD}
1_{\sst[k]\times[k]}\ ,
\elabel{replace}\end{equation}
where the $N^{1/2}$ dependence has already been noted.
To this end, consider a general insertion with a string of such
combinations $\bar\nu^{A_1}\nu^{B_1}\otimes\cdots\otimes\bar\nu^{A_p}\nu^{B_p}$. 
We must insert this expression into the measure {\it before\/} the
$\nu$ integrals have been performed, i.e.~into
\eqref{dmudef}. Performing the $\nu$ integrals as in \eqref{intoutnu}
in the presence of the insertion leads to a modified expression
involving factors of $\chi^{-1}$ which
can be derived by considering
\begin{equation}\begin{split}
&{\partial\over\partial x^T_{A_1B_1}}\otimes
\cdots\otimes{\partial\over\partial x^T_{A_pB_p}}
\left({\rm det}_{4k}x\right)^{N-2k}\Big\vert_{x=\chi}\\
&\qquad\qquad\qquad=N^p\left({\rm det}_{ 4k}\chi\right)^{N-2k}(\chi^{-1})_{B_1A_1}\otimes
\cdots\otimes(\chi^{-1})_{B_pA_p}+\cdots\ ,
\end{split}\end{equation}
where $T$ acts on instanton indices and 
the ellipsis represent terms of lower order in $N$. This shows
that, after performing the $\nu$ integrals, a term of the form $\bar\nu^A\nu^B$ is
replaced by $(gN/\sqrt8\pi i)(\chi^{-1})_{BA}$, to leading order. Now we rescale
$\chi$ as in \eqref{E56}, and replace it with its saddle-point
value \eqref{specsol} to give \eqref{replace}.
The replacement shows that dependence on the position
on $S^5$ is associated on the Yang-Mills side as dependence on
combination $\bar\nu^A\nu^B$, where we recall from Sec.~IV.3 that the
$\{\nu^A,\bar\nu^A\}$ are superpartners of the iso-orientation of the multi-instanton.

We now consider, a class of operators which depend on the $\nu^A$ and
$\bar\nu^B$ in an essential way.
On the supergravity side, for simplicity, we will restrict our attention to a 
single supergravity field, the dilaton, 
$\phi(x,\rho, \sfc)$. 
Although the dilaton is a single massless scalar field in ten dimensions, 
it yields an infinite tower of massive fields in five dimensions via 
Kaluza-Klein reduction on $S^5$. This dimensional reduction amounts 
to expanding $\phi$ in partial waves as, 
\begin{equation}
\phi(x,\rho,\sfc)=\sum_{p=0}^{\infty} 
\phi^{(p)}_{a_{1}a_{2}\cdots a_{p}}(x,\rho) Y^{(p)}_{a_{1}a_{2}
\cdots a_{p}}(\sfc)\, ,
\elabel{partial}
\end{equation}
where the scalar spherical harmonics for $S^{5}$ are defined in terms of the 
unit vector $\sfc_{a}$ as,   
\begin{equation}  
Y^{(p)}_{a_{1}a_{2}
\ldots a_{p}}(\sfc)=
\sfc_{a_{1}}\sfc_{a_{2}}\cdots \sfc_{a_{p}}\,-\, {\rm Traces}\ .       
\elabel{spherical}
\end{equation}
The field $\phi^{(n)}_{a_{1}a_{2}\ldots a_{p}}$ has five dimensional mass 
$m^{2}=p(p+4)L^{-2}$ and transforms in the $p^{\rm th}$ symmetric traceless tensor 
representation of $SO(6)$. 

The AdS/CFT correspondence suggests that each of these 
Kaluza-Klein modes is associated with an operator 
${\cal O}^{(p)}_{a_{1}a_{2}\cdots a_{p}}$ in ${\cal N}=4$
supersymmetric Yang-Mills theory which 
transforms in the same irreducible representation of $SO(6)_{R}$ and has
scaling dimension $\Delta=p+4$. 
The required operators were identified in \cite{WIT150,GKP} as, 
\begin{equation}
{\cal O}^{(p)}_{a_{1}a_{2}\cdots a_{p}}(x)= {\rm tr}_N\,\big[
(RA)_{a_{1}}(RA)_{a_{2}}\cdots (RA)_{a_{p}} v_{mn}^{2} \big] \, -\,{\rm
Traces}\ .
\elabel{on}
\end{equation}  
Here, $R$ is an $SO(6)$ rotation such that $\sfc=R^{-1}\sfc^{(0)}$
where $\sfc^{(0)}$ is a reference point (e.g.~the north pole).
The explicit relation \eqref{corr} between gauge theory and Type IIB 
supergravity correlators suggests \cite{BGKR} that we should compare the 
Yang-Mills instanton contributions to the gauge theory correlators, 
$\langle {\cal O}^{(p_{1})}(x_{1})
\cdots {\cal O}^{(p_{r})}(x_{r}){\cal O}_{\cal F}\rangle$ with the D-instanton 
contributions to the following boundary correlation functions on the 
supergravity side of the correspondence, 
\begin{equation}
G_{p_{1}p_{2}\ldots p_{r}} = \left\langle \phi^{(p_{1})}(x_{1},\rho=0,
\sfc^{(1)}) 
\cdots\phi^{(p_{r})}(x_{r},\rho=0,\sfc^{(r)})
{\cal F}\right\rangle\, .
\elabel{btob} 
\end{equation}
Here, ${\cal O}_{\cal F}$ represents an additional fermionic operator insertion
which is present to saturate the exact sixteen
zero modes of the multi-instanton and ${\cal F}$ are the corresponding 
insertions in the supergravity correlator. Another requirement for these
latter 
operator insertions is that they do not contain $\nu^A$ or $\nubar^A$ modes
(otherwise the complicated issue of cross terms with the other insertions
arises). 
For instance, ${\cal O}_{\cal F}$
could be a product of sixteen composite operators $\Lambda^A_\alpha$
considered in Sec.~VI.1.\footnote{It is trivial to see that $\Lambda^A_\alpha$ does
not depend on $\nu^A$ or $\bar\nu^A$: since it is a gauge-invariant operator
that is linear in the Grassmann parameters, it could only depend on the
inner products $\wbar\nu^A$ or $\bar\nu^A w$, but these vanish by definition;
see Eq.~\eqref{E45}.}
The appropriate bulk-to-boundary propagator for  a  
field which transforms in a non-trivial representation $SO(6)$, and has
scaling dimension $\Delta=p+4$, is that which we would expect for an
$SO(6)$ singlet operator, as in \eqref{prop}, augmented by the appropriate 
spherical harmonic for the representation in question. Hence the 
external leg corresponding to the field insertion 
$\phi^{(p)}_{a_{1}a_{2}\cdots a_{p}}(x,\rho=0, \sfc^{(p)})$ at the
boundary, with the interior point $(X,\rho,\sfc)$,
comes with the propagator 
\begin{equation}
K_{p+4}(X,\rho; x,0)Y^{(p)}_{a_{1}a_{2}
\cdots a_{p}}(R^{(p)}\sfc)\, ,
\elabel{prop2}
\end{equation}  
where we have defined the $SO(6)$ rotation $R^{(p)}$ such that
$\sfc^{(p)}=(R^{(p)})^{-1}\sfc^{(0)}$.
As before, we can now identify the degrees-of-freedom of a multi-instanton at 
large $N$, with those describing a D-instanton on $AdS_{5}\times
S^{5}$. Since, we have already demonstrated the equality between the
measures, it remains only to 
demonstrate that replacing the operators ${\cal O}^{(p)}$ by their classical 
values in the multi-instanton background at large $N$ is equivalent to the D-instanton 
``Feynman rule'' \eqref{prop2} for the dual operators $\phi^{(p)}$. 

Taking into account that
the supersymmetric and superconformal modes are
saturated by the unspecified insertion ${\cal O}_{\cal F}$, let us set
these variables to zero in the scalar field. Consequently, at leading order,
\begin{equation}
iA^{AB}=\bar U\cdot \begin{pmatrix}
{1\over2\sqrt2(y^2+\rho^2)}(\nu^A\bar\nu^B-\nu^B\bar\nu^A) & 0_{\sst[N]\times[2k]}\\
0_{\sst[2k]\times[N]}& {1\over4\sqrt2\rho^2}(\bar\nu^A\nu^B-\bar\nu^B\nu^A)\end{pmatrix}\cdot U\ ,
\elabel{losf}\end{equation}
where we have made use of the large-$N$ behavior of $\BL$ in \eqref{spff}.
We can now find the appropriate form for the operator \eqref{on} 
by taking the trace of a string of scalar fields and
$v_{mn}^2$. Notice that in the replacement \eqref{replace}, the
combination $\bar\nu^A\nu^B$ is ${\cal O}(N^{1/2})$, hence the leading
order term in the operator comes from the product of $p$
bottom-right-hand corners of \eqref{losf}. Note that the product of $p$
top-left-hand corners of \eqref{losf}, which potentially gives a
contribution of the same order, is canceled against the 0 in the
top-left-hand corner of the ADHM matrix in \eqref{lofs}.
So at leading order
\begin{equation}
{\cal O}^{(p)}_{a_{1}\cdots a_{p}}=
-{48\rho^4\over(2i)^p(y^2+\rho^2)^{p+4}}
{\rm tr}_k
\big[(R^{(p)}\BL^{-1}\cdot\hat\Lambda)_{a_1}\cdots(R^{(p)}
\BL^{-1}\cdot\hat\Lambda)_{a_p}\,-\, {\rm Traces}\big]\ ,
\end{equation}
where $\hat\Lambda^{AB}$ was defined in \eqref{E51} and, as before,
\begin{equation}
\hat\Lambda^{AB}={1\over\sqrt8}\bar\Sigma^{AB}_a\hat\Lambda_a\ .
\end{equation}
We can now make the replacement \eqref{replace}, which can be written
in the present context as
\begin{equation}
\hat\Lambda_a\rightarrow {2ig\rho\sqrt N\over\pi}\sfc_a1_{\sst[k]\times[k]}\ ,
\end{equation}
to give finally
\begin{equation}\begin{split}
{\cal O}^{(p)}_{a_1\cdots a_p}(x)\ &=\ -48k\left({g\sqrt N\over\pi}\right)^p
\cdot{\rho^{p+4}\over(y^2+\rho^2)^{p+4}}\big[(R^{(p)}\sfc)_{a_1}
\cdots(R^{(p)}\sfc)_{a_p}\,-\, {\rm Traces}\big]\\
&\qquad\qquad\qquad\qquad\qquad\propto K_{p+4}(X,\rho; x,0)Y^{(p)}_{a_1\cdots a_p}(R^{(p)}\sfc)\, .
\elabel{on3}\end{split}
\end{equation}      
Thus, at leading order in $N$, we have reproduced the ``Feynman rule'' 
\eqref{prop2} for the equivalent operator insertion on the 
the supergravity side of the correspondence.

\rsen\section{Comments and Conclusions}

\subsection{Living large in a large-$N$ world}

Prior to the recent applications to supersymmetric theories, the
multi-instanton program has had a checkered history (see Sec.~I of
Ref.~\cite{OSB}). On the positive side, the most noteworthy technical
achievements include the ADHM construction of the full space of
$k$-instanton solutions \cite{ADHM} as well as the fermion zero modes
and propagators
in the ADHM background \cite{CGTone}. On the negative side, the ADHM
construction has the major drawback that the collective coordinate
matrix $a$ is highly overcomplete, and must be accompanied by the
nonlinear constraints \eqref{fconea} which are only explicitly
resolvable for $k\le3$ \cite{CWS,KS}.  A closely related problem was
the lack of a collective coordinate measure beyond $k=2$ \cite{OSB}.
Largely as a consequence of these difficulties, the ADHM construction
languished for many years as a largely mathematical achievement with
virtually no practical application to quantum 
field theory.  Instead, much of the
multi-instanton work in the physics community focused on the dilute
instanton gas regime where single-instanton methods suffice and these
obstructions do not appear. Unfortunately---by any measure---successful
phenomenological applications to QCD require an instanton density much
greater than this regime allows, as well as considerations of configurations of instantons
and anti-instantons \cite{SSI}.

The addition of supersymmetry improves this situation
substantially. On the one hand, as reviewed in Sec.~IV.1, the fact
that the bosonic and fermionic small-fluctuations determinants cancel,
together with the stringent requirement of supersymmetric invariance,
means that the collective coordinate integration measure can be fixed
uniquely \cite{measure1,DHKM,KMS}. On the other hand, a host of new
exact solutions to the low-energy dynamics of models with extended
supersymmetry have been proposed \cite{SeibWitt,SW3D} which have the
property that physical measurables receive contributions from all
orders in multi-instantons. These testable predictions have provided
fresh impetus for explicit ADHM calculations \cite{MO-I,MO-II,AOYAMA}
which in certain cases have suggested emendations to the original
solutions.

As the present paper demonstrates, the situation improves even more
dramatically when supersymmetric ADHM calculus is accompanied by the
large-$N$ limit.  Common lore holds that instanton physics is
unimportant in this limit \cite{WITOLD}; after all, a single instanton
is weighted by the factor $\exp(-8\pi^2/g^2)=\exp(-8\pi^2N/(g^2N))$,
which vanishes rapidly as $N\rightarrow\infty$ with fixed 't~Hooft
coupling $g^2N$. In the present paper, we have been considering
quantities which get all their contributions from instantons, except
for tree and one-loop effects. They are indeed tiny contributions at large $N$, as the standard
lore would have it, but nevertheless non-zero and with a precise, calculable, form
fixed by the AdS/CFT correspondence.

As we have seen, with the additional ingredient of large $N$, the
multi-instanton series can be evaluated explicitly, for the first time
in a four-dimensional theory.  We envision that the simplifications in
multi-instanton calculus in this limit will have widespread
applications, and are therefore worth emphasizing. Let us enumerate
the various lessons we have learned at large $N$:

(i) A significant calculational advantage of the gauge-invariant
measure introduced in Sec.~IV.3 is that the ADHM constraints
\eqref{fconea}, which are quadratic constraints in terms of the
original collective coordinates $w,$ become \it linear \rm constraints
(hence trivially resolvable) in terms of the gauge-invariant variables $W.$
However, the gauge-invariant form of the measure is only
available for $k\le N/2$ (see Eqs.~\eqref{Kdef} and \eqref{uptri}).
The large-$N$ limit (with $k$ fixed, or at least $k$ growing more
slowly than $N/2$) is therefore the regime in which these nonlinear
constraints---which, we reiterate, have been the main technical
impediment to progress in ADHM calculus---entirely disappear.

(ii) The detailed saddle-point analysis of Secs.~V.2-V.3 confirms
ones initial intuition regarding $k$-instanton configurations at large
$N$, namely, that the dominant configurations consist of $k$ single
instantons embedded in mutually commuting $SU(2)$ subgroups of
$SU(N)$.  Since this conclusion follows from statistics alone, we
expect it to hold equally for other gauge theories at large $N$. In
this respect the large-$N$ solutions are dilute-gas-like as, in the
absence of other component fields, instantons that live in mutually commuting
$SU(2)$'s do not interact.

(iii) More surprising---and perhaps special to the $\N=4$ theory at
large $N$---is that these $k$ instantons live on top of one another in
space-time, and share a common scale size. In this respect, and in
sharp contrast to (ii), the large-$N$ limit is the \it opposite \rm
of the dilute instanton gas regime.  From the seemingly schizophrenic
features (ii) and (iii) there follows a pleasing simplification
unique to large $N$, discussed in Sec.~VI.1: classical
insertions into gauge-invariant correlators can simply be replaced by
$k$ times the corresponding one-instanton insertion.

(iv) Another remarkable feature of large $N$ is that the auxiliary
variables $\chi_a$, $a=1,\ldots,6$, introduced merely as a
mathematical convenience in order to bilinearize the four-fermion
interaction \eqref{Skquadef}, become confined to $S^5$ in the limit
$N\rightarrow\infty$. As such, they provide a ``window'' from the
Yang-Mills theory into the ten-dimensional geometry of the string
theory. 

(v) Finally, the $1/N$ expansion justifies the truncation at the
quartic level of the small-fluctuations action about the $AdS_5\times
S^5$ saddle-point. One of the principal results of this paper is that
this truncated action precisely describes ${\cal N}=1$ ten-dimensional
Yang-Mills theory dimensionally reduced to zero dimensions. Since the
latter theory is also the D-instanton measure \eqref{ukpfn}, this is
yet another unexpected window from large-$N$ Yang-Mills theory into
ten-dimensional string theory.

\subsection{Comments on the nonrenormalization theorem, and on 
higher-order corrections in $1/N$ and in $g^2$}

As discussed in Sec.~I.1, the comparison between the Yang-Mills and
supergravity pictures elucidated herein can be quantitative if and
only if there exists a nonrenormalization theorem that allows one to
relate the small $g^2N$ to the large $g^2N$ behavior of chiral
correlators such as $G_n$, as has been suggested in
Refs.~\cite{DKMV}.  In the absence of such a theorem the best one
can hope for is that qualitative features of the agreement persist
beyond leading order while the exact numerical factor in each
instanton sector does not, in analogy with the mismatch in the
numerical prefactor between weak and strong coupling results for
black-hole entropy \cite{GKP}.  In our view, however, our present
results provide strong evidence in favor of such a nonrenormalization
theorem for the correlators $G_n$, for the following reason.  Consider
the planar diagram corrections to the leading semiclassical 
(i.e.~$g^2N\rightarrow0$) result for, say, $G_{16}$, Eq.~\eqref{finalans}.
In principle, these would not only modify the above result by an
infinite series in $g^2N,$ but also, at each order in this expansion,
and independently for each value of $k$, they would produce a
different function of space-time. The fact that the leading
semiclassical form for $G_{16}$ that we obtain is not only
$k$-independent, but already reproduces the space-time dependent of the
D-instanton/supergravity prediction exactly, strongly suggests that such 
diagrammatic corrections (planar and otherwise) must  vanish.
Nevertheless there are necessary subleading corrections, both in $1/N$
and in $g^2$, to our leading semiclassical results, as follows.  

In the one-instanton sector, the complete series expansion in $1/N$,
at fixed leading order in $g$, is encapsulated in the exact expression
\eqref{CNdef}. At the multi-instanton level, the analogous $1/N$
corrections are encoded instead in the exact formula for the measure
in \eqref{gaugeinvmeas}. It may be that the variables $W^0$ can be
integrated out (as at large $N$) to leave a dimensional reduction of a
ten-dimensional ${\cal N}=1$ supersymmetric Yang-Mills theory with a
generalized action that is some non-trivial function of $N$. The
action could then be expanded in $1/N$ with a leading term that is the
conventional $SU(k)$ Yang-Mills action of \eqref{sukpart}. It is conceivable
that the more general action is of the Born-Infeld type \cite{TSEY}.

Finally let us consider the source of $g^2$ corrections. As is clear
from Eqs.~\eqref{fexpand}-\eqref{Gdef}, these corrections are
absolutely necessary if (as we fully expect) the Yang-Mills expression
for the correlator $G_n$ is to sum to the complete modular form
$f_n(\tau,\bar\tau)$, and not merely reproduce the leading
semiclassical approximation thereto. Yet how can such corrections be
compatible with the nonrenormalization theorem postulated above? We
believe that the $g^2$ corrections arise, not from conventional
Feynman diagrams in the multi-instanton background, but rather from
the fact that the instanton supermultiplet constructed herein is \it
not \rm an exact solution to the coupled Euler-Lagrange
equations. This important point, which we stressed at the beginning of
Sec.~II, stems from the fact that we have included for convenience all
the fermion zero modes in the collective coordinate matrix $\M^A$, and
not just the exact supersymmetric and superconformal modes. As
discussed in Sec.~II, this approach has the calculational advantage
that the quadrilinear action \eqref{Skquadef} is automatically
generated at leading order.  But it also implies that the coupled
Euler-Lagrange equations must be solved iteratively, order by order in
$g$. In particular, at the next order, the antigauginos $\lambdabar^A$
turn on, as do the auxiliary superfield components $D$ and $F$; these,
in turn, back-react on the classical component fields already
constructed in Sec.~II, resulting in corrections to the
multi-instanton action. We expect these corrections to sum to the
appropriate modular forms, but such checks lie beyond the scope of the
present paper. However, it seems likely that these corrections will
be much simpler than ordinary perturbation theory in the instanton
background.


\section*{Acknowledgments}

The UKSQCD collaboration thanks Eva Silverstein,
Michael Peskin, Tom Banks and Michael Green for comments and discussions.
Much of the review material in Sec.~II, included here for the reader's
convenience, was taken from Ref.~\cite{KMS} which was written in
collaboration with Matt Slater.
ND, VVK and MPM acknowledge a NATO Collaborative Research Grant,
ND, TJH and VVK acknowledge the TMR network grant FMRX-CT96-0012
and SV acknowledges a PPARC Fellowship for support.

\startappendix

\rsen
\Appendix{Clifford Algebras and fermions in 4, 6 and 10 dimensions}

In the appendix we establish connections between the Clifford algebras
and fermions in various dimensions of interest to us. Our conventions
for such representations are taken from \cite{STRATHDEE}.

To begin with we establish representations for the $d=4$ and $d=6$
Clifford algebras in Euclidean space.\footnote{As such we shall not
distinguish between upper and lower vector indices.} 
First of all, in $d=4$ we take the representation of the gamma
matrices from Wess and Bagger \cite{WB}:
\begin{equation}
\gamma_n=\begin{pmatrix}0&\sigma_n\\
\bar\sigma_n&0\end{pmatrix},\qquad\gamma_5=\begin{pmatrix}1&0\\
0&-1\end{pmatrix}\, ,
\end{equation}
where we implicitly assume the analytic continuation to Euclidean
space $(\sigma^0,\sigma^i)\rightarrow(\sigma^0,i\sigma^i)$.
In $d=6$ we introduce an analogous representation of the form 
\begin{equation}
\hat\gamma_a=\begin{pmatrix}0&\Sigma_a\\
\bar\Sigma_a&0\end{pmatrix},\qquad\hat\gamma_7=\begin{pmatrix}-1&0\\
0&1\end{pmatrix}\, ,
\end{equation}
where the $4\times 4$ dimensional matrices $\Sigma_a$ and
$\bar\Sigma_a$, $a=1,\ldots,6$, are most conveniently defined in terms
of the `t~Hooft eta symbols \cite{tHooft}:
\begin{equation}\begin{split}
\Sigma^a_{AB}&=\big(\eta^c_{AB},i\bar\eta^c_{AB}\big),\\
\bar\Sigma^{AB}_a&=\big(-\eta^c_{AB},i\bar\eta^c_{AB}\big),
\end{split}\end{equation}
where \cite{tHooft} for $c=1,2,3$: 
\begin{align}
&\bar\eta_{AB}^c=\eta_{AB}^c=\epsilon_{cAB}\qquad A,B\in\{1,2,3\}, \notag\\
&\bar\eta_{4A}^c=\eta_{A4}^c=\delta_{cA},\elabel{E15.1}\\
&\eta_{AB}^c=-\eta_{BA}^c,\qquad\bar\eta_{AB}^c=-\bar\eta_{BA}^c.\notag
\end{align}

In both $d=4$ and $d=6$, we can define the associated quantities
\begin{equation}\begin{split}
\gamma_{nm}&=\tfrac{i}4[\gamma_n,\gamma_m]=i\begin{pmatrix}\sigma_{mn}&0\\
0&\bar\sigma_{nm}\end{pmatrix}\, ,\\
\hat\gamma_{ab}&=\tfrac{i}4[\hat\gamma_a,\hat\gamma_b]=i\begin{pmatrix}\Sigma_{ab}&0\\
0&\bar\Sigma_{ab}\end{pmatrix}\, . 
\end{split}\end{equation}
The charge conjugation matrices are
\begin{equation}\begin{split}
C^{(4)}&=\gamma_1\gamma_3=\begin{pmatrix}-i\sigma^2&0\\
0&-i\sigma^2\end{pmatrix}\, ,\\
C^{(6)}&=\hat\gamma_4\hat\gamma_5\hat\gamma_6=\begin{pmatrix}0&-i\\ -i&0
\end{pmatrix} \, .
\end{split}\end{equation}

{}From these representations in 4 and 6 dimensions, we wish to build
representations of the $d=10$ Clifford. We shall need two different
representations for Secs.~4.2 and 5.2.
First of all we must explain some of the subtleties of defining
Majorana-Weyl fermions in $d=10$.
Majorana fermions can only be properly defined in Minkowski space in
$d=10$, so we start with a Hermitian representation $\Gamma_\mu$,
$\mu=0,\ldots,9$, of the $d=10$ Clifford algebra.
The coupling of a 32 components
fermion to a vector is given by $\bar\Psi\Gamma_\mu\Psi$. 
The Majorana condition on fermions is
$\Psi=C\bar\Psi^T$, with $\bar\Psi=\Psi^\dagger\Gamma_0$,
where $C$ is the charge conjugation matrix defined such that for the 
Euclidean space gamma matrices ($\Gamma_0\rightarrow-i\Gamma_0$) 
we have $\Gamma_\mu^*=-C^{-1} \Gamma_\mu C$. At this stage we can continue back to
Euclidean space and simply treat the Euclidean Majorana fermion $\Psi$
as real (since we never have to make use of its complex conjugate).
In Euclidean space the coupling of a Majorana fermion to a vector is
\begin{equation}
\Psi^T(C^{-1})^T\Gamma_\mu\Psi.
\end{equation}
This coupling is not actually
real, and the resulting Hamiltonian is not Hermitian. However, this is
not inconsistent because the relevant requirement in Euclidean quantum
field theory is Osterwalder-Schrader reflection positivity rather than
Hermiticity. From now on we work in Euclidean space.

The charge conjugation matrix in $d=10$ is
\begin{equation}
C=C^{(6)}\otimes C^{(4)}=\begin{pmatrix}0&-i\\ -i&0
\end{pmatrix}\otimes
\begin{pmatrix}-i\sigma^2&0\\ 0&-i\sigma^2\end{pmatrix}\, ,
\end{equation}
The representation of the  $d=10$
Clifford algebra needed in Sec.~4.2, is built as follows:
\begin{equation}
\Gamma_{n}= 1_{\sst [8]\times [8]} \otimes \gamma_{n},
\qquad  
\Gamma_{a+3}= \hat{\gamma}_{a} \otimes \gamma_{5}, 
\elabel{gamma}
\end{equation}
where $n=0,\ldots,3$ and $a=1,\ldots,6$. In this representation 
\begin{equation} 
\Gamma_{11}=\hat{\gamma}_{7} \otimes \gamma_{5}\, .
\elabel{gamma11}
\end{equation}
A Majorana-Weyl spinor of positive chirality in $d=10$ satisfies 
$\Gamma_{11}\Psi=\Psi$. In this basis, $\Psi$ can be decomposed as, 
\begin{equation}
\Psi = \sqrt{\frac{\pi}{2}}
\begin{pmatrix} 0 \\ 1\end{pmatrix}\,\otimes\, 
\begin{pmatrix} {\cal M}^{\prime A}_{\alpha}  \\ 
0 \end{pmatrix} \,\,\, +\,\,\, \sqrt{\frac{\pi}{2}}
\begin{pmatrix} 1 \\ 0\end{pmatrix}\,\otimes\, 
\begin{pmatrix} 0  \\ 
\lambda_{A}^\aD \end{pmatrix}\, ,
\elabel{cov}
\end{equation}  
where ${\cal M}'$ and $\lambda$
are Weyl spinors both of $SO(4)$ and of $SO(6)$. ${\cal M'}$ and
$\lambda$ have positive and negative chirality with respect to both groups, respectively.

The representation of the $d=10$ Clifford algebra that is needed in
Sec.~5.3 is somewhat different. The fermion coupling \eqref{fermc} 
can be written precisely as in Eq.~\eqref{actymf} with
a rotated representation of the six-dimensional gamma matrices that depends on
$\sfc_a$. We define an $\SO(6)$ rotation matrix $R$,
$RR^T=1$, such that $\sfc'_a=R_{ab}\sfc_b$ lies entirely along, say, the first
direction, i.e.~$\sfc'_a\propto\delta_{a1}$. In the new basis, we have a
representation of the $d=6$ Clifford algebra $\hat\gamma'_a=R_{ab}\hat\gamma_b$.

In the rotated basis, we can construct a representation of the 
$d=10$ Clifford algebra as follows:
\begin{equation}
\Gamma_n=\hat\gamma'_1\otimes\gamma_n,\qquad
\Gamma'_{3+a}=\hat\gamma'_a\otimes\big(\delta_{a1}\gamma_5+(1-\delta_{a1})
1_{\sst[4]\times[4]}\big)\, ,
\elabel{rotca}\end{equation}
where $n=0,\ldots,3$ and $a=1,\ldots,6$.
The representation of the $d=10$ Clifford algebra that appears in the
text is then found by un-doing the rotation on the $d=6$ subspace:
\begin{equation}
\Gamma_{3+a}=(R^{-1})_{ab}\Gamma'_{3+b}\, .
\end{equation}
In this basis 
\begin{equation}
\Gamma_{11}=\hat\gamma_7\otimes 1_{\sst [4]\times [4]},
\end{equation}
and a positive chirality Weyl fermion in $d=10$ is decomposed 
as a positive chirality Weyl fermion in $d=6$ which is a Dirac fermion in
$d=2$. With the correct normalization to reproduce the fermion
coupling in the text, the $d=10$ Majorana-Weyl fermion $\Psi$ has
components 
\begin{equation}
\Psi=N^{1/8}\sqrt{\pi\over
2g}\left({0\atop1}\right)\otimes\left({\rho^{-1/2}\hat{\cal
M}^{\prime A}_\alpha\atop\rho^{1/2}\hat\zeta^{\aD A}}\right).
\end{equation}


\begin{thebibliography}{99}

\bibitem{TH1} G. 't Hooft, Nucl. Phys. {\bf B72} (1974) 461 

\bibitem{MAL}{J.~Maldacena, 
Adv. Theor. Math. Phys. {\bf 2:231} (1998), {\tt hep-th/9711200}}


\bibitem{GKP}
S.~Gubser, I.~Klebanov and A.~Polyakov, 
Phys. Lett. {\bf B428} (1998) 105, {\tt hep-th/9802109}

\bibitem{WIT150}E.~Witten, 
Adv. Theor. Math. Phys. {\bf 2:253} (1998),
{\tt hep-th/9802150}


\bibitem{FFone}S. Ferrara and C. Fronsdal, 
Class. Quant. Grav. {\bf15} (1998) 2153, {\tt hep-th/9712239}\\
S. Ferrara, C. Fronsdal and A. Zaffaroni, Nucl. Phys. {\bf B532} (1998) 153,
{\tt hep-th/9802203}


\bibitem{AFone} L. Andrianopoli and S. Ferrara, 
Phys. Lett. {\bf B430} (1998) 248,  {\tt hep-th/9803171}

\bibitem{HO} G Horowitz and H. Ooguri, Phys. Rev. Lett. {\bf 80} (1988) 4116,
{\tt hep-th/9802116}


\bibitem{FZ1}S. Ferrara and A. Zaffaroni, {\sl Bulk gauge fields in
AdS supergravity and supersingletons\/}, {\tt hep-th/9807090}


\bibitem{ADHM} M.  Atiyah, V.  Drinfeld, N.  Hitchin and
Yu.~Manin, Phys. Lett. A65 (1978) 185


\bibitem{LETT} N. Dorey, T. J. Hollowood, V. V. Khoze, M. P. Mattis
and S. Vandoren, {\sl Multi-instantons and Maldacena's Conjecture\/},
{\tt hep-th/9810243}


\bibitem{N4finite}M. Sohnius and P.C. West, Phys. Lett. {\bf B100}
(1981) 245\\
M. Grisaru and W. Siegel, Nucl. Phys. {\bf B201} (1982) 292\\
S. Mandelstam, Nucl. Phys. {\bf B213} (1983) 149\\
K.S. Stelle and P.K. Townsend, Nucl. Phys. {\bf B214} (1983) 519;
Nucl. Phys. {\bf B236} (1984) 125\\
L. Brink, O. Lindgren and B. Nilsson, Phys. Lett. {\bf B123} (1983) 323

\bibitem{MOOS}C. Montonen, D. Olive, Phys. Lett. {\bf B72} (1977)
117\\
H. Osborn, Phys. Lett. {\bf B83} 321,(1979) 321 

\bibitem{SEN}A. Sen, Phys. Lett. {\bf B329} (1994) 217, {\tt hep-th/9402032}\\
J.P. Gauntlett and D.A. Lowe, Nucl. Phys. {\bf B472}
(1996) 194, {\tt hep-th/9601085}\\ 
E.J. Weinberg and P. Yi, Phys. Lett. {\bf B376} (1996) 97, {\tt
hep-th/9601097}\\
C. Fraser and T.J. Hollowood, Phys. Lett. {\bf B402} (1997) 106,
{\tt hep-th/9704011}\\
N. Dorey, C. Fraser, T.J. Hollowood and M.A.C. Kneipp,
Phys. Lett. {\bf B383} (1996) 422, {\tt hep-th/9605069}


\bibitem{EFS}E. D'Hoker, D.Z. Freedman and  W. Skiba, {\sl Field theory
tests for correlators in the AdS/CFT correspondence\/}, {\tt hep-th/9807098}


\bibitem{HW}S. Howe and P.C. West, {\sl Nonperturbative Green functions
in theories with extended superconformal symmetry\/}, {\tt hep-th/9509140};
Phys. Lett. {\bf B400} (1997) 307, {\tt hep-th/9611075}



\bibitem{OSII}H. Osborn, {\sl N=1 superconformal symmetry in
four-dimensional quantum field theory\/},  {\tt hep-th/9808041}


\bibitem{INT}K. Intriligator, {\sl Bonus symmetry of N=4
superYang-Mills correlation functions via AdS daulity\/}, {\tt hep-th/9811047}


\bibitem{SV}M.A. Shifman and A.I. Vainshtein and
V.I. Zakharov, Nucl. Phys. {\bf B277} (1986) 456

\bibitem{Seib} N. Seiberg, Phys. Lett. {\bf B206} (1988) 75


\bibitem{Shifman}  M. Shifman,
Prog. Part. Nucl. Phys. {\bf 39} (1997) 1, {\tt hep-th/9704114}

\bibitem{IS} K. Intriligator and N. Seiberg, {\sl Lectures on Supersymmetric
Gauge Theories and Electric-magnetic Duality\/},
{\tt hep-th/9509066}


\bibitem{Ads}I. Affleck, M. Dine and N. Seiberg, Nucl. Phys. {\bf
B241} (1984) 493;  Nucl. Phys. {\bf
B256} (1985) 557



\bibitem{Cordes}S. Cordes, Nucl. Phys. {\bf B273} (1986) 629


\bibitem{NSVZbeta}V.A. Novikov, M.A. Shifman, A.I. Vainshtein and
V.I. Zakharov, Nucl. Phys. {\bf B229} (1983) 381

\bibitem{SeibWitt}N. Seiberg and E. Witten, Nucl. Phys. {\bf B426} (1994) 19,
(E) {\bf B430} (1994) 485, {\tt hep-th/9407087} 


\bibitem{MO-I} N.~Dorey, V.V.~Khoze and M.P.~Mattis,
Phys. Rev. {\bf D54} (1996) 2921, {\tt hep-th/9603136}

\bibitem{SW3D} N. Seiberg and E. Witten, 
{\sl Gauge Dynamics and Compactification to Three Dimensions}, 
in {\it The Mathemetical Beauty of Physics},
p.333, Eds. J. M. Drouffe and J.-B Zuber (World Scient., 1997),
{\tt hep-th/9607163} 

\bibitem{DKMTV} N. Dorey, V. V. Khoze, M. P. Mattis, D. Tong and 
S. Vandoren, Nucl. Phys. {\bf B502} (1997) 59. 

\bibitem{PP} J. Polchinski and P. Pouliot, Phys. Rev. {\bf D56} (1997) 
6601, {\tt hep-th/9704029}

\bibitem{DKM3D} N. Dorey, V. V. Khoze and M. P. Mattis, Nucl. Phys. 
{\bf B502} (1997) 94, {\tt hep-th/9704197}

\bibitem{PSS} S. Paban, S. Sethi and M. Stern, {\sl Summing up Instantons 
in Three-Dimensional Yang-Mills Theory}, {\tt hep-th/9808119}

\bibitem{VW} C. Vafa and E. Witten, {  Nucl. Phys.} {\bf B431} (1994) 3

\bibitem{BG}{T.~Banks and M.B.~Green, J. High Energy Phys. {\bf05:002}
(1998), {\tt hep-th/9804170}}

\bibitem{BGKR}{M.~Bianchi, M.B.~Green, S.~Kovacs and G.~Rossi, J. High
Energy Phys. {\bf9808:013} (1998), {\tt hep-th/9807033}}

\bibitem{BK}M. Bianchi and S. Kovacs, {\sl Yang-Mills instantons
vs. type IIB D-instantons\/}, {\tt hep-th/9811060}


\bibitem{GG1}M.B.~Green and M.~Gutperle, 
Nucl. Phys. {\bf B498} (1997) 195, {\tt hep-th/9701093}

\bibitem{GG2}M.B.~Green and M.~Gutperle, 
J. High Energy Phys. {\bf 9801:005} (1998),  {\tt hep-th/9711107}

\bibitem{GG3}M.B.~Green and M.~Gutperle, Phys. Rev. {\bf D58:046007} (1998), 
{\tt hep-th/9804123}

\bibitem{GG4}M.B. Green and M. Gutperle, Phys. Lett. {\bf B398} (1997)
69, {\tt hep-th/9612127}

\bibitem{GGK}M.B.~Green, M.~Gutperle and H. Kwon,
Phys. Lett. {\bf B421} (1998) 149,
{\tt hep-th/9710151}

\bibitem{KP1}A. Kehagias and H. Partouche, Phys. Lett. {\bf B422} (1998) 109, {\tt
hep-th/9710023}

\bibitem{KP2}A. Kehagias and H. Partouche, Int. J. Mod. Phys. {\bf A13} (1998) 5075,
{\tt hep-th/9712164}

\bibitem{BRW} E. Bergshoeff, M. de Roo and B. de Wit, 
{  Nucl. Phys.} {\bf B182} (1981) 173

\bibitem{DKMV}{N.~Dorey, V.V.~Khoze, M.P.~Mattis and S.~Vandoren,
Phys. Lett. {\bf B442} (1998) 145, {\tt hep-th/9808157}}


\bibitem{KMS}{V.V.~Khoze, M.P.~Mattis and J.~Slater, Nucl. Phys. {\bf
B536} (1998) 69, {\tt hep-th/9804009}}


\bibitem{CGTone}E. Corrigan, D. Fairlie, P. Goddard and S. Templeton,
Nucl. Phys. B140 (1978) 31\\ 
E. Corrigan, P. Goddard and S. Templeton,
Nucl. Phys. B151 (1979) 93

\bibitem{MO-II} N. Dorey, V. V. Khoze and M. P. Mattis, Phys. Rev.
{\bf D54} (1996) 7832, {\tt hep-th/9607202}

\bibitem{DKMn4}{N.~Dorey, V.V.~Khoze and M.P.~Mattis, 
Phys. Lett. {\bf B396} (1997) 141, {\tt hep-th/9612231}}

\bibitem{DHKM}{N.~Dorey, T.J.~Hollowood, V.V.~Khoze and M.P.~Mattis,
Nucl. Phys. {\bf B519} (1998) 470, {\tt hep-th/9709072}}

\bibitem{D1} M. Douglas, {\sl Branes within Branes\/}, {\tt hep-th/9512077}

\bibitem{D2} M. Douglas, {\sl Gauge Fields and D-branes}, 
{\tt hep-th/9604198} 

\bibitem{W1} E, Witten, Nucl. Phys. {\bf B460} (1996) 541, {\tt hep-th/9511030}


\bibitem{AKH} E. T. Akhmedov, {\sl D-instantons Probing D3-branes and 
the AdS/CFT correspondence\/}, {\tt hep-th/9812038}


\bibitem{WIT112}{E.~Witten, J. High Energy Phys. {\bf 9807:006} (1998),
 {\tt hep-th/9805112}}


\bibitem{GN} D. J. Gross and A. Neveu, {  Phys. Rev.} {\bf D10} 
(1974) 3235\\
K. Wilson, Phys. Rev. {\bf D7} (1973) 2911


\bibitem{KNS}{W.~Krauth, H.~Nicolai and M.~Staudacher,
Phys. Lett. {\bf B431} (1998) 31, {\tt hep-th/9803117}\\
W.~Krauth and M.~Staudacher, Phys. Lett. {\bf B435} (1998) 350 , {\tt hep-th/9804199}} 

\bibitem{MNS}{G.~Moore, N.~Nekrasov and S.~Shatashvili, {\sl D
particle Bound States and Generalized Instantons\/}, {\tt
hep-th/9803265}}

\bibitem{part}I.K. Kostov and P. Vanhove, Phys. Lett. {\bf B444}
(1998) 196, {\tt hep-th/9809130}

\bibitem{IKKT}N. Ishibashi, H. Kawai, Y. Kitazawa and A. Tsuchiya,
Nucl. Phys. {\bf B498} (1997) 467, {\tt hep-th/9612115} 

\bibitem{AFGJ}D. Anselmi, D. Freedman, M. Grisaru and A. Johansen,
Phys. Lett. {\bf B394} (1997) 329, {\tt hep-th/9708042}


\bibitem{FMMR}D.Z. Freedman, S.D. Mathur, A. Matusis and L. Rastelli,
{\sl Correlation functions in the $CFT(D)/ADS(D+1)$ correspondence},
{\tt hep-th/9804058}; {\sl Comments on 4-point functions in the CFT/AdS
correspondence}, {\tt hep-th/9808006}





\bibitem{LMRS}S. Lee, S. Minwalla, M. Rangamani and N. Seiberg, {\sl
Three point functions of chiral operators in D=4, N=4 SYM at large N\/},
{\tt hep-th/9806074}


\bibitem{CNSS} G. Chalmers, H. Nastase, K. Schalm and R. Siebelink, {\sl R current
correlators in N=4 super Yang-Mills theory from anti-de Sitter
gravity\/}, {\tt hep-th/9805105}

\bibitem{LT}H. Liu and A.A. Tseytlin, {\sl On four point functions in the
CFT/AdS correspondence\/}, {\tt hep-th/9807097}


\bibitem{MV}W. Muck and K. S. Viswanathan, Phys. 
Rev. {\bf D58:041901} (1998), {\tt hep-th/9804035};
Phys. Rev. {\bf D58:106006} (1998), {\tt hep-th/9805145}


\bibitem{BGut}J.H. Brodie and M. Gutperle, {\sl String corrections to
four point functions in the AdS/CFT correspondence\/}, {\tt hep-th/9809067} 


\bibitem{AF1}G.E. Arutyunov and S.A. Frolov, {\sl On the origin of
supergravity boundary terms in the AdS/CFT correspondence\/}, {\tt hep-th/9806216}


\bibitem{SOLOD}S.N. Solodukhin, Nucl. Phys. {\bf B539} (1999) 403,
{\tt hep-th/9806004}


\bibitem{GHEZ}A.M. Ghezelbash, Phys. Lett. {\bf B435} (1998) 291,
{\tt hep-th/9805162}



\bibitem{GSI}M.B. Green and S. Sethi, {\sl Supersymmetry constraints on
type IIB supergravity\/}, {\tt hep-th/9808061}



\bibitem{HS1}M. Henningson and K. Sfetsos, Phys. Lett. {\bf B431}
(1998) 63, {\tt hep-th/9803251}


\bibitem{WB}{J.~Wess and J.~Bagger, {\sl Supersymmetry and
Supergravity\/}, Princeton University Press, 1992}


\bibitem{tHooft}G.~'t~Hooft, Phys. Rev. {\bf D14} (1976) 3432; ibid.
(E) {\bf D18} (1978) 2199


\bibitem{Affleck}{I. Affleck, Nucl. Phys. {\bf B191} (1981) 429}


\bibitem{CWS}{N.H. Christ, E.J. Weinberg and N.K. Stanton,
Phys. Rev. {\bf D18} (1978) 2013}


\bibitem{OSB}H. Osborn, Ann. of Phys. {\bf135} (1981) 373

\bibitem{BCGW}C. Bernard, N.H. Christ, A. Guth and E.J. Weiberg,
Phys. Rev. {\bf D16} (1977) 2967

\bibitem{BPST}A. Belavin, A. Polyakov, A.Schwartz and Y. Tyupkin,
Phys. Lett. {\bf B59} (1975) 85

\bibitem{NSVZ}V.A. Novikov, M.A. Shifman, A.I. Vainshtein and
V.I. Zakharov, Nucl. Phys. {\bf B229} (1983) 394; 
Nucl. Phys. {\bf B229} (1983) 407;
 Nucl. Phys. {\bf B260} (1985) 157 


\bibitem{Corrup}{E. Corrigan, {\it unpublished}}

\bibitem{AOYAMA}H. Aoyama, T. Harano, M. Sato and S. Wada,
Phys. Lett. {\bf B388} (1996) 331, {\tt hep-th/9607076}


\bibitem{measure1}{N.~Dorey, V.V.~Khoze and M.P.~Mattis,
Nucl. Phys. {\bf B513} (1998) 681, {\tt hep-th/9708036}}

\bibitem{dadda} A. D'Adda and P. Di Vecchia, 
 Phys. Lett.  {\bf 73B} (1978) 162 


\bibitem{Bernard}{C.~Bernard, Phys. Rev. {\bf D19} (1979) 3013}

\bibitem{P}J. Polchinski, {\sl Notes on D-branes\/}, {\tt
hep-th/9602052}

\bibitem{W135} E, Witten, Nucl. Phys. {\bf B460} (1996) 335, 
{\tt hep-th/9510135}



\bibitem{CHS} C. Callan, J. Harvey and A. Strominger, 
{  Nucl. Phys.} {\bf B367} (1991) 60 

\bibitem{A} O. Aharony, M. Berkooz, S. Kachru, N. Seiberg and E. Silverstein, 
Adv. Theor. Math. Phys. {\bf 1} (1998) 148, {\tt hep-th/9707079}

\bibitem{W2} E. Witten, J. High Energy Phys. {\bf 07} (1997) 003, {\tt hep-th/9707093}

\bibitem{GILMORE}R. Gilmore, {\sl Lie groups, Lie algebras and some of
their applications\/}, Wiley-Interscience 1974



\bibitem{AG} Alvarez-Gaum\'e,
Commun. Math. Phys. {\bf 90} (1983) 161


\bibitem{KS}{V.E. Korepin and S.L. Shatashvili,
Sov. Phys. Dokl {\bf 28} (1983) 1018}


\bibitem{SSI}E. Shuryak and T. Schafer,
Nucl. Phys. Proc. Suppl. {\bf53} (1997) 472


\bibitem{WITOLD}E. Witten, Nucl. Phys. {\bf B149} (1979) 285


\bibitem{TSEY}A.A. Tseytlin, Nucl. Phys. {\bf B501} (1997) 41, {\tt hep-th/9701125 }


\bibitem{STRATHDEE}{J.~Strathdee, Int. J. Mod. Phys. {\bf A2} (1987) 273}







\end{thebibliography}
\end{document}